%% file: paper.tex
\documentclass[a4paper,fleqn,usenatbib]{mnras}

\usepackage{newtxtext}
\usepackage{newtxmath}

\usepackage[T1]{fontenc}
\usepackage{ae,aecompl}

\usepackage{graphicx}	
\usepackage{amsmath}	
\usepackage{scrextend}

\usepackage{ulem}

\title[QUIJOTE study of W49, W51 and IC443]{QUIJOTE scientific results -- V. The microwave intensity and polarisation spectra of the Galactic regions W49, W51 and IC443}

\input{authors_and_institutions.tex}

\date{Accepted 2022 November 25. Received 2022 November 20; in original form 2022 July 28}

\pubyear{2021}

\begin{document}
\label{firstpage}
\pagerange{\pageref{firstpage}--\pageref{lastpage}}
\maketitle

\begin{abstract}
	We present new intensity and polarisation maps obtained with the QUIJOTE experiment towards the Galactic regions W49, W51 and IC443, covering the frequency range from 10 to 20\,GHz at $\sim1\,\text{deg}$ angular resolution, with a sensitivity in the range 35--79\,$\mu\text{K}\,\text{beam}^{-1}$ for total intensity and 13--23\,$\mu\text{K}\,\text{beam}^{-1}$ for polarisation. For each region, we combine QUIJOTE maps with ancillary data at frequencies ranging from 0.4 to 3000\,GHz, reconstruct the spectral energy distribution and model it with a combination of known foregrounds. We detect anomalous microwave emission (AME) in total intensity towards W49 at $4.7\sigma$ and W51 at $4.0\sigma$ with peak frequencies $\nu_{\rm AME}=(20.0\pm1.4)\,\text{GHz}$ and $\nu_{\rm AME}=(17.7\pm3.6)\,\text{GHz}$ respectively; this is the first detection of AME towards W51. The contamination from ultra-compact HII regions to the residual AME flux density is estimated at 10\% in W49 and 5\% in W51, and does not rule out the AME detection. The polarised SEDs reveal a synchrotron contribution with spectral indices $\alpha_{\rm s}=-0.67\pm0.10$ in W49 and $\alpha_{\rm s}=-0.51\pm0.07$ in W51, ascribed to the diffuse Galactic emission and to the local supernova remnant respectively. Towards IC443 in total intensity we measure a broken power-law synchrotron spectrum with cut-off frequency $\nu_{\rm 0,s}=(114\pm73)\,\text{GHz}$, in agreement with previous studies; our analysis, however, rules out any AME contribution which had been previously claimed towards IC443. No evidence of polarised AME emission is detected in this study.
\end{abstract}





\section{Introduction}
\label{sec:introduction}

During the past two decades the study of the cosmic microwave background (CMB) 
anisotropies allowed to estimate cosmological parameters to the per cent 
precision level, and led to the establishment of the Lambda-Cold Dark Matter 
($\Lambda$CDM) cosmology as the most widely accepted paradigm describing our 
Universe~\citep{hinshaw13, planck_18_i}. The most recent endeavors are focussing 
on the characterisation of CMB polarisation, in particular on the search for the 
inflationary B-mode anisotropies~\citep{kamionkowski97, zaldarriaga97}. However, 
to date only upper limits on the level of primordial B-modes anisotropies are 
available~\citep{planck_18_vi,bicep21,bicep_planck21}, as this signal is subdominant 
compared to other polarised Galactic and extra-Galactic emissions. An accurate 
modelling of these foregrounds becomes then of paramount importance, not only to 
produce clean CMB maps for their cosmological exploitation, but also to better 
characterise the physical properties of the interstellar medium (ISM).

In total intensity, the main foregrounds affecting any CMB observations are 
synchtrotron radiation generated by cosmic ray electrons spiralling in the 
Galactic magnetic field, free-free emission from electron thermal 
\textit{bremsstrahlung}, thermal radiation from insterstellar dust, and the so-called 
anomalous microwave emission~\citep[AME,][]{planck_15_xxv}; to this list of continuum 
emissions we can also add the contamination from molecular lines (mostly CO). While the 
mechanisms responsible for the first three continuum foregrounds are physically well 
understood, the nature of AME is still under debate~\citep[for a detailed review we 
redirect to][]{dickinson18}. This emission, first discovered in the mid-90s~\citep{leitch97}, 
appears at frequencies $\sim$10--60\,GHz as a dust-correlated signal that cannot be 
explained in terms of the other foreground components. The most accredited model 
explaining its origin is the dipole radiation from small, fast spinning dust grains in 
the ISM~\citep{draine98, ali_haimoud09, hoang10, ysard10, silsbee11, ali_haimoud13}; 
this scenario is referred to as spinning dust emission (SDE). 
A different model known as magnetic dust emission (MDE) has also been proposed, 
according to which the thermal emission from magnetised ISM dust grains is the 
mechanism responsible for AME~\citep{draine99, draine13}. More recently, models based 
on thermal emission from amorphous dust grains proved effective in yielding the excess
AME signal at microwave frequencies~\citep{nashimoto20}. Many observational efforts 
have tackled the detection of AME in the attempt of better characterising its 
nature~\citep{de_oliveira_costa98,de_oliveira_costa99,finkbeiner02,watson05,
casassus06,davies06,ami_consortium09,dickinson09,murphy10,tibbs10,planck_er_xx,vidal11,
genova_santos11,planck_ir_xv,battistelli15,vidal20}. However, up to date the results are 
not clear; a promising way of distinguishing between different AME models is the observation 
of its polarisation.

Out of the aforementioned foregrounds, the free-free emission is known to be 
practically unpolarised\footnote{Strictly speaking, the emission from one accelerated 
charge is intrinsically polarised; the observed free-free foreground, however, is 
generated by the superposition of the emissions from a population of electrons with 
random velocity directions, which erases the overall polarisation 
fraction.}~\citep{rybicki79, trujillo_bueno02}, whereas synchrotron and 
thermal dust emissions yield respectively polarisation fractions $\Pi$ up to 40\% and 20\% 
in some regions of the sky~\citep{kogut07, vidal15, planck_15_xxv, planck_18_xi, planck_18_xii}. 
On the contrary, the 
polarisation properties of AME are still under debate. Theoretically, predictions have 
been made for the spinning dust model; although the dipole emission from an 
individual dust grain is expected to be linearly polarised, the presence of a detectable 
polarisation fraction in the observed SDE strongly depends on the spatial alignment 
of ISM grains. Studies of the conditions for grain alignment have been addressed 
in the literature~\citep{lazarian00, hoang13} and generally place upper limits for 
the SDE polarisation fraction at the per cent level. The work by~\citet{draine16},
however, suggests that at $\nu\gtrsim 10\,\text{GHz}$ quantum effects prevent the 
dissipation of the grain rotational energy, thereby erasing the alignment and 
resulting in negligible polarisation levels ($\Pi \ll 1\%$). The theoretical study of
MDE polarisation has also been tackled:~\citet{draine99} and~\citet{draine13} found 
that in the case of free-flying magnetized grains the resulting polarisation fraction
can be as high as 30\%; a more recent study described in~\citet{hoang16}, however, 
set the upper limits for the polarised emission fraction from free-flying iron 
nanoparticles at $\sim5\%$. Similar constraints were obtained in the case of magnetic
inclusions within larger, non-magnetic grains~\citep{draine13}. The MDE is 
then expected to provide higher polarisation fractions compared to the SDE scenario.
The amorphous dust models, finally, found that the AME polarisation is suppressed 
by the contribution from amorphous carbon dust, accounting for nearly half of the 
ISM grains~\citep{nashimoto20}.

Observationally, there are no clear detections of AME in polarisation to date yet. 
Polarised AME has been searched for with dedicated observations towards a number of 
different environments, such as supernova remnants (SNRs), planetary nebulae or molecular 
clouds~\citep{rubino_martin12a}.~\citet{battistelli06} reported a 
tentative detection towards the Perseus molecular complex using the COSMOlogical 
Structures On Medium Angular Scales (COSMOSOMAS) experiment, with an observed 
polarisation fraction at $\Pi=3.4^{+1.5}_{-1.9}\%$ that could be proceeding from AME.
Other works employing different facilites only determined upper limits, usually in 
the range $\Pi\lesssim 1$--6\% ~\citep{casassus08, mason09, macellari11, lopez_caraballo11, 
dickinson11, battistelli15}. These constraints agree in the observed lack of a strong polarised 
AME component, but are still too loose to clearly discriminate between the 
different AME models.

\begin{table*} 
\centering
\caption{Summary of the three sources considered in this study. For 
	each one we report the effective central Galactic coordinates $(l_{\rm c},b_{\rm c})$, 
	the possible substructures, the diametral angular size, the 
	estimated distance and the age of the associated SNR. The last three columns quote the 
	relevant quantities for the aperture photometry analysis described in Section~\ref{sec:seds}, 
	namely the radii of the chosen aperture ($r_{\rm ap}$) and of the internal ($r_{\rm int}$) 
	and external ($r_{\rm ext}$) boundaries of the background annulus.}
\label{tab:sources_info}
\begin{tabular}{ccccccccc}
\hline
\hline
	Source 	& ($l_{\rm c}$,$b_{\rm c}$) & Substructures & Size & Distance  &  SNR age & $r_{\rm ap}$ & $r_{\rm int}$ & $r_{\rm ext}$ \\
	&	[deg]	& & [arcmin] & [kpc] & [kyr]  &  [arcmin]    &  [arcmin]  &  [arcmin]	\\
\hline
\hline
	W49	&  (43.20, $-0.10$) & W49A (thermal), W49B (SNR)       & 30 & $\sim11$ & 1--4 & 60 & 80 & 100 \\
	W51	&  (49.20, $-0.35$) & W51A, W51B (thermal), W51C (SNR) & 50 & $\sim5$  &  30  & 80 & 100 & 120 \\
	IC443	&  (189.06, 3.24) & -			             & 45 & $\sim2$  &  3--30  & 70 & 90 & 110 \\
\hline
\hline
\end{tabular}
\end{table*}  

This observational landscape emphasizes the need for accurate polarised sky surveys, 
in order to assess the foreground level of contamination to existing and future CMB 
observations. Polarised maps from the Wilkinson Microwave Anisotropy 
Probe~\citep[\textit{WMAP},][]{kogut07} and \textit{Planck}~\citep{planck_18_i} have been 
used to derive dust and synchrotron polarised sky models over frequencies ranging from 
22.7\,GHz to 353\,GHz~\citep{Krachmalnicoff16}. Ground-based facilities are capable of 
complementing this range with observations in the lower frequency portion of the microwave 
spectrum; for instance, the Cosmology Large Angular Scale Surveyor (CLASS) is designed to 
observe at 40, 90, 150 and 220\,GHz~\citep{watts15}, while the C-Band All Sky 
Survey~\citep[C-BASS,][]{irfan15} is currently mapping the full sky at 5\,GHz. This  
work focusses on data from the multi-frequency instrument (MFI), the first instrument of the 
Q-U-I JOint TEnerife (QUIJOTE) experiment~\citep{rubino_martin17}, spanning the frequency 
range from 10 to 20\,GHz. QUIJOTE-MFI data have already been used to characterise polarised 
foregrounds;~\citet{genova_santos15b} provided the constraints $\Pi < 10.1\%$ at 12\,GHz and 
$\Pi < 3.4\%$ at 18\,GHz for the AME polarisation fraction (95 per cent C.L.), using QUIJOTE 
data in combination with ancillary observations towards the Perseus molecular complex. A 
similar study conducted over the W43 molecular complex provided the tightest constraints on 
AME polarisation to date, $\Pi <0.39\%$ at 17\,GHz and $\Pi <0.22\%$ at 41\,GHz (95 per cent 
C.L.); these results are reported in~\citet{genova_santos17} together with the detection of 
AME in total intensity towards the molecular complex W47 and the modelling of the synchrotron 
polarised emission from the SNR W44. QUIJOTE data were also employed 
in~\citet{poidevin19} for the characterisation of AME towards the Taurus molecular cloud and 
the L1527 dark cloud nebula, enabling to set upper limits for the polarised fraction at 
$\Pi <3.8\%$ and $\Pi <4.5\%$ (95 per cent C.L.) respectively at 28.4\,GHz. 
Finally,~\citet{cepeda_arroita21} combined QUIJOTE and C-BASS data to study the morphology of 
AME emission in total intensity towards the $\lambda$ Orionis ring. As confirmed by these results, 
data already acquired by QUIJOTE represent a valuable addition to existing radio and microwave 
surveys when modelling both polarised and non-polarised foregrounds.

This work, the fifth in the series of QUIJOTE scientific papers, aims at characterising 
in intensity and polarisation the emission towards three Galactic regions: the SNRs 
with associated molecular clouds W49 and W51, and the SNR IC443. The 
goal is to provide a characterisation of the local synchrotron emission (in particular of its 
spectral index), and to investigate AME intensity and possible polarisation towards SNRs. This 
work is then intended as a continuation of the studies presented
in~\citet{genova_santos15b,genova_santos17} and~\citet{poidevin19}, dedicated to the 
characterisation of astrophysically relevant Galactic regions using QUIJOTE data.

This paper is outlined as follows. In Section~\ref{sec:regions} we describe the Galactic 
regions we consider for our study. Section~\ref{sec:quijotedata} presents the new data obtained 
with the QUIJOTE experiment, describing the observations and discussing the resulting maps. In 
Section~\ref{sec:ancillary} we detail the ancillary data set that we employ in our analysis. 
Section~\ref{sec:seds} is dedicated to the measurement of the relevant quantities associated with 
the intensity and polarised emission proceeding from the three regions, while in Section~\ref{sec:fits} 
we discuss the modelling and physical interpretation of these findings in terms of the known 
foreground emission mechanisms. Finally, Section~\ref{sec:conclusions} presents the conclusions.


\section{The Galactic regions W49, W51 and IC443}
\label{sec:regions}

\begin{figure} 
\centering
\includegraphics[trim= 0mm 0mm 0mm 0mm, scale=0.35]{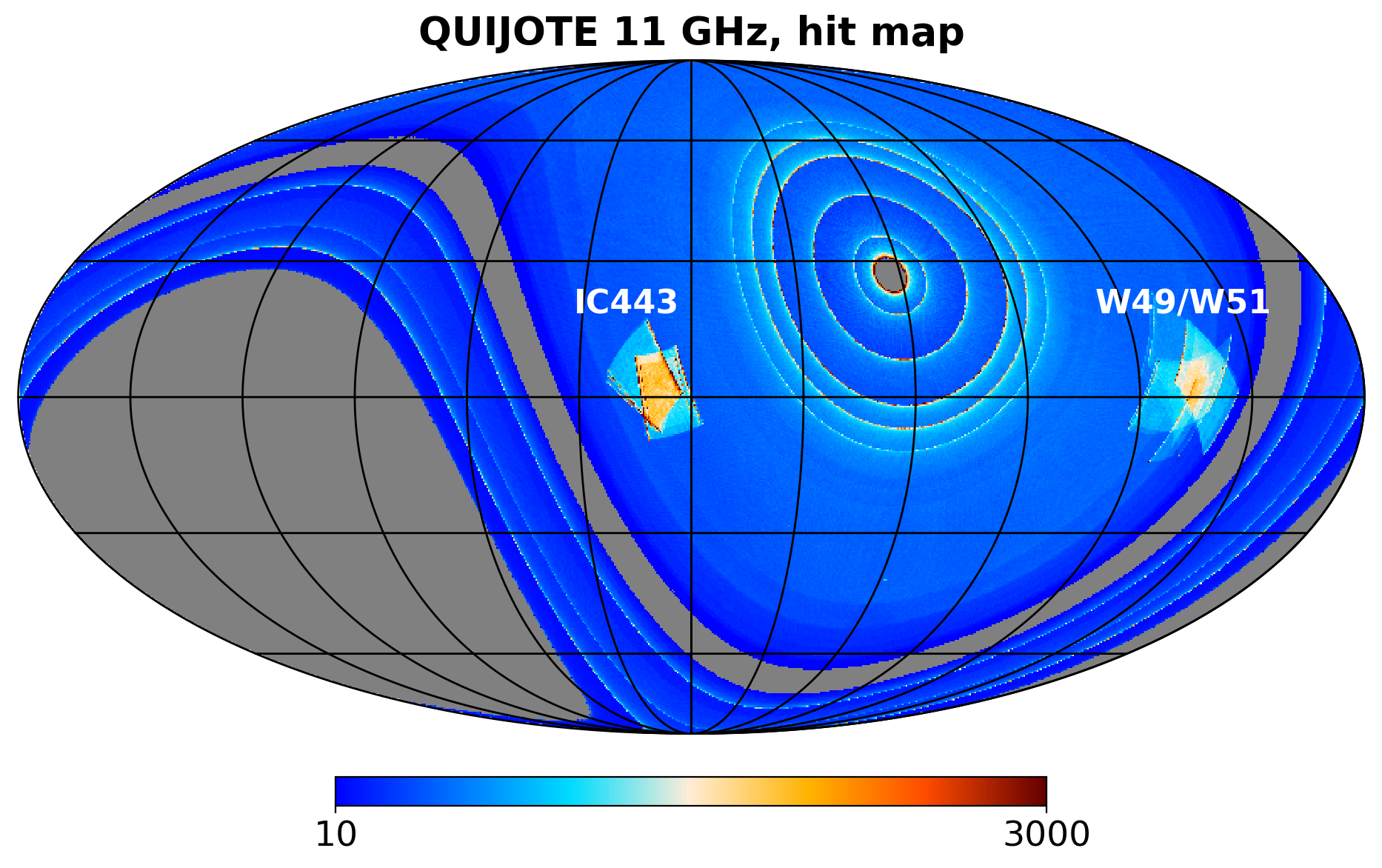} 
	\caption{Map showing the number of hits per pixel (i.e., the number of $40\,\text{ms}$ samples 
	in each $N_{\rm side}=512$ pixel) for the combined wide survey and 
	raster scan data set (see Section~\ref{ssec:maps} for details). 
	The inclusion of raster scans determines a local increase 
	in sensitivity towards the IC443 region and the area surrounding W49 and W51. 
	The map is centred at Galactic coordinates $(l,b)=(180^{\circ},0^{\circ})$ to 
	better display the source areas. This map is obtained using 11\,GHz data in total intensity, but 
	the hit maps for the other frequency bands are qualitatively similar. }
\label{fig:hitmap}
\end{figure}

We dedicate this section to a description of the three Galactic regions considered 
in the current work. W49 and W51 are the first important complexes encountered  
east of the W47 region that was studied in the second QUIJOTE scientific 
paper~\citep{genova_santos17}. Both these 
regions host SNRs; it is then interesting to combine their analysis 
with IC443, which is a relatively more isolated SNR. This allows to assess the 
influence of molecular clouds and star forming regions on the AME signal towards 
SNRs. The position on the sky of these three sources is shown in Fig.~\ref{fig:hitmap}, while in Table~\ref{tab:sources_info} we report a summary 
of their relevant information.

\subsection{W49}
\label{ssec:w49}
W49 is a Galactic radio source discovered in the 22 cm survey of~\citet{westerhout58}; 
it is located in \textit{Aquila} on the plane of the Milky Way at Galactic 
coordinates\footnote{Nominal coordinates quoted in the text are taken from the SIMBAD 
Astronomical Database at \url{http://simbad.u-strasbg.fr/simbad/}. However, due to the 
extended nature of these sources, the SIMBAD coordinates are often representative only 
of specific sub-regions. The mean coordinates for the full regions were assessed considering 
studies in the literature at different wavelengths; in such cases, coordinates are labelled 
with the subscript ``eff'', to stress that they mark an effective central position.} 
$(l_{\rm eff},b_{\rm eff})=(43.2^{\circ}, -0.1^{\circ})$, with an angular size of 
$\sim 30'$. The radio emission from this region can be separated into a thermal component 
designated W49A, centred at $(l,b)=(43.17^{\circ}, 0.00^{\circ})$, and a non-thermal 
component labelled W49B, located at $(l,b)=(43.27^{\circ}, -0.19^{\circ})$; this composite 
structure of W49 soon became evident from the results of high angular resolution continuum 
observations in the radio domain~\citep{mezger67}. 

W49A (G43.2+0.0) is one of the largest and most active sites of star formation in our 
Galaxy; it is embedded in a giant molecular cloud of estimated mass $10^6\,\text{M}_{\odot}$ 
over a total extension of 100\,pc~\citep{sievers91,simon01,galvan_madrid13}. Star formation 
is concentrated in a central region of $\sim20\,\text{pc}$ extension, which hosts several 
ultra-compact HII (UCHII) regions. The W49A stellar population has been extensively studied 
using infrared (IR) observations~\citep{wu16,eden18}, while observations in the radio domain 
have contributed to map the complex kinematics of the local 
gas~\citep{brogan01,roberts11,galvan_madrid13,de_pree18,rugel19,de_pree20}; high-energy 
$\gamma$-ray emission towards the region has also been detected~\citep{brun11}. From the proper 
motion of $\text{H}_2\text{O}$ masers, \citet{gwinn92} estimated the distance of this region 
from the Sun at $11.4\pm1.2\,\text{kpc}$, which was later refined by \citet{zhang13} as 
$11.11^{+0.79}_{-0.69}\,\text{kpc}$. Because of its large distance, the region is optically 
obscured by intervening interstellar dust, which accounts for the lack of W49 optical observations. 
Given its total extension and gas mass, W49A is comparable to extra-Galactic giant star-forming 
regions, and as such it is the ideal environment to study massive star formation in starburst 
clusters, and the early evolution of HII regions~\citep{wu16,de_pree18}.

W49B (G43.3-0.2) is a young SNR which has also been extensively studied in the 
literature with observational data at different wavelengths. \citet{pye84} compared X-ray and 
radio observations of the region, revealing an anti-correlation between the corresponding 
morphologies: while the X-ray brightness profile is centrally peaked, the radio image is typical 
of a shell-like remnant with no indication of a central energy source. This morphology 
was confirmed by later studies~\citep{rho98,hwang00,keohane07} and ascribed to a jet-driven
supernova triggered by the collapse of a supermassive Wolf-Rayet star, with subsequent interaction 
with the circumstellar medium~\citep{lopez13,gonzalez_casanova14,siegel20}. The supernova event occurred 
in between 1 and 4\,kyr ago~\citep{hwang00,zhou11}; the distance of W49B from the Sun was set at 
$\sim 8\,\text{kpc}$ by~\citet{moffett94}. However, radio observations by~\citet{brogan01} pointed 
out that the morphology of HI gas towards the southern edge of W49B suggests interaction with the 
nearby W49A molecular cloud, thus allowing for a distance range of 8 to 12\,kpc. By using a combination 
of radio and IR data,~\citet{zhu14} later suggested a distance of $\sim10\,\text{kpc}$; in contrast, 
observations of H$_2$ emission lines pointed again towards a shorter 
distance of $7.5\,\text{kpc}$~\citep{lee20}. The more recent analysis presented in~\citet{sano21},
based on the observation of CO emission lines, finally favours the value $11.0\pm0.4\,\text{kpc}$, 
ascribing the smaller value obtained with H$_2$ to the latter tracing only a small portion of 
the associated molecular cloud; as the H$_2$ is thermally shocked and accelerated, distance 
measurements based on its velocity are likely biased. Finally, emission in $\gamma$ rays proceeding 
from W49B has also been observed~\citep{brun11,hess_collab18}, confirming the nature of a SNR 
interacting with molecular clouds.

W49 has been considered as a possible candidate for AME in~\citet{planck_ir_xv}, where 
an emission excess was indeed found at typical AME frequencies. However, W49 was finally 
discarded from the list of significant AME sources due to the presence of local UCHII regions, 
which could contribute to the observed excess.

\subsection{W51}
\label{ssec:w51}
The W51 region is located on the Galactic plane east of W49, at central coordinates 
$(l_{\rm eff},b_{\rm eff})=(49.20^{\circ},-0.35^{\circ})$. It was also initially discovered 
in~\citet{westerhout58} radio survey, as another bright source in \textit{Aquila}, 
and initially classified as an HII region. Subsequent radio observations identified 
the source as a molecular cloud~\citep{penzias71} and allowed to further separate 
the region into three main substructures~\citep{kundu67,mufson79}: W51A and W51B, 
responsible for the observed thermal emission, and a non-thermal component 
W51C~\citep[for a detailed review of W51 morphology see][]{ginsburg17}.

W51A and W51B are located at Galactic coordinates 
$(l,b)=(49.48^{\circ}, -0.33^{\circ})$ and 
$(l_{\rm eff},b_{\rm eff})=(49.1^{\circ}, -0.3^{\circ})$, 
respectively; these regions are embedded in a giant molecular cloud of 
$M>10^6\,\text{M}_{\odot}$~\citep{carpenter98}, with W51A the main star-forming 
component. W51A is one of the most studied regions of massive star formation in our 
Galaxy; its rich morphology allows to distinguish two main components, named 
G49.5-0.4 and G49.4-0.3~\citep{wilson70}. These components are further resolved into 
several regions~\citep{martin72, mehringer94, okumura00}, among which we can cite 
the protoclusters IRS1 and IRS2~\citep{ginsburg16}. Several hyper-compact HII regions 
have also been detected with radio data~\citep{ginsburg20,rivera_soto20}. This complex structure, combined 
with the local richness in gas and dust content, makes this region the ideal 
environment to study high-mass star formation~\citep{saral17,ginsburg17a,goddi20}. Despite 
all the observational endeavours, however, W51A stellar population is likely not 
completely catalogued~\citep{binder18}; recent studies of W51A using IR observations 
can be found in~\citet{lim19} and~\citet{bik19}, who confirmed the complexity of 
the ongoing star formation processes. The region also shows a rich 
astrochemistry~\citep{vastel17,watanabe17}, which has been the object of studies 
searching for prebiotic molecules in star-forming 
regions~\citep{rivilla16,rivilla17,jacob21}. The parallax distance to this molecular cloud 
has been estimated from maser observations as 
$d=5.41^{+0.31}_{-0.28}\,\text{kpc}$~\citep{sato10} and as 
$d=5.1^{+2.9}_{-1.4}\,\text{kpc}$~\citep{xu09}, depending on the considered 
subregion; these estimates place W51 in the Carina-Sagittarius arm. 

W51B appears as 
a filamentary structure populated by UCHII regions; however, unlike 
W51A, most of the star formation in W51B seems to have already taken place, as 
suggested by the lower fraction of dense gas~\citep{ginsburg15}.

W51C is an extended source of non-thermal radio emission located west of W51A and 
mostly south of W51B at $(l_{\rm eff},b_{\rm eff})=(49.2^{\circ}, -0.7^{\circ})$, 
which has been identified as a SNR~\citep{seward90} with an estimated 
age of 30\,kyr~\citep{koo95}. This source has been extensively observed in radio 
wavelenghts~\citep{zhang17,ranasinghe18}, X-rays~\citep{koo05,hanabata13} and 
$\gamma$-rays~\citep{abdo09, aleksic12}. The distance of this SNR was first placed 
at 4.1\,kpc~\citep{sato73}; later,~\citet{koo95} estimated the distance to be 
$\sim 6\,\text{kpc}$, although with a high uncertainty given that this result is 
based on the association of the SNR with a molecular cloud which may extend over 
1.5\,kpc. Observations of the HI 21-cm line in absorption led to an estimated 
distance of 4.3\,kpc~\citep{tian13} which has been recently re-evaluated as 
5.4\,kpc~\citep{ranasinghe18}. The latter result is dictated by the observed interaction 
between W51C and W51B~\citep{koo97,brogan13}, which constrains the SNR to be at the 
same distance as the other W51 HII regions. 

\begin{table*} 
\centering
\caption{Summary of the information on the QUIJOTE-MFI observations centred on each source; 
	the last column reports the mean fractional wide survey (WS) time contribution to the final 
	maps, computed with respect to the total time obtained combining the selected raster 
	data with the nominal mode data. See~\ref{ssec:observations} for more details.}
\label{tab:observations}
\begin{tabular}{ccccccc}
\hline
\hline
Source 	& Observation type & Observing dates & Covered area & Total observing time & Selected time fraction & WS time fraction \\
\hline
\hline
W49   & 193 drift scans (73\,min)  & June-August 2015          & 616\,deg$^2$ & 235.6\,h & 78\% & 20\% \\	
W51   & 170 drift scans (74\,min)  & October-December 2016     & 708\,deg$^2$ & 209.1\,h & 73\% & 20\% \\
IC443 & 552 raster scans (29\,min) & October 2014 -- June 2015 & 474\,deg$^2$ & 269.1\,h & 66\% & 21\% \\
\hline
\hline
\end{tabular}
\end{table*} 

Overall, W51 has a luminosity to mass ratio similar to W49~\citep{eden18}; it 
subtends a region with diameter $\sim 50'$ on the sky~\citep{bik19}. The search for 
AME in the region has been tackled by~\citet{demetroullas15}, where it is shown that 
a spinning dust component is not required for a proper modelling of W51 intensity spectrum.
To our knowledge, no clear evidence of AME emission towards W51 has been reported in
the literature.

\subsection{IC443}
\label{ssec:ic443}
IC443 (\textit{Index Catalogue 443}) is a SNR located in the 
constellation of \textit{Gemini} at Galactic coordinates 
$(l,b)=(189.06^{\circ}, 3.24^{\circ})$; it is therefore found close to the 
Galactic anticentre, with a lower emission from the neighbouring background compared 
to W49 and W51, which are much closer in projection to the Galactic bulge. This 
source has been extensively observed throughout the whole spectrum, including 
radio~\citep{kundu72,green86,castelletti11,mitra14,planck_ir_xxxi,egron16,egron17}, 
optical~\citep{fesen84,ambrocio_cruz17}, 
X-ray~\citep{petre88,matsumura17,greco18,hirayama19,zhang18} 
and $\gamma$-ray~\citep{torres03,tavani10} frequencies. IC443 angular size is 
$\sim 45'$~\citep{green14}; the related supernova event occured in between 
3~\citep{petre88} and 30\,kyr ago~\citep{olbert01}, with more recent estimates 
favouring $\sim20$\,kyr~\citep{lee08}. The distance to IC443 is still under debate; 
estimates range in between 0.7 and 1.5\,kpc~\citep{lozinskaya81,fesen84}, with the 
latter value being the most commonly used in the literature. A higher distance of 
1.5-2\,kpc has been proposed due to IC443 association with the HII region 
Sh2-249~\citep[see also Fig.~\ref{fig:ic443_mask}]{gao11}; a more recent estimate 
by~\citet{ambrocio_cruz17} locates IC443 at 1.9\,kpc from the Sun.

At radio frequencies IC443 shows two roughly spherical sub-shells of synchrotron 
emission, with different centres and radii~\citep{onic17}; the region appears 
brighter in its north-eastern limb in equatorial coordinates. These shells 
traditionally define the boundary of IC443; a fainter and larger shell has been 
detected, but it is not clear whether it is a different SNR, nor if it is actually 
interacting with IC443~\citep{lee08}. Radio thermal emission towards the region has 
also been observed, and ascribed to \textit{bremmstrahlung}~\citep{onic12}. X-ray 
observations, instead, reveal a centrally-filled profile, which makes IC443 a 
mixed-morphology or thermal-composite SNR~\citep{rho98,rajwade16}. IC443 evolution 
and observed structure is the result of the interaction between the SNR and 
different phases of the surrounding interstellar medium~\citep{lee08,ritchey20,koo20,ustamujic21}, with 
observational evidence of interaction with molecular gas in the southeast and 
northwest~\citep{su14,yoshiike17} and with an atomic cloud in the 
northeast~\citep{kokusho13}. Observations in $\gamma$-rays towards IC443 are 
relevant for the study of Galactic cosmic rays; in particular, $\gamma$-ray emission 
deriving from the decay of pions (the ``pion-decay bump'') has been detected 
by~\citet{ackermann13}. These pions are produced in collisions between accelerated 
cosmic-ray protons and interstellar material, so that the detection of the pion bump
provides a direct evidence for the occurrence of proton acceleration in SNRs~\citep{huang20}.

Finally, IC443 has already been a target in the search for AME.~\citet{onic17} 
analised the radio and microwave spectrum of the region, reporting hints of AME 
detection and a bending of the synchrotron spectrum; however, they stressed 
that further data in the range 10 to 100\,GHz are required to confirm these results. 
A similar conclusion was reached by~\citet{loru19}, who detected a clear bump in the 
range 20--70\,GHz in IC443 intensity spectrum.

\begin{figure} 
\centering
\end{figure}
\begin{figure*}
\centering
\includegraphics[trim= 0mm 0mm 0mm 0mm, scale=0.6]{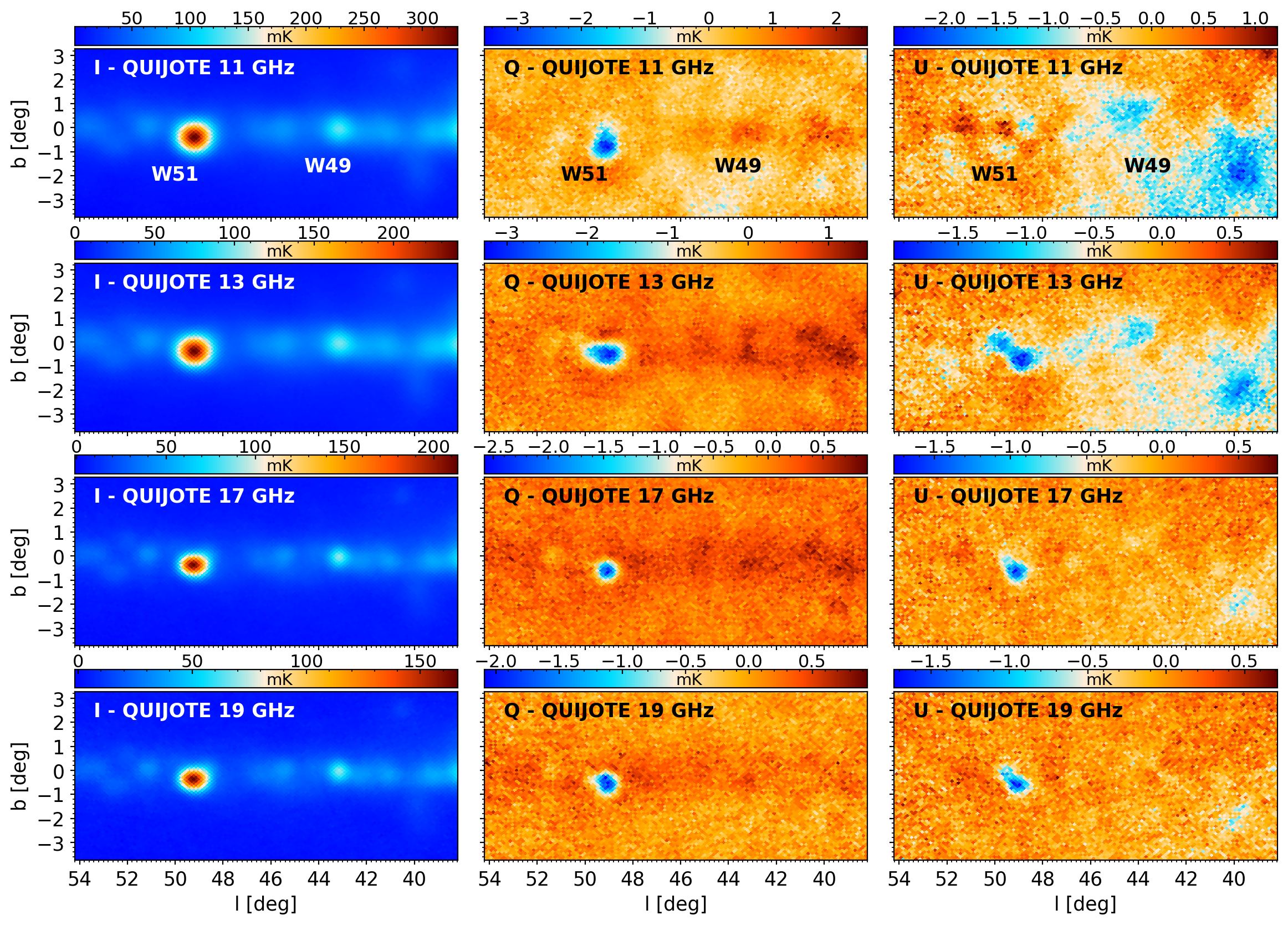} 
\caption{Intensity and polarisation destriped maps of the sky region 
	including the sources W49 and W51, for the four QUIJOTE frequencies. 
	All maps are shown here at their original resolution.  }
\label{fig:ws_qjt_maps}
\end{figure*} 

\begin{figure*} 
\centering
\includegraphics[trim= 0mm 0mm 0mm 0mm, scale=0.6]{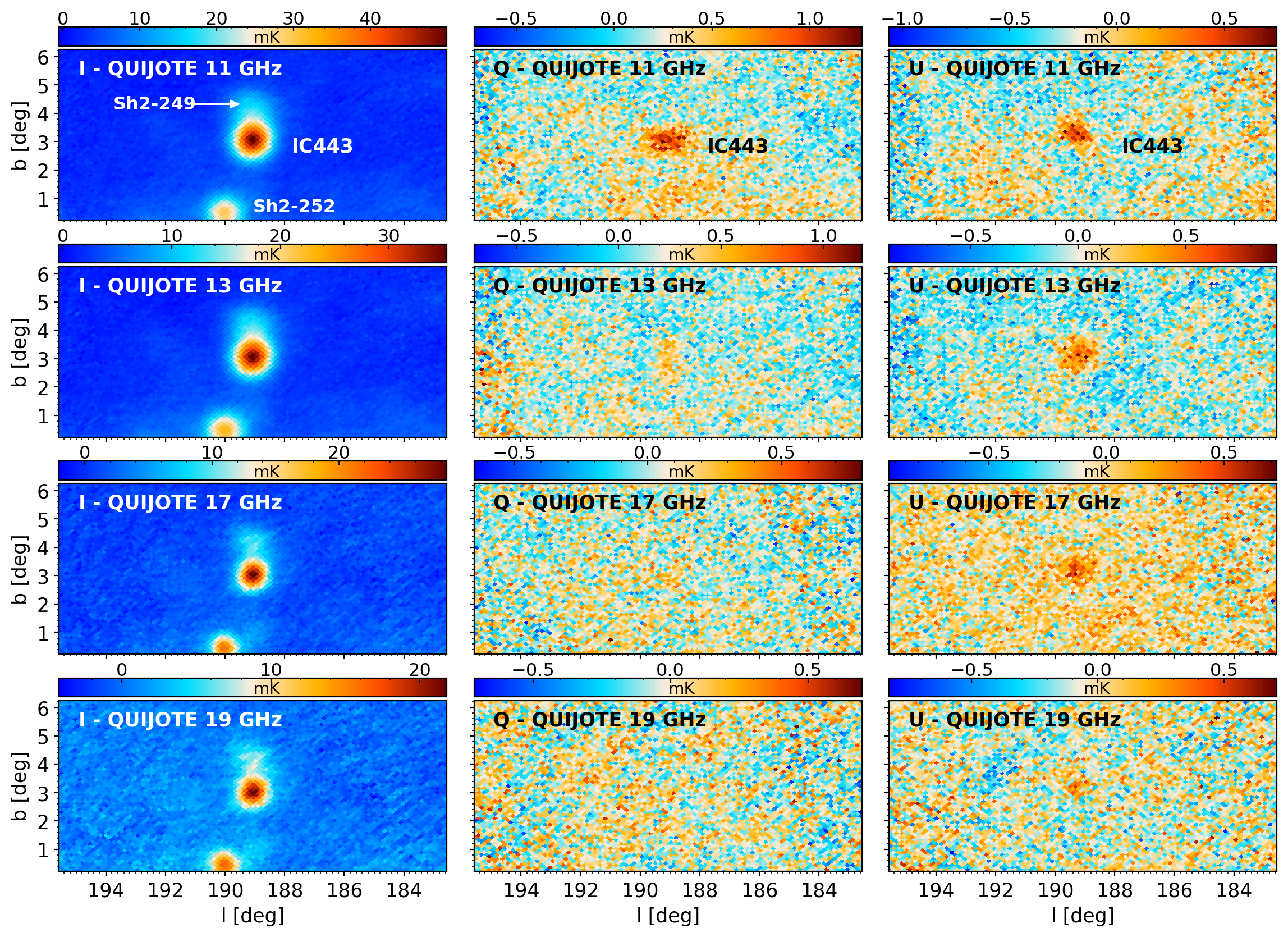} 
\caption{Same as in Fig.~\ref{fig:ws_qjt_maps}, but showing this time the 
	IC443 region. The 11\,GHz intensity map also marks the position of the two HII complexes surrounding the SNR.}
\label{fig:ic443_qjt_maps}
\end{figure*} 


\section{QUIJOTE data}
\label{sec:quijotedata}

We present in this work data acquired with the QUIJOTE experiment~\citep{genova_santos15a, 
rubino_martin17}, a scientific collaboration aimed at characterising the polarisation of 
the CMB, and other Galactic and extra-Galactic physical processes, in the frequency 
range 10--40\,GHz and at angular scales larger than 1 degree.  QUIJOTE consists of 
two telescopes and three instruments, operating from the Teide Observatory (Tenerife, 
Spain), a site that provides optimal atmospheric conditions for CMB observations in 
the microwave range~\citep{rubino_martin12b}.

The data used in this work were acquired with the QUIJOTE multi-frequency 
instrument~\citep[MFI,][]{hoyland12}, which consists of four conical corrugated feedhorns, 
each feeding a novel cryogenic on-axis polar modulator that can be rotated in steps of 
$22.5^{\circ}$. The output from each horn consists of eight channels carrying different 
linear combinations of the radiation Stokes parameters, which in turn are separated by 
the subsequent data analysis pipeline. Horns 1 and 3 both observe at frequency bands 
centred at 11 and 13\,GHz, whereas horns 2 and 4 both observe at frequency bands 
centred at 17 and 19\,GHz, all with a 2\,GHz bandwidth. The full width 
at half-maximum (FWHM) is equal to 52\,arcmin for the two low-frequency horns and to 
38\,arcmin for the two high-frequency horns. Overall, the MFI provides eight different maps 
of the sky in intensity and polarisation, and is primarily devoted to 
the characterisation of the Galactic emission. A full description of the MFI data reduction 
pipeline and map-making is provided in~\citet{genova_santos22}.

\subsection{Observations}
\label{ssec:observations}

The analysis presented in this paper is based on the combination of different 
QUIJOTE observations. We first consider data from the MFI wide survey, which 
covers more than $25,000\,\text{deg}^2$ including most of the northern sky, 
and as such encompasses the three regions targeted in this study.
Observations were performed at fixed elevation by letting the telescope spin 
continuously in azimuth, and took place between May 2013 and June 2018 for a total 
observing time of $\sim9200\,\text{h}$. The removal of bad data performed by 
the pipeline yields an effective total time of $\sim8500\,\text{h}$ in intensity 
and in the range $\sim4700\,\text{h}$ to $\sim6800\,\text{h}$ in polarisation, depending 
on the chosen horn; the resulting data were used to 
produce the QUIJOTE wide survey legacy maps. An extensive description of the MFI
wide survey and its associated data products can be found in~\citet{rubino_martin22}. 
On top of this large scale survey, we also consider a set of dedicated raster scan 
observations focussed on each region, which we describe in the following; we 
stress that the raster scan observations constitute the major source of data used
in the current analysis. A summary 
of the related information is also reported in Table~\ref{tab:observations}. 

W49 was mapped in 193 dedicated observations performed between June and August 
2015; these observations were drift scans in which the telescope elevation was 
maintained fixed and the pointing was switched back and forth in azimuth with an 
$\sim18/\cos{(\text{EL})}$\,deg amplitude for each scan; each observation took on 
average 73\,min, and covered a final mean area of $408\,\text{deg}^2$ per horn on 
the sky. Notice that the MFI focal plane configuration determines a different 
pointing for each horn for the same observation, with separations up to $\sim 5^{\circ}$; 
this extends the sky area covered by W49 rasters to $616\,\text{deg}^2$. The total 
observing time is 235.6\,h; after the inspection and removal of bad data, which is 
performed for each horn independently\footnote{Apart from the data flagging pipeline 
described in~\citet{genova_santos22}, we also performed a visual inspection of 
local maps centred on each source. These maps were generated for each observation and each one 
of the 32 MFI channels, and were used to discard observations where the local background was too 
noisy in at least one channel.}, the effective observing times for the four 
horns are 171.4, 194.1, 177.0 and 189.5\,h, corresponding to a an average of $\sim78\%$ 
of the total. After combining the good raster data with the nominal mode data, the 
latter account for 13.4\%, 29.0\%, 19.0\% and 19.8\% of the total observing time in 
each horn; on average, the wide survey fractional time contribution to the final maps 
in the W49 region is $\sim 20\%$.

W51 is located $\sim 6\,\text{deg}$ east of W49 on the Galactic plane, so it was also 
captured by the drift scans we just described. However, due to the different horn sky 
coverage, W51 was only partially mapped in several observations. For this reason it was 
decided to perform an independent set of observations centred on W51: they were conducted 
between October and December 2016, as a set of 170 fixed-elevation scans with a 
$\sim20/\cos{(\text{EL})}$\,deg azimuth amplitude, each lasting 74\,min on average. The 
total observing time is 209.1\,h, which after the good data selection amounts to 140.9, 
152.7, 158.8 and 155.3\,h for the four horns, corresponding to an average of $\sim73\%$ of 
the total time. In this case, the nominal mode data contribute for 14.7\%, 21.3\%, 
25.7\% and 17.3\% of the total observing time in each horn after combination with raster 
data; again, the mean wide survey time contribution to the final maps 
is $\sim 20\%$. Each horn covers on average $534\,\text{deg}^2$ on the sky, for a total 
$708\,\text{deg}^2$ mapped by the MFI with this set of observations. The overall sky area 
covered by the joint set of observations of W49 and W51 amounts to $891\,\text{deg}^2$, 
thus enabling an extended reconstruction of the Galactic plane region surrounding the 
two molecular complexes. 

Finally, IC443 was targeted by 552 observations conducted between October 2014 and 
June 2015. In this case the observations consisted in raster scans with an amplitude of 
$\sim 12/\cos{(\text{EL})}$\,deg in azimuth, with the telescope being stepped 
$\sim0.1$\,deg in elevation at the end of each scan, over a total $\sim 12$\,deg range. 
Each observation lasted 29\,min on average, for a total observing time of 269.2\,h. The 
selection of good data yielded 174.3, 192.8, 187.6 and 159.5\,h for the four horns, averaging 
at $\sim66\%$ of the overall dedicated time. Out of the combination with nominal mode data, 
the latter account for 10.2\%, 22.3\%, 24.0\% and 27.9\% of each horn observing time, for 
an average time fraction of $\sim21\%$. The mean area covered 
by each horn amounts to $360\,\text{deg}^2$, for a total $474\,\text{deg}^2$ mapped by the 
MFI in this region.

\subsection{Maps}
\label{ssec:maps}

Time-ordered data from the raster scan observations are processed adopting the standard 
QUIJOTE pipeline described in~\citet{genova_santos22}. Subsequently, these data are 
combined with data from the wide survey to produce the associated sky maps. The map-making 
is based on the \texttt{PICASSO} code~\citep{guidi21} with the same set of destriper 
parameters (noise priors and baseline length) adopted in~\citet{rubino_martin22} when 
producing maps of the wide survey data alone. The resulting destriped map is provided 
using a {\sc HEALPix} pixelisation~\citep{healpix} with $N_{\rm side}=512$, 
corresponding to a 6.9\,arcmin pixel size; this resolution is enough given that the 
MFI beam FWHM values are larger than 0.5$^{\circ}$. The code provides full sky maps, 
with increased sensitivities towards the relevant regions thanks to the inclusion of 
raster scan data; this is visible in the hit map shown in Fig.~\ref{fig:hitmap}, 
where we can clearly distinguish two regions with higher integration time, 
one centred on the Galactic plane around W49 and W51, and one centred on IC443.
In the following we will focus on sub-maps centred on these specific areas. 
We remind that the wide survey maps presented in~\citet{rubino_martin22} are corrected 
for residual radio frequency interference contamination by subtracting the 
median value of pixels in rings of constant declination (FDEC correction, see Sec. 2.4.2
in the aforementioned reference). However, as in this analysis we will only perform 
aperture photometry measurements (Sec.~\ref{sec:seds}), this correction is unnecessary 
and is not applied to our combined wide survey plus raster scan 
maps. A quantitative estimation of the impact of FDEC on aperture photometry measurements 
is presented in appendix B1 in~\citet{rubino_martin22}, and is well 
below the uncertainties associated with our measurements presented in Sec.~\ref{sec:seds}.

Local maps for W49/W51 and IC443 are plotted in Fig.~\ref{fig:ws_qjt_maps} and in 
Fig.~\ref{fig:ic443_qjt_maps} respectively, in the three Stokes parameters $I$, $Q$ and 
$U$, for the four MFI frequencies. Although in our subsequent analysis we will consider 
all maps smoothed to a common resolution of $1\,\text{deg}$, these plots show the maps 
in their original resolution. As two horns cover each frequency band, their individual 
maps are combined to yield the final map of the corresponding band. However, for the low 
frequency bands (11 and 13\,GHz) the maps we employ are the ones obtained from horn 3 
data alone; this is due to a fault in the polar modulator of horn 1, which has been fixed 
to the same position since September 2012, thus making data obtained with this horn not 
reliable\footnote{Although we discard horn 1 data in this paper analysis, its maps in total 
intensity were nonetheless generated using the standard pipeline; hence, for completeness, 
we still included the description of its contribution to the acquired data in 
Section~\ref{ssec:observations} and in Table~\ref{tab:observations}.}. For the case of 17 
and 19\,GHz, instead, the maps we show are obtained from a linear combination of horn 2 and 
horn 4 maps, each weighted by a uniform weight computed from the white noise in the power 
spectrum of the associated wide survey map~\citep[for details see][]{rubino_martin22}.
\input{tables/nulltest_table.tex}  

In Fig.~\ref{fig:ws_qjt_maps} we show the QUIJOTE maps for the Galactic plane region 
encompassing the sources W49 and W51. The Galactic plane is easily recognisable in the 
intensity maps, as a stripe of diffuse emission surrounding the compact objects; this diffuse 
emission is also visible in polarisation, with the expected\footnote{In this work we follow 
the {\sc HEALPix} definition of the Stokes parameters, usually referred to as the ``COSMO 
convention'', which differs from the IAU convention in a sign flip of $U$.} imprint of positive 
$Q$ and nearly zero $U$. In intensity both W49 and W51 are clearly visible; at the western edge 
of the intensity maps it is possible to see the border of the source W47, which was studied 
in~\citet{genova_santos17}. W51 appears particularly bright in these maps; it saturates the 
chosen colour scales in intensity, while in polarisation it shows a clear emission with both 
negative $Q$ and $U$, which makes it stand out from the diffuse Galactic plane polarisation 
signal. As commented in Section~\ref{ssec:w51}, W51 size is $\sim50'$, which means it is partially 
resolved with QUIJOTE beams; indeed, a mismatch is visible between the centres of the intensity 
and polarisation emissions, with $Q$ and $U$ peaks being slightly displaced towards the south-east. 
A possible explanation for this is that while the bulk of the intensity emission comes from the 
molecular cloud hosting W51A, the polarised emission is from the SNR W51C only. W49 appears 
fainter in intensity, and does not reveal a clear hint of polarised emission, although the diffuse 
Galactic plane polarisation shows an increment in positive $Q$ at the source position. Unlike W51, 
there is no particular internal structure visible for this source; its angular size of $\sim30'$ 
implies it is not resolved by QUIJOTE.

\begin{figure} 
\centering
\includegraphics[trim= 3mm 9mm 0mm 0mm, scale=0.4]{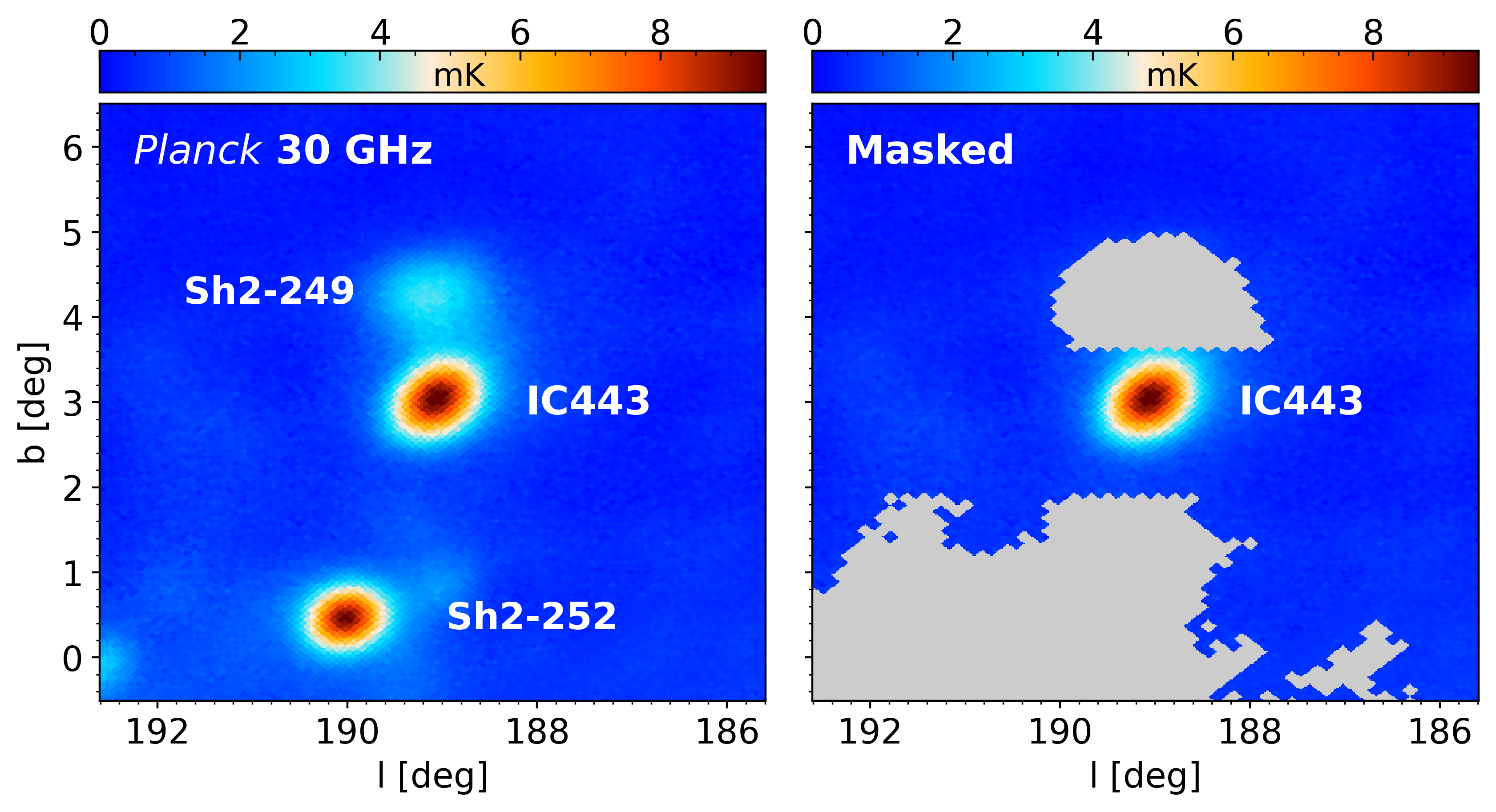} 
	\caption{\textit{Left}. The area surrounding IC443 shown on the 
		\textit{Planck} 30\,GHz map (2015 data release). \textit{Right}. 
		The same area after applying a tailored mask which excises the 
		contribution from the HII regions Sh2-249 and Sh2-252.}
\label{fig:ic443_mask}
\end{figure}

IC443 maps are shown in Fig.~\ref{fig:ic443_qjt_maps}. As commented in Section~\ref{ssec:ic443}, 
the source is located towards the Galactic anticentre and at a Galactic latitude of $\sim3^{\circ}$, 
therefore the Galactic plane emission is not visible in these maps; its outskirts, however, still 
enter the maps as a background emission gradient that increases towards the south. In addition, 
two radio sources are visible in the vicinity of IC443: a bright source towards the southeast, 
which we can identify as the HII region Sh2-252~\citep[also referred to as NGC2174,][]{bonatto11} 
and a fainter feature just north of the SNR, which produces an apparent elongation of IC443 in that 
direction and can be ascribed to the HII region Sh2-249~\citep{dunham10, gao11}. The emission from 
these sources can affect the flux density estimates in this region; therefore, we apply to all IC443 
maps an \textit{ad hoc} mask built to block the contribution of the neighbouring objects. The details 
of the mask are visible in the right panel of Fig.~\ref{fig:ic443_mask}. In the QUIJOTE intensity maps 
the source is clearly detected, although the background at 19\,GHz is noisier compared to the maps 
obtained for W49 and W51; in polarisation the maps show a positive $U$ emission and hints of a positive 
$Q$ emission (with a more significant detection at 11\,GHz) towards the source, 
which can be ascribed to synchrotron emission proceeding from the SNR. 
 
\subsection{Null-test maps and noise levels}
\label{ssec:nulltest}

We use a null test map built from QUIJOTE data to evaluate the contribution 
of instrumental noise. For each observed region, we split the available data in 
two halves with the same number of observations, in such a way that each half has 
a sky coverage as similar as possible as the other; for the wide survey data we 
use the same halves as in~\citet{rubino_martin22}.
In order to allow a proper reconstruction of the polarised signal from each half, 
care is taken to split observations in such a way that the distribution of observations 
in each polar modulator position is the same in both halves. 
For each half we then generate the corresponding maps adopting the same 
processing pipeline that was used for the full data set. We then obtain the 
null-test maps as:
\begin{equation}
	\label{eq:nulltest}
	M_{\rm NT} = \frac{1}{2}\left( M_{\rm H1} - M_{\rm H2} \right),
\end{equation}
where $M_{\rm H1}$ and $M_{\rm H2}$ represent the first and second half maps.
By construction, the resulting map $M_{\rm NT}$ should be free from the sky 
signal, which cancels out in the subtraction, and be representative of the 
instrumental contribution only. We compute the instrumental noise component as the 
rms of the pixel values found in a 1-degree radius aperture on each null map.
The resulting quantity represents the rms of individual pixels; to be consistent 
with other QUIJOTE works, we convert it into the rms of a Gaussian beam with 
FWHM=$1^{\circ}$, by dividing the original rms by the square root of the number 
of pixels entering the beam solid angle. 
We place our aperture far from the sources, centred at $(l,b)=(46^{\circ},-2^{\circ})$ for the 
W49/W51 map, and at $(l,b)=(186^{\circ},2.5^{\circ})$ for the IC443 map; this 
choice allows us to compare this rms with the same quantity evaluated on the 
full data set maps. The two values of rms are directly comparable thanks to the 
factor $1/2$ in equation~\eqref{eq:nulltest}, which cancels the $\sqrt{2}$ factor 
coming from the linear combination of the two half-maps, and the additional 
$\sqrt{2}$ factor coming from the splitting of the available statistics, that 
increases the rms of each half-map. The comparison for the Stokes parameters 
$I$, $Q$ and $U$ is reported in Table~\ref{tab:nulltest} for both the W49/W51 and  
the IC443 maps. 

We see that in polarisation the real maps and the null-test maps have similar 
rms values, which means that the main contribution to the polarisation maps 
used in this study is noise. For intensity, instead, the rms is 
considerably higher when evaluated in the full data maps than in the null-test 
maps. This is indicative of an important contribution from the sky emission in 
the estimation of the noise. The effect is stronger for the W49/W51 map due to the proximity of the 
regions to the Galactic plane, which introduces gradients in the background signal. 
The polarisation rms values quoted in Table~\ref{tab:nulltest} are about a factor 
2 lower than the ones obtained from the wide survey maps only~\citep{rubino_martin22};
this shows the improvement in the map quality achieved by including the 
dedicated raster scan observations towards our sources. In fact, the sensitivity 
in polarisation averages around $\sim15\,\mu\text{K}/\text{beam}$, which is the 
deepest level reached by published QUIJOTE data so far. 

The last column in Table~\ref{tab:nulltest} reports the instantaneous sensitivity 
of the maps in polarisation, computed by multiplying the quoted $Q$ and $U$ rms values
per beam by the square root of the mean observing time per 1-degree beam; the results 
for $Q$ and $U$ are then averaged (in quadrature) to yield the values of $\sigma_{Q,U}$
reported in the table. In general, these estimates range from 1.0 to 
$1.4\,\text{mK}\,\text{s}^{1/2}$; this is expected, since the nominal instrument sensitivity 
in polarisation is $\sim 1\,\text{mK}\,\text{s}^{1/2}$. Our findings are consistent with 
previously reported values for the instantaneous sensitivity~\citep{genova_santos15b,genova_santos17}. 
This test confirms the good quality of the polarisation maps, and that the instrumental 
noise contribution to the maps is controlled and within the expected limits.


\begin{table*} 
\centering
\caption{Summary of the ancillary data detailed in Section~\ref{sec:ancillary}, which 
	are used in combination with QUIJOTE data to broaden the considered frequency range. 
	For each map we report the frequency, the reference telescope(s) or satellite, which 
	Stokes parameters are available, the original map resolution and units, the applied 
	calibration uncertainty and the reference in the literature.}
\label{tab:ancillary}	
\setlength{\tabcolsep}{0.5em}
\begin{tabular}{ccccccc}
\hline
\hline
Frequency & Facility & Stokes & Original resolution  & Original  & Calibration & Reference \\
	$[\text{GHz}]$  &    &        &  [arcmin]            &  units      & uncertainty [\%] &\\
\hline
\hline
	0.408   & Jodrell Bank MkI, MkIA-76\,m, & $I$ & 51 & K      & 10 & \citet{haslam82,platania03} \\
		& Bonn-100\,m, Parkes-64\,m  &  &  &  &  \\
	0.820	& Dwingeloo-25\,m	 & $I$     & 72    & K      & 10 & \citet{berkhuijsen72} \\
	1.410   & DRAO-25.6\,m		 & $Q,U$   & 36	   & mK     & 10 & \citet{wolleben06} \\	
	1.420	& Stockert-25\,m, Villa-Elisa-30\,m & $I$ & 36 & K  & 10 & \citet{reich86,platania03} \\
	2.326	& HartRAO-26\,m	 	 & $I$     & 20    & K      & 10 &\citet{jonas98,platania03} \\
	4.800	& Urumqi-25\,m		 & $I,Q,U$ & 9.5   & mK     & 10 & \citet{gao10,sun11} \\
	22.8	& \textit{WMAP} 9-yr 	 & $I,Q,U$ & 50.7  & mK     & 3 &\citet{bennett13} \\
	28.4	& \textit{Planck}-LFI	 & $I,Q,U$ & 32.29 & K      & 3 &\citet{planck_15_i,planck_ir_lvii} \\	
	33.0	& \textit{WMAP} 9-yr 	 & $I,Q,U$ & 38.8  & mK     & 3 & \citet{bennett13} \\
	40.6	& \textit{WMAP} 9-yr 	 & $I,Q,U$ & 30.6  & mK     & 3 & \citet{bennett13} \\
	44.1	& \textit{Planck}-LFI	 & $I,Q,U$ & 27.00 & K      & 3 & \citet{planck_15_i,planck_18_i} \\	
	60.8	& \textit{WMAP} 9-yr 	 & $I,Q,U$ & 20.9  & mK     & 3 & \citet{bennett13} \\
	70.4	& \textit{Planck}-LFI	 & $I,Q,U$ & 13.21 & K      & 3 & \citet{planck_15_i,planck_18_i} \\
	93.5	& \textit{WMAP} 9-yr 	 & $I,Q,U$ & 14.8  & mK     & 3 & \citet{bennett13} \\	
	100	& \textit{Planck}-HFI	 & $I,Q,U$ & 9.68  & K      & 3 & \citet{planck_15_i,planck_18_i} \\
	143	& \textit{Planck}-HFI	 & $I,Q,U$ & 7.30  & K      & 3 & \citet{planck_15_i,planck_18_i} \\
	217	& \textit{Planck}-HFI	 & $I,Q,U$ & 5.02  & K      & 3 & \citet{planck_15_i,planck_18_i} \\
	353	& \textit{Planck}-HFI	 & $I,Q,U$ & 4.94  & K      & 3 & \citet{planck_15_i,planck_18_i} \\
	545	& \textit{Planck}-HFI	 & $I$ 	   & 4.83  & MJy/sr & 6.1  & \citet{planck_15_i} \\
	857	& \textit{Planck}-HFI	 & $I$ 	   & 4.64  & MJy/sr & 6.4  & \citet{planck_15_i} \\
	1249	& \textit{COBE}-DIRBE	 & $I$	   & 37.1  & MJy/sr & 11.9 & \citet{hauser98} \\
	2141	& \textit{COBE}-DIRBE	 & $I$	   & 38.0  & MJy/sr & 11.9 & \citet{hauser98} \\
	2998	& \textit{COBE}-DIRBE	 & $I$	   & 38.6  & MJy/sr & 11.9 & \citet{hauser98} \\
\hline
\hline
\end{tabular}
\end{table*} 

\section{Ancillary data}
\label{sec:ancillary}
In order to better characterise the emission from the Galactic regions considered 
in this study, we also employ ancillary data that extend the information provided 
by QUIJOTE observations to both lower and higher frequencies. 
We use all maps in a common {\sc HEALPix} pixelisation of $N_{\rm side}=512$, 
and at a common angular resolution of $1\,\text{deg}$. With the exception of \textit{WMAP} 
maps, which are already available at this angular resolution at the LAMBDA 
website\footnote{Legacy Archive for Microwave Background Data Analysis,
\url{http://lambda.gsfc.nasa.gov/}.}, in all the other cases 
the smoothed maps are obtained by deconvolving the original beam (the nominal survey 
FWHM was used in each case and a Gaussian beam shape was assumed) and convolving 
with a 1\,deg FWHM Gaussian beam. QUIJOTE maps are also degraded to this coarser resolution 
for the photometric analysis. The relevant properties of all the ancillary maps 
used in this study are summarised in Table~\ref{tab:ancillary}.

\begin{figure*} 
\centering
\includegraphics[trim= 0mm 0mm 0mm 0mm, scale=0.6]{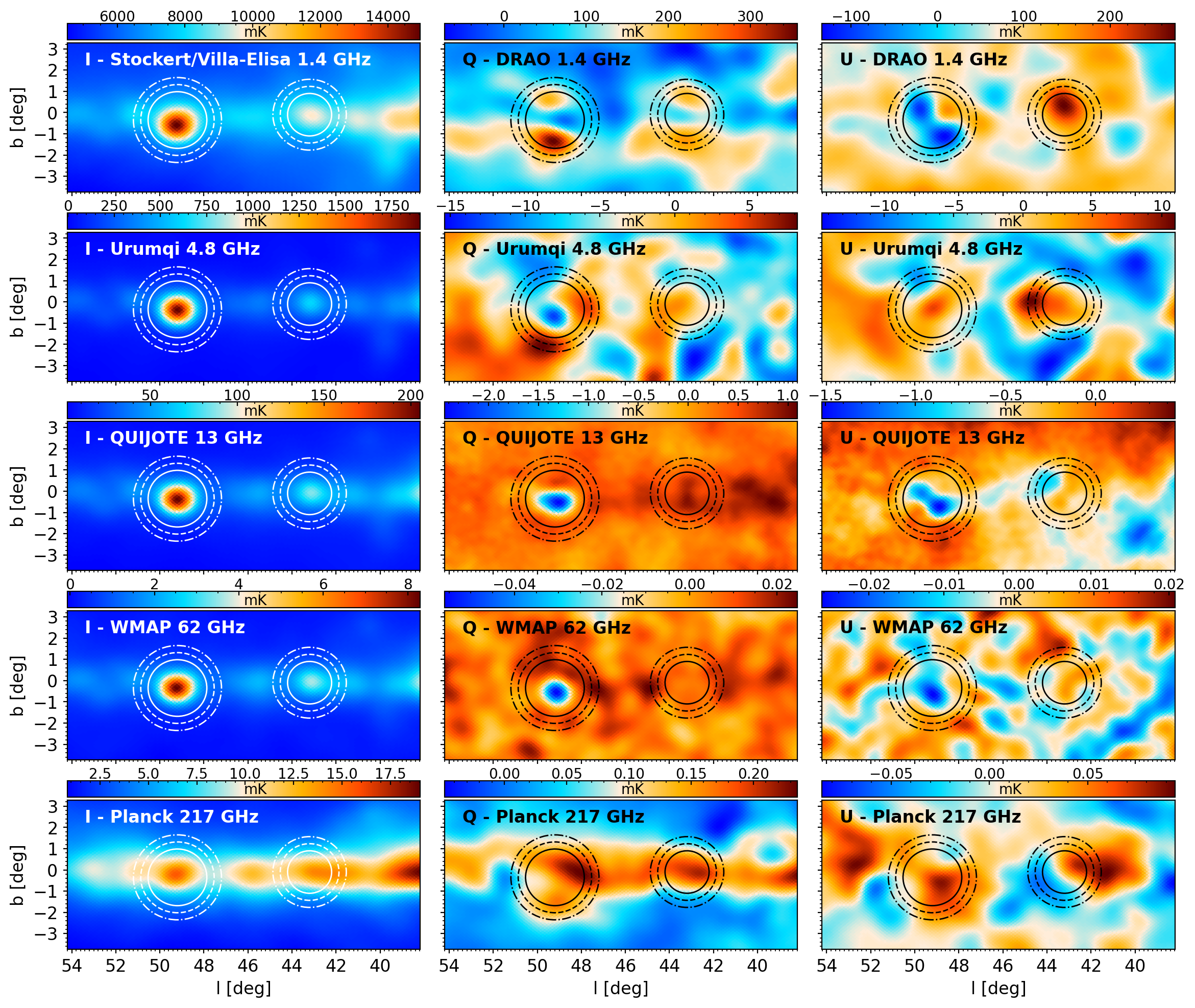} 
\caption{Sample intensity and polarisation maps of the sky region including the 
	sources W49 and W51, degraded to a 1\,deg resolution, 
	across the frequency range employed for the SED study. The apertures and 
	background annuli 
	used for the aperture photometry analysis are overplotted to each panel.}
\label{fig:ws_smt_maps}
\end{figure*} 

\begin{figure*} 
\centering
\includegraphics[trim= 0mm 0mm 0mm 0mm, scale=0.6]{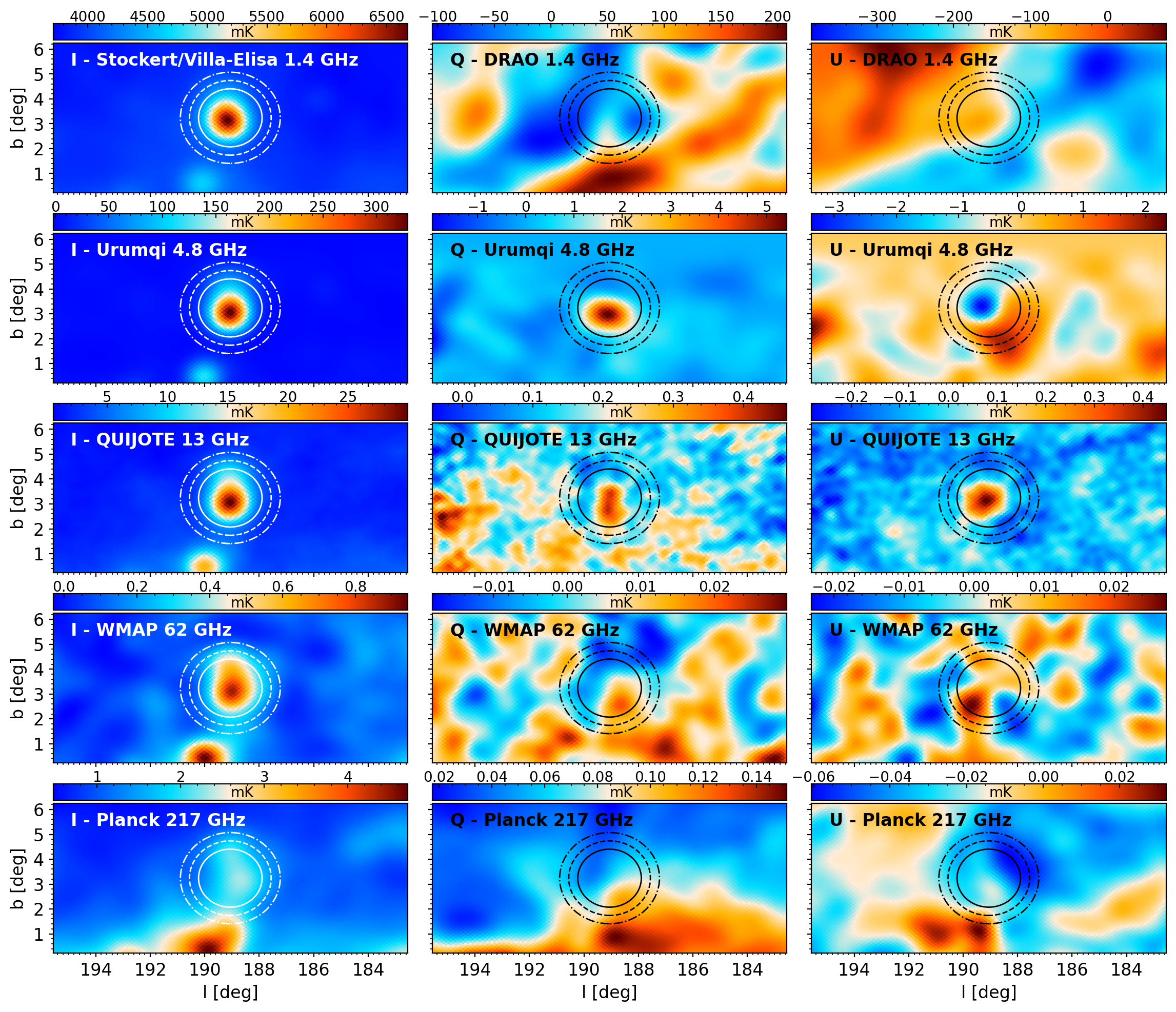} 
\caption{Same as in Fig.~\ref{fig:ws_smt_maps}, but showing this time 
	the IC443 region. The mask is not included to better show the evolution of 
	the background with frequency.}
\label{fig:ic443_smt_maps}
\end{figure*} 

\subsection{Lower frequency ancillary data}
\label{ssec:lowfreq}
In the radio and microwave domains 
we consider the map at 408\,MHz from~\citet{haslam82} survey, which covers the whole sky 
with observations from different facilities (Jodrell Bank MkI and 
MkIA-76\,m, Bonn-100\,m and Parkes-64\,m telescopes), the~\citet{berkhuijsen72} map at 
820\,MHz, obtained from a survey with the Dwingeloo-25\,m telescope, 
the~\citet{reich86} survey at 1420\,MHz conducted with the Stockert-25\,m 
and the Villa-Elisa-30\,m telescopes, and the~\citet{jonas98} survey at 2326\,MHz obtained
with the HartRAO-26\,m telescope.

We actually use the 408, 1420 and 2326\,MHz maps provided by~\citet{platania03}, 
which are corrected for the artifacts produced by observational strategy and 
instrumental gain drifts; the 820\,MHz map is instead generated by projecting into 
{\sc HEALPix} pixelisation the corresponding observational data retrieved from 
the MPIfR's Survey Sampler\footnote{\label{mpifr}\url{http://www3.mpifr-bonn.mpg.de/survey.html}.}.
These radio maps only provide intensity data and are calibrated referring to the 
full-beam solid angle; as explained in~\citet{reich88}, the sidelobe contribution 
would lead to an underestimation of the flux density for sources which are small 
compared to the main beam. This effect is particularly important for the~\citet{reich86} 
survey, and in the aforementioned reference it is argued that a correction factor of 
1.55 can be used to account for the flux density loss. Since the FWHM for the Stockert telescope 
is $\sim35'$, this factor can be applied for W49, but not for W51 and IC443, whose 
angular sizes are larger than the survey beam; in these cases, we employed a more 
conservative value of 1.25. These multiplicative factors are employed to correct the 
flux densities extracted from the 1420\,MHz map in the aperture photometry analysis described 
in Section~\ref{sec:seds}. Although a similar issue affects the 2326\,MHz map, we do 
not apply any explicit correction as the HartRAO beam is considerably smaller (20')
than the size of our sources. It is worth mentioning that the HartRAO map, which we 
use for the total intensity signal, actually carries the combination $I+Q$, and can 
consequently yield biased estimates of $I$ flux densities towards polarised regions.
However, as it is clear from the photometric analysis results in section~\ref{sec:seds}, 
the polarised flux density is typically $<2\%$ of the flux density in total intensity, and the 
conservative 10\% calibration uncertainty adopted for the HartRAO measurement
(see again Section~\ref{sec:seds}) is enough to account for this bias. 

In polarisation we consider the survey carried out by the Dominion Radio Astronomy 
Observatory~\citep[DRAO, ][]{wolleben06}, which provides Stokes $Q$ and $U$ maps for 
the northern sky at 1.41\,GHz. We stress that the DRAO maps are 
delivered\footref{mpifr} following the IAU 
convention for the definition of the Stokes parameters; hence, we adapt it to the convention 
chosen in this work by flipping the sign of the $U$ map. 
Finally, we employ the intensity and polarisation maps from the 
Sino-German $6\,\text{cm}$ survey of the Galactic plane~\citep{gao10,sun11}, 
conducted with the Urumqi-25\,m telescope at 
4.8\,GHz\footnote{\url{http://zmtt.bao.ac.cn/6cm/surveydata.html}.}.

\begin{figure*} 
\centering
\includegraphics[trim= 5mm 0mm 0mm 0mm, scale=0.48]{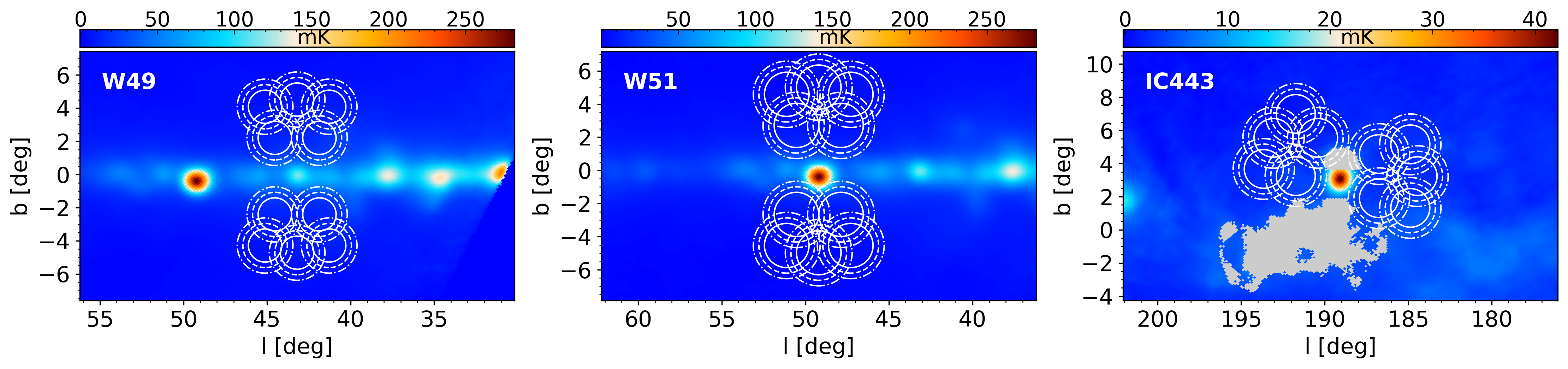} 
\caption{Random apertures employed to estimate the uncertainties on the 
	source flux densities, overplotted for all three regions onto QUIJOTE 
	11\,GHz maps smoothed to a 1\,deg resolution. The IC443 tailored 
	mask is also shown in the third panel. The size of the central 
	apertures and the background annuli are the same as the 
	ones used for measuring the source flux density.  }
\label{fig:err_apertures}
\end{figure*} 

\subsection{Higher frequency ancillary data}
\label{ssec:highfreq}

In the microwave and far-infrared (FIR) range we employ data from the CMB satellite 
missions \textit{WMAP}, \textit{Planck} and \textit{COBE}. 
We consider the five \textit{WMAP} bands at 23, 33, 41, 61, 
and 94\,GHz, using the corresponding maps in the Stokes parameters $I$, $Q$ and $U$ from 
the 9-year \textit{WMAP} data release~\citep{bennett13}, available at the LAMBDA 
data base. 

As per the \textit{Planck} data, we use the maps from the second data 
release~\citep{planck_15_i}, which are publicly available at the \textit{Planck} 
Legacy Archive (PLA) webpage\footnote{\url{https://pla.esac.esa.int/\#home}.}. 
The website provides the nine individual frequency survey maps in intensity, 
at 30, 44 and 70\,GHz from the Low Frequency Instrument (LFI), and at 100, 143, 
217, 353, 545 and 857\,GHz from the High Frequency Instrument (HFI). The 100, 217 
and 353 maps are contaminated by the CO rotational transition lines (1--0), (2--1) 
and (3--2) respectively, so these maps were first corrected using the 
\textit{Planck}-released Type 1 CO maps~\citep{planck_co}, also available in the 
PLA webpage. In polarisation, instead, we use the available maps up to 353\,GHz
from the third data release~\citep{planck_18_i}, whose main improvement with respect 
to the second data release is precisely in the polarisation data. Since after a visual 
inspection of the maps
we detected artifacts affecting the 30\,GHz $U$ map towards W51, just for this frequency 
we adopt instead the polarisation maps obtained from the \texttt{NPIPE} 
pipeline~\citep{planck_ir_lvii}.

Finally, we use the Zodi-Subtracted Mission Average 
maps from the DIRBE instrument on the \textit{COBE} mission~\citep{hauser98} at 1249, 2141 
and 2998\,GHz; these maps are available in {\sc HEALPix} format at the CADE data 
base\footnote{\textit{Centre d'Analyse de Donn\'ees Etendues}, \url{http://cade.irap.omp.eu/dokuwiki/doku.php?id=dirbe}.}. Maps for these frequencies are in total intensity only.


\section{SED of sources}
\label{sec:seds}
The characterisation of the emission from our sources is primarily achieved by analysing their spectral 
energy distribution (SED), i.e., the frequency dependence of the observed flux densities in 
intensity and polarisation. Flux densities are measured with the 
aperture photometry technique, following the implementation 
described in~\citet{genova_santos15b} and also adopted 
by~\citet{genova_santos17} and~\citet{poidevin19} for the study of other 
compact sources. Apart from the previous QUIJOTE studies, aperture photometry has 
already been exploited for SED reconstruction in previous 
works~\citep{lopez_caraballo11, dickinson11, genova_santos11, rubino_martin12a, demetroullas15, planck_ir_xxxi}. 
The technique 
consists in estimating the mean map temperature inside a suitable aperture centred on the 
source, and in removing a background level estimated as the median signal within a 
surrounding annulus; the result is then converted into flux density using the analytical 
conversion factor between temperature units and Jy, and the
angular size subtended by the aperture. The sizes of the aperture and annulus depend 
on the considered source. We adopt the values reported in Table~\ref{tab:sources_info} 
which suit the different angular extents of the three regions; the table also reports 
the reference source position onto which all circles are centred.
This choice is adopted for all the QUIJOTE and ancillary 
maps smoothed to a 1\,deg resolution, and for measuring both the intensity and 
polarisation SEDs.
The corresponding circles are 
overplotted to the intensity and polarisation maps shown in Figs~\ref{fig:ws_smt_maps}
and~\ref{fig:ic443_smt_maps}.   
Notice that although in other works the background is estimated as the mean value of 
map pixels in the annulus, in our case the regions are close 
to the Galactic plane, implying that strong signal variation on scales comparable 
to our apertures may bias the estimation of the background level. The choice of 
the median of the annulus pixels is then made to minimize this 
effect, as discussed in~\citet{rubino_martin12a}.
\input{tables/w49_fluxtable.tex}  
\input{tables/w51_fluxtable.tex}  
\input{tables/ic443_fluxtable.tex} 

The uncertainties on the measured flux densities can be computed as explained 
in~\citet{genova_santos15b}, by using the rms of the pixels located in the 
background annulus (which is not biased by the source emission) and considering the 
contribution from the number of independent pixels in the aperture and in the annulus. 
We found that this method, although providing reasonable error estimates in total 
intensity, tends to underestimate the uncertainties in the $Q$ and $U$ flux densities. 
We then choose to compute the flux density errors, both in intensity and in polarisation, as 
the rms of the flux densities computed 
in 10 apertures randomly located around the source, in such a way as 
to avoid the diffuse emission from the Galactic plane region and other intervening 
compact sources. The chosen apertures are shown in Fig.~\ref{fig:err_apertures} for 
all three regions; the radii for the central aperture and the background annulus are the 
same as the ones employed for computing the source flux densities. These error estimates capture 
the typical fluctuations of the background at the aperture scale, and represent 
therefore a more reasonable estimate for the flux density uncertainties.  

We also include a calibration error \textit{a posteriori}, which is taken as 
a fixed fraction of the measured flux density and added in quadrature to the rms of the random 
apertures. This additional component is chosen to include all possible systematics 
affecting our flux density measurements. We adopt a conservative calibration error equal to 
10\% of the flux density for all the low frequency points, from 408\,MHz to 4.8\,GHz; this 
choice is consistent with previous QUIJOTE studies, as in~\citet{genova_santos17} and~\citet{poidevin19}.
Indeed, these points have a high statistical weight in the modelling of the source emission, 
since they anchor the total intensity level of synchrotron and free-free amplitudes at low 
frequencies, and as such affect the required level of AME in the microwave range. 
Similarly, the DRAO and Urumqi points are the only available low-frequency polarised flux densities, 
and their statistical weight is high in fixing the slope of the polarised synchrotron power law.
However, these points are also affected by important systematic effects (for instance, the main-beam 
calibration issue described in Section~\ref{ssec:lowfreq}), so that a significative calibration error is 
required to mitigate them. For the specific case of the Urumqi point we also consider an additional contribution to the error in the $Q$ and $U$ flux densities, taken as a fraction ($\sim$1\%) of the measured flux density in total intensity; this is done in view of the irregular trend shown by Urumqi polarised maps in Figs~\ref{fig:ws_smt_maps} and~\ref{fig:ic443_smt_maps}, in order to account for possible intensity-to-polarisation leakages. Finally, following again a similar convention as in other QUIJOTE-MFI studies, 
we include conservative calibration uncertainties of 5\% for QUIJOTE points, 3\% for \textit{WMAP} and 
\textit{Planck} up to 353\,GHz, 6.1\% for 545\,GHz, 6.4\% for 857\,GHz and 11.9\% for DIRBE 
points~\citep{planck_er_xx,planck_ir_xv,poidevin19,cepeda_arroita21}.

Apart from the Stokes parameters $Q$ and $U$, it is interesting to investigate the frequency 
dependence of the total polarised intensity $P$. The flux density for the latter can be evaluated 
as $P=(Q^2+U^2)^{1/2}$, and the corresponding uncertainty $\sigma_P$
is obtained by propagating the uncertainties $\sigma_Q$ and $\sigma_U$ on the Stokes parameters, 
measured on the corresponding maps. 
We also consider the polarisation fraction, defined as $\Pi=P/I$, 
whose uncertainty is determined by propagating the errors on $I$ and $P$. It is important to 
notice that these definitions of $P$ and $\Pi$ result in their posterior distributions being 
non-Gaussian; as detailed in~\citet{rubino_martin12a}, this may result in the final polarisation 
estimates being positively biased, especially when the signal-to-noise ratio of the detection 
is low. Since studies of AME polarisation generally quote upper limit on $\Pi$, this issue is 
particularly relevant in our analysis. Hence, we apply a debiasing correction according to 
the methodology described in~\citet{rubino_martin12a}, whereby 
the most likely value for the polarised flux density and fraction is obtained 
by integrating the posterior distributions on $P$ and $\Pi$, respectively. 
The posterior distribution on $P$ has been derived analytically in the 
literature~\citep{vaillancourt06}, and we adopt it to debias the polarisation flux densities. 
For the case of $\Pi$, instead, the posterior is 
reconstructed with a Monte Carlo approach, by drawing random values for $Q$ and $U$; 
values are sampled from a normal distribution whose width is given by the measured uncertainties 
on the Stokes parameters. The same approach was also adopted in~\citet{genova_santos17}.

The measurements of the polarised Stokes parameters $Q$ and $U$ also allow to estimate the 
polarisation angle $\gamma$ for the three regions at different frequencies. We 
adopt the definition $\gamma=-0.5\arctan{(U/Q)}$,
where it is understood that the arctangent is to be evaluated element-wise, taking into 
account the individual signs of $Q$ and $U$, in order to have the final angle defined 
in $[-\pi/2,\pi/2]$. Although we adopt the {\sc HEALPix} convention for the sign of the Stokes parameters, 
this expression still ensures the angle is measured positive north through east.

Finally, we apply colour corrections (CC) to our flux density estimates, in order to account for 
the signal alteration due to the finite instrumental bandpass. This issue is particularly 
relevant for QUIJOTE, \textit{WMAP}, \textit{Planck} and DIRBE, whereas for the lower 
frequency surveys the bandwidth is typically narrow enough to make this correction unnecessary.
Information on the bandpasses for the aforementioned satellite-based data was retrieved 
from the LAMBDA and PLA archives, whereas for QUIJOTE the measured instrumental bandpasses 
were employed~\citep[see][ for details]{genova_santos22}. The CC computation requires 
the bandpass integration of a suitable model for the spectral dependence of the source 
flux density; for this we employ the model we fit on the data (Section~\ref{sec:fits}) in an iterative 
procedure. The model initially fitted on the uncorrected flux densities is used to obtain a first 
CC estimation; these CCs are used to correct the initial flux densities and their uncertainties 
before repeating the fit. 
This procedure is repeated until convergence, which is always reached within the third 
iteration. The magnitude of the final colour corrections is 
typically $\lesssim2\%$ for QUIJOTE, $\lesssim3\%$ for \textit{WMAP}, 
$\lesssim1.5\%$ for \textit{Planck}-LFI and $\lesssim10\%$ for \textit{Planck}-HFI and DIRBE.

\begin{figure*} 
\centering
\includegraphics[trim= 0mm 0mm 0mm 0mm, scale=0.39]{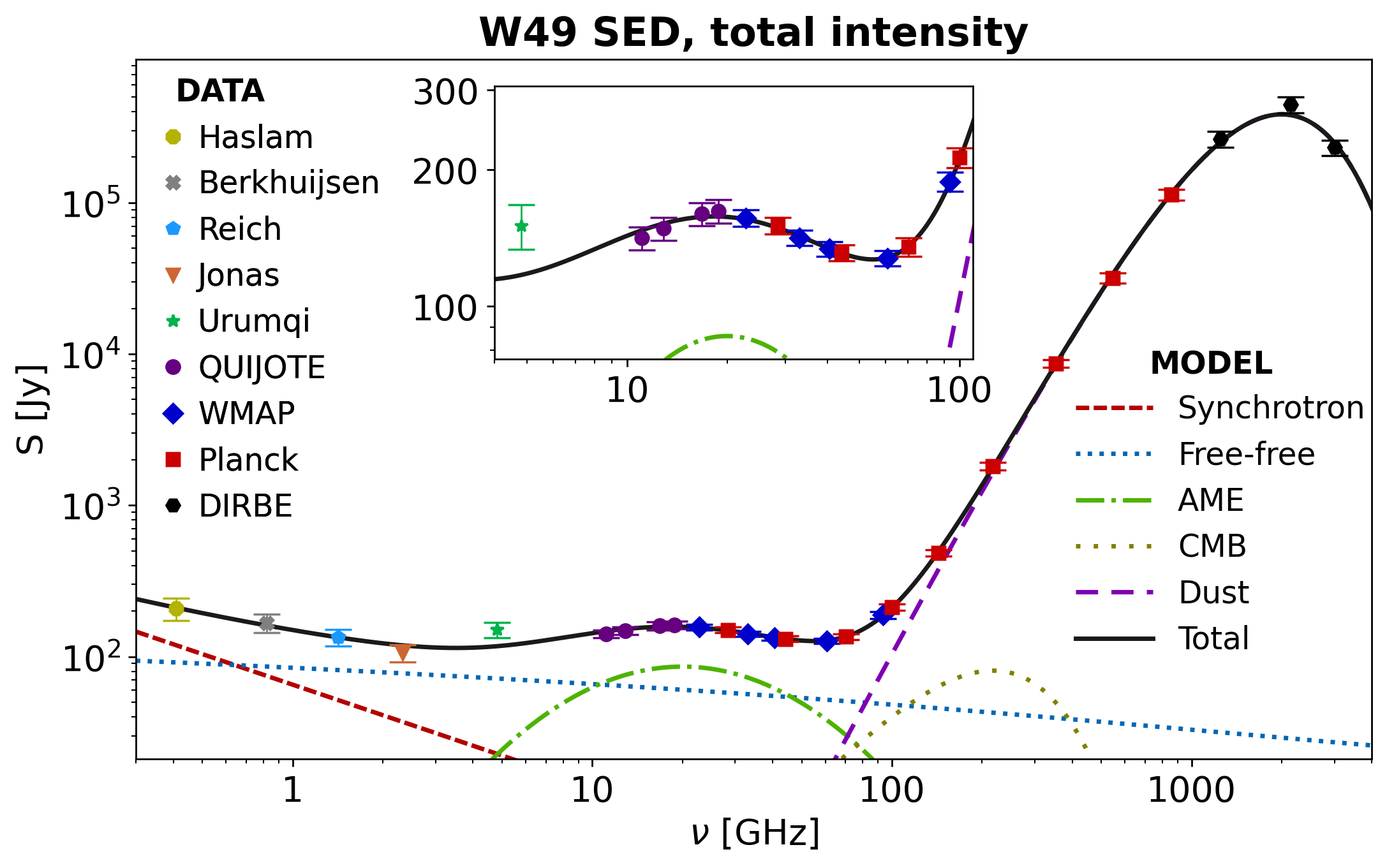}
\includegraphics[trim= 0mm 0mm 0mm 0mm, scale=0.39]{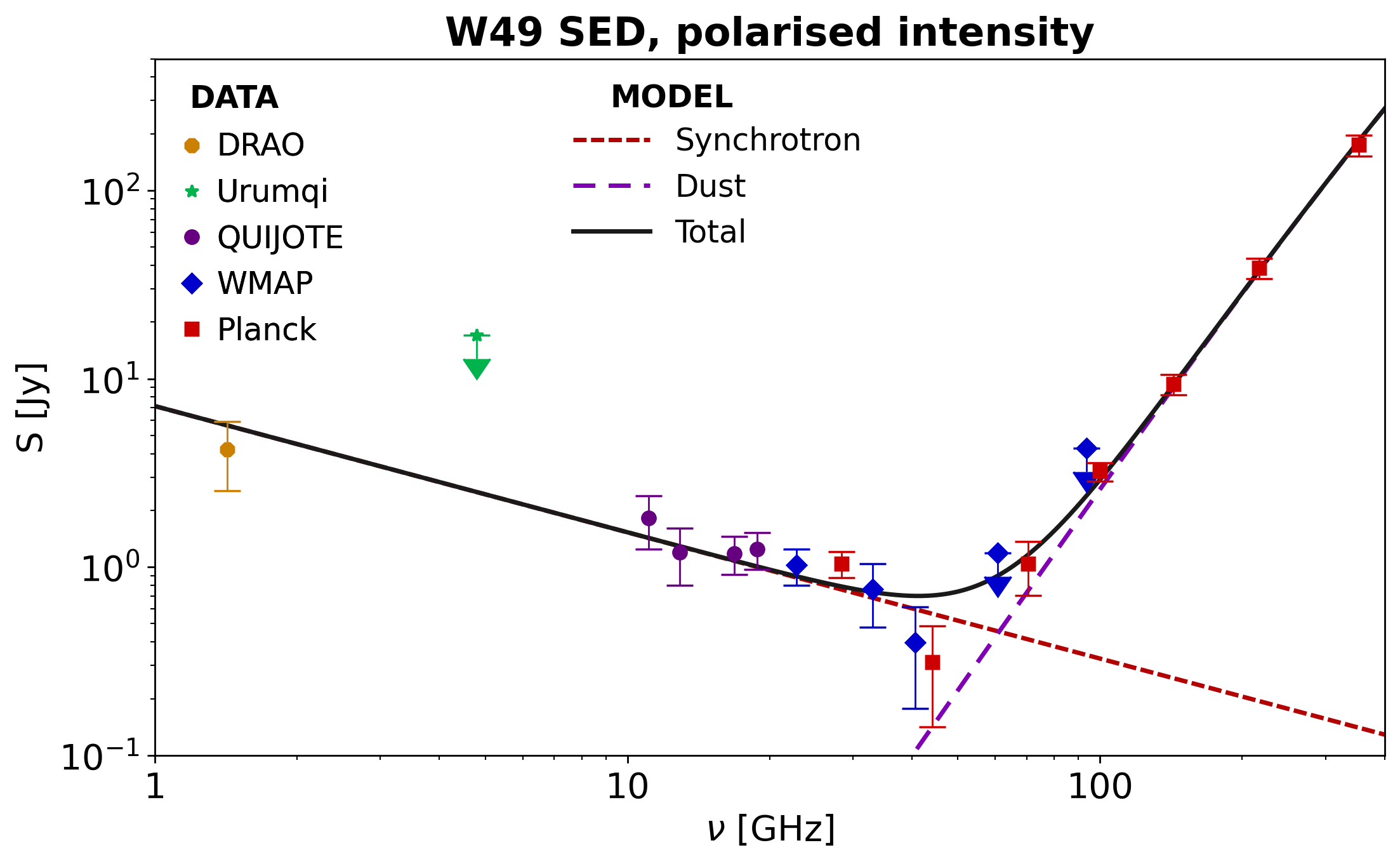} 
\includegraphics[trim= 0mm 0mm 0mm 0mm, scale=0.39]{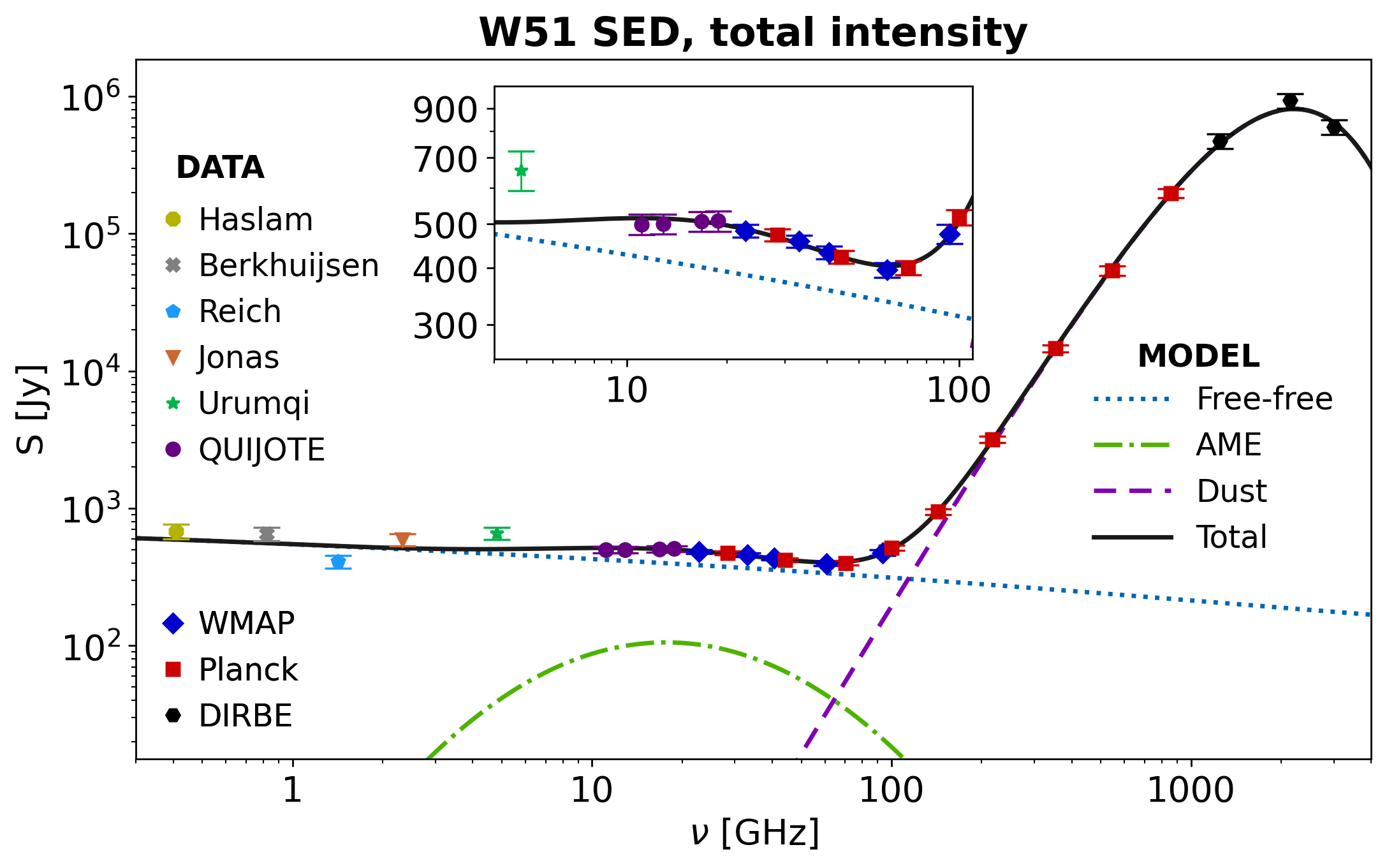}
\includegraphics[trim= 0mm 0mm 0mm 0mm, scale=0.39]{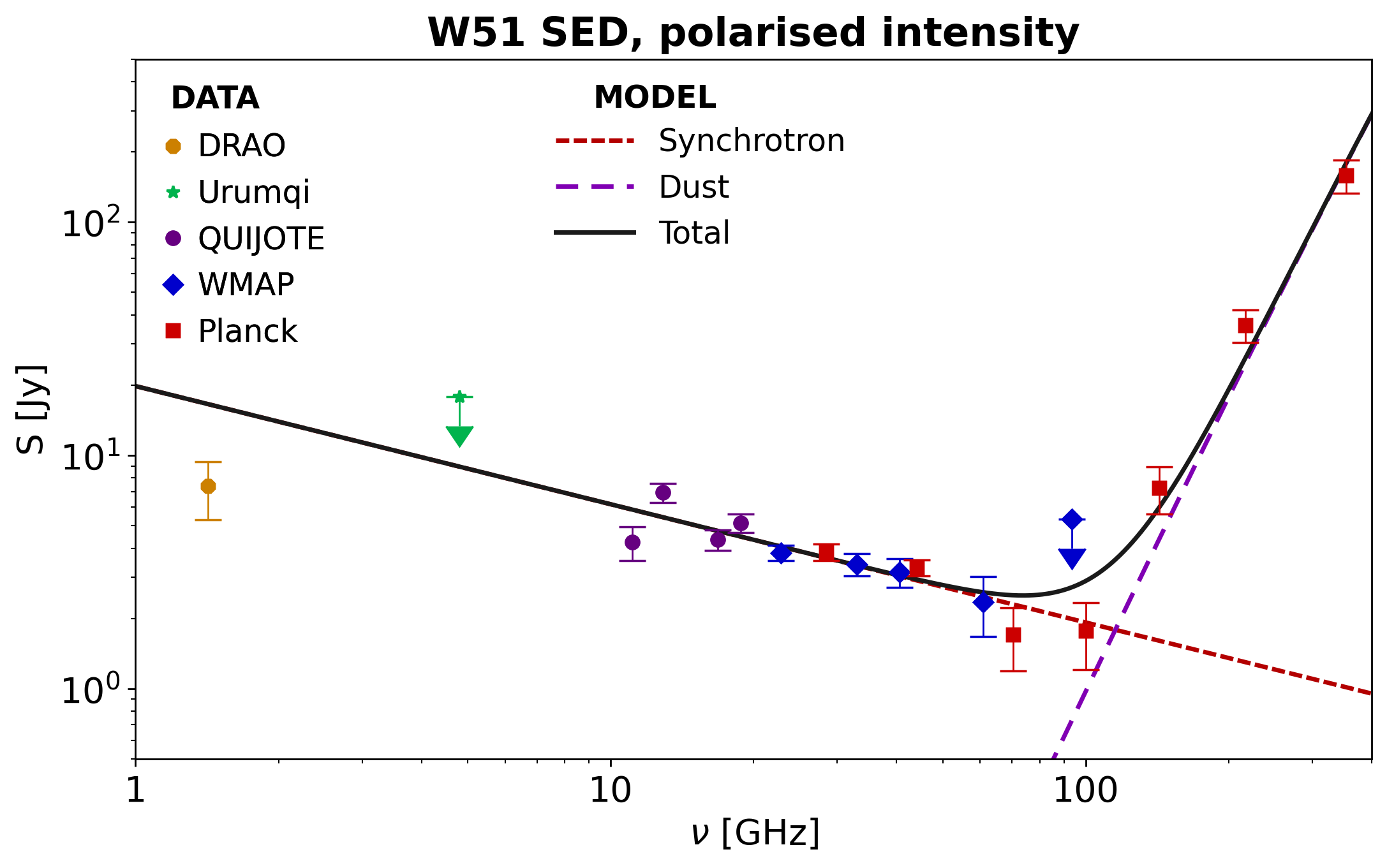} 
\includegraphics[trim= 0mm 0mm 0mm 0mm, scale=0.39]{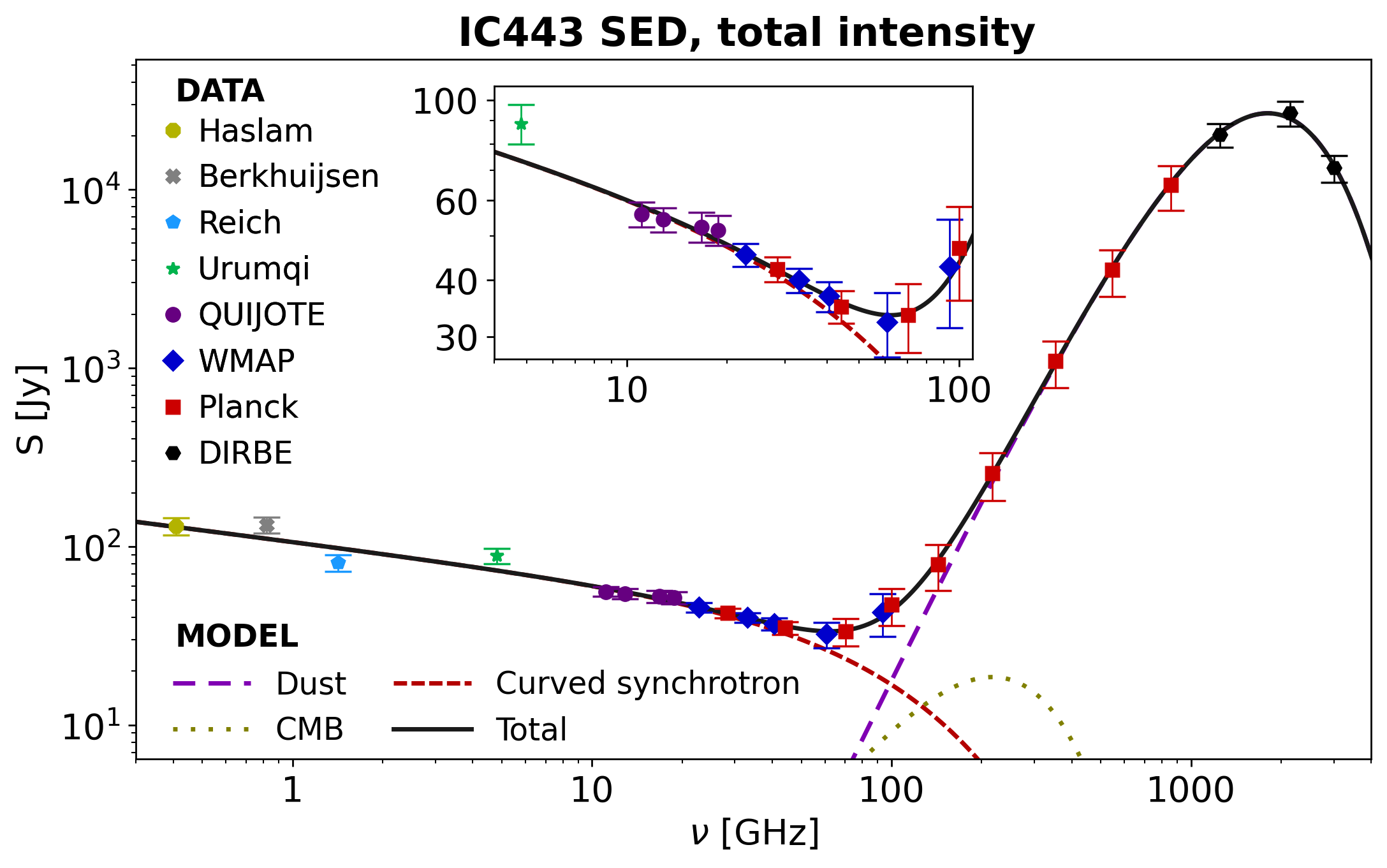}
\includegraphics[trim= 0mm 0mm 0mm 0mm, scale=0.39]{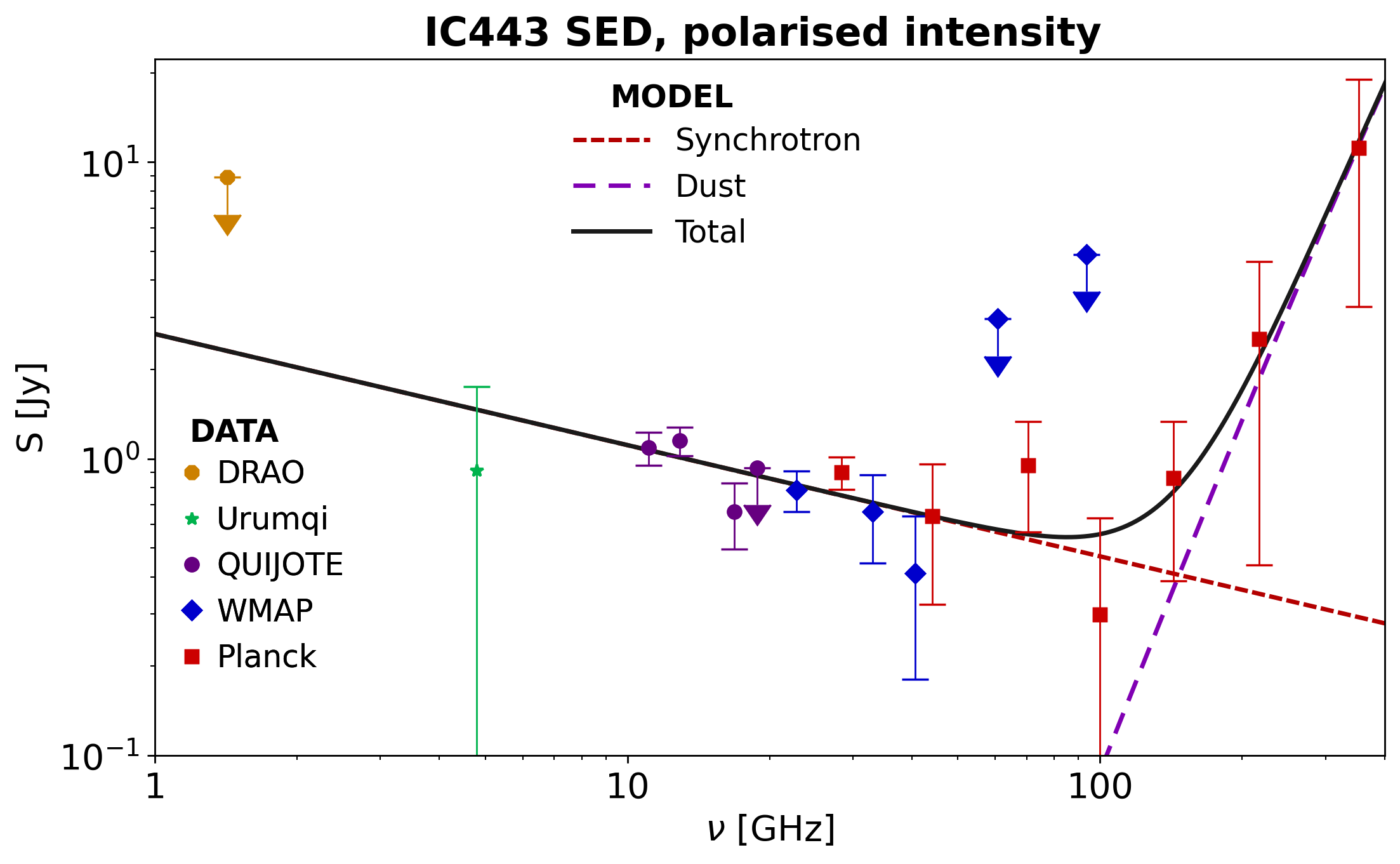} 
\caption{SEDs for W49, W51 and IC443, obtained from the combination of QUIJOTE 
	and ancillary data, for the total intensity $I$ (left panels) 
	and the total polarised intensity $P$ (right panels). The flux densities 
	have been obtained via aperture photometry in a 1-degree radius aperture 
	centred in the source, as detailed in Section~\ref{sec:seds}, considering 
	all maps smoothed at a 1\,deg resolution. QUIJOTE points are the only available 
	data for these regions covering the range between 10 and 20\,GHz. Overplotted to 
	the data point we show the best-fit models discussed in Section~\ref{ssec:apply_fit};
 	the foregrounds considered for each fit are quoted in the plot legends.
	In total intensity the model plotted here for W49 is the one including a synchrotron 
	component with a fixed spectral index (third column in Table~\ref{tab:w49_fit}), 
	the model for W51 is the one with no synchrotron component (second column in 
	Table~\ref{tab:w51_fit}), and for IC443 we show the model with a broken 
	power-law synchrotron spectrum (last column of Table~\ref{tab:ic443_fit}). 
	When fitting for a dust component in the polarised SEDs, the dust temperature is 
	always fixed to the value obtained in total intensity; for W51, the DRAO point was 
	not included in the polarisation fit.}
\label{fig:ip_seds}
\end{figure*}

\begin{figure} 
\centering
\includegraphics[trim= 0mm 0mm 0mm 0mm, scale=0.38]{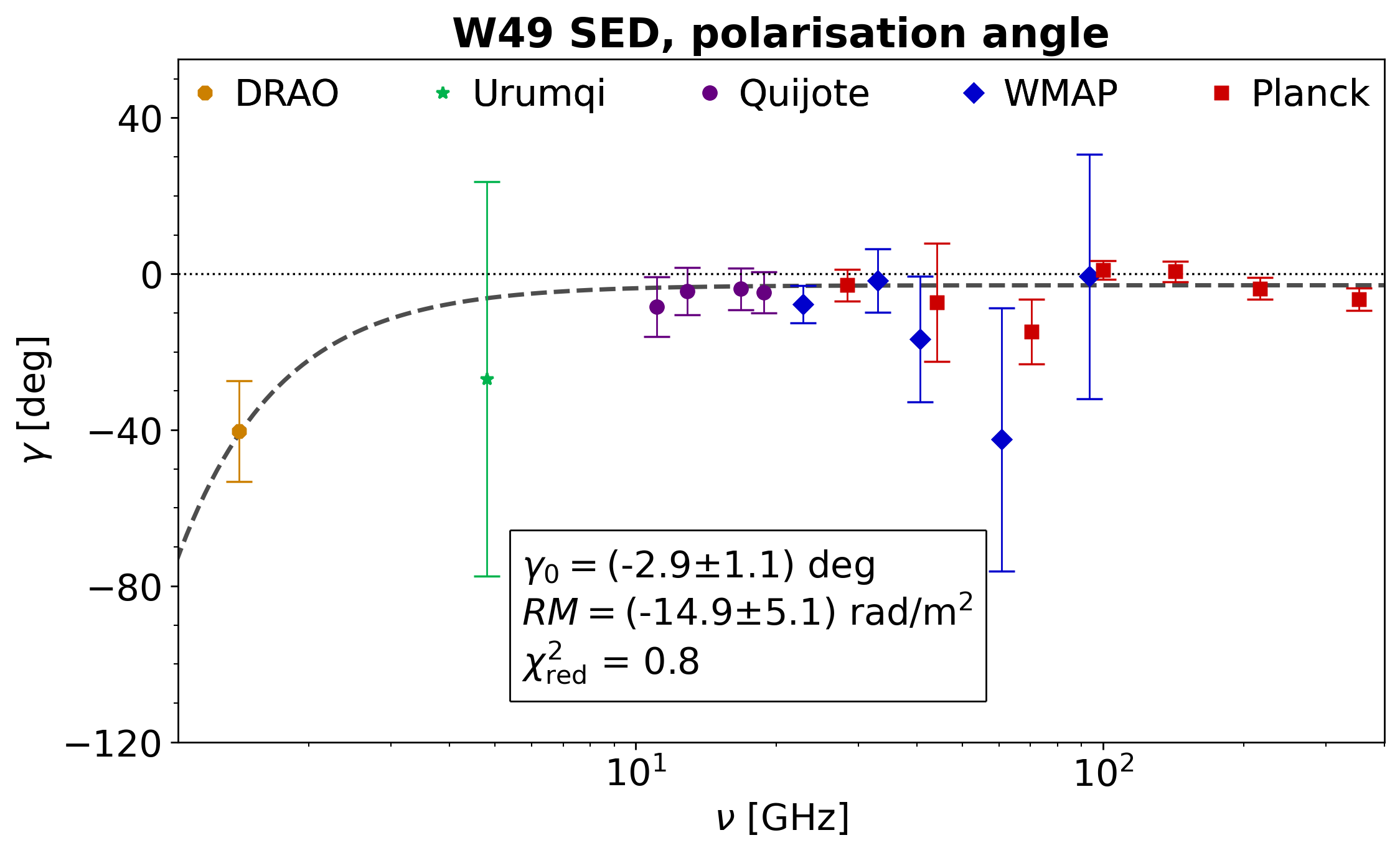} 
\includegraphics[trim= 0mm 0mm 0mm 0mm, scale=0.38]{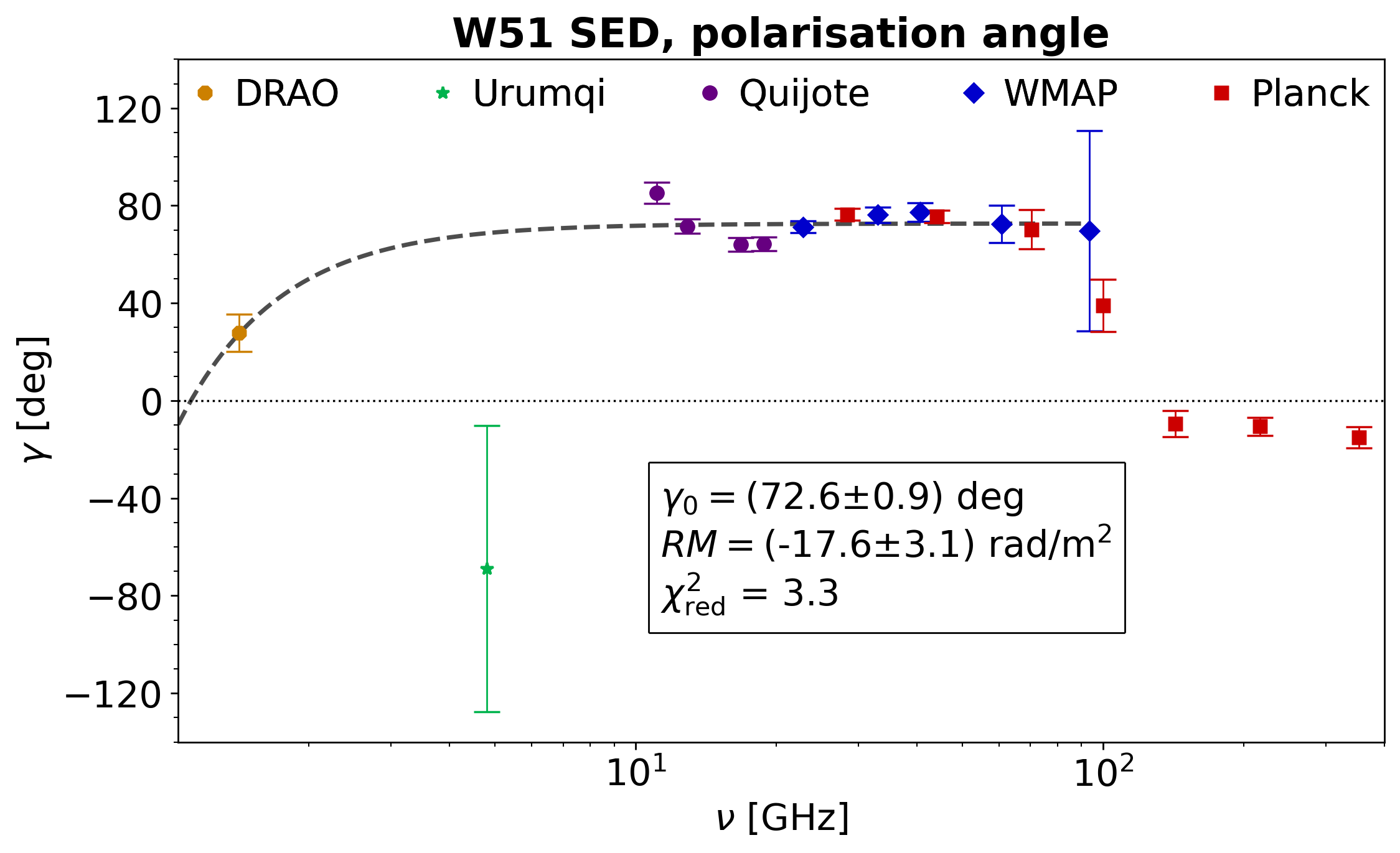} 
\includegraphics[trim= 0mm 0mm 0mm 0mm, scale=0.38]{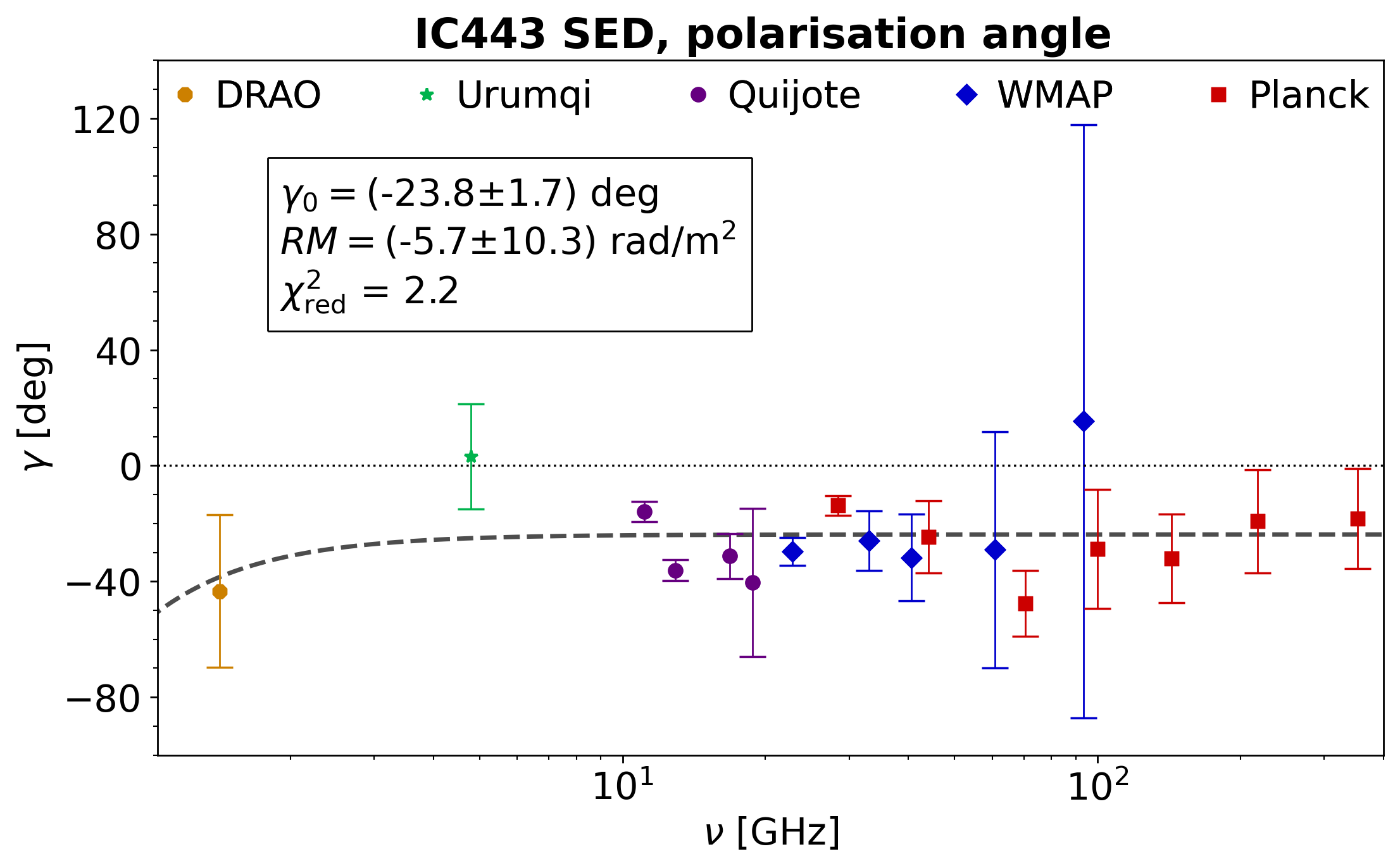} 
\caption{Polarisation angle $\gamma$ as a function of frequency, towards the three regions. 
	We also overplot to the data points the best-fit Faraday rotation model from 
	equation~\eqref{eq:rm}, and report the resulting parameter estimates in each panel.}
\label{fig:polang}
\end{figure}

The final estimates for the parameters $I$, $Q$, $U$, $P$, $\Pi$ and $\gamma$ are reported in 
Tables~\ref{tab:w49_fluxes},~\ref{tab:w51_fluxes} and~\ref{tab:ic443_fluxes} for the regions 
W49, W51 and IC443 respectively. Notice that the 2326\,MHz point is not available for IC443 as the 
source is located outside the survey footprint. The DRAO polarised fraction $\Pi$ is computed 
using the total intensity value $I$ measured from the~\citet{reich86} map at the same frequency. 
Finally, the debiasing employed to compute $P$ can result in a final null flux density, especially when 
the signal-to-noise ratio for $Q$ and $U$ is low; in that case we quote the 95\% confidence level 
as the upper limit for the polarised flux density. 
The same considerations apply to the polarisation fraction $\Pi$.
Notice that the tables report the flux densities as they are obtained from the 
photometric measurements, prior to any colour correction; the CC coefficients are reported in 
the last column of each table. For all three regions, the corresponding SEDs in $I$ and $P$ are 
plotted in Fig.~\ref{fig:ip_seds}, while the frequency dependence of the polarisation angle is 
shown in Fig.~\ref{fig:polang}. Unlike the tables, the plots show 
the final colour-corrected SEDs from the last iteration of the multi-component fit described in 
Section~\ref{ssec:fit_method}. 
 
\section{Modelling the source emissions}
\label{sec:fits}

This section is devoted to the physical interpretation of the flux densities measured towards 
the three sources considered in this work, taking into account both the total intensity
and the polarisation SEDs.

\subsection{Methodology}
\label{ssec:fit_method}

Our modelling of the source emissions considers the four continuum foreground mechanisms listed in 
Section~\ref{sec:introduction}. At low frequency, the contribution from the synchrotron 
emission is modelled with a power law:
\begin{equation}
\label{eq:sync}
	S_{\rm sync}(\nu) = A_{\rm s} \left( \frac{\nu}{\nu_{\rm p}} \right)^{\alpha_{\rm s}},
\end{equation}
with a pivot frequency $\nu_{\rm p}=1\,\text{GHz}$; we fit in this case for the synchrotron 
amplitude $A_{\rm s}$ at 1\,GHz, and for the synchrotron spectral index $\alpha_{\rm s}$. 
We will consider the synchrotron contribution both in intensity and polarisation. 

The free-free flux density as a function of frequency is given by~\citep[see for example][]{wilson09}:
\begin{equation}
	S_{\rm ff}(\nu) = \frac{2k_{\rm B}\nu^2\Omega}{c^2}T_{\rm e}\left( 1- e^{-\tau_{\rm ff}(\nu)} \right), 
\end{equation}
which is the radiation transfer equation for the free-free emission, with the first 
fraction representing the conversion factor between brightness temperature and flux density in the 
Rayleigh-Jeans limit; 
$k_{\rm B}$ is the Boltzmann constant, $c$ the speed of light, $T_{\rm e}$ the electron 
temperature and $\Omega$ the solid angle 
subtended by the chosen photometry aperture. All the information on the 
emission is encoded in the free-free opacity $\tau_{\rm ff}$, for which we adopt the parametrisation 
provided by~\citet{draine11}:
\begin{equation}
\label{eq:tau_ff}
	\tau_{\rm ff}(\nu) = 5.468\times10^{-2} \left(\frac{T_{\rm e}}{\text{K}}\right)^{-1.5} \left(\dfrac{\nu}{\text{GHz}}\right)^{-2} \left(\frac{\text{EM}}{\text{pc}\,\text{cm}^{-6}}\right) g_{\rm ff}(\nu),
\end{equation}
where the frequency $\nu$, the electron temperature $T_{\rm e}$ and the emission measure $\text{EM}$ 
are expressed in convenient units, 
and $g_{\rm ff}$ is the Gaunt factor. For the latter we also choose the parametrisation 
 given in~\citet{draine11}, namely:
\begin{align}
	&g_{\rm ff}(\nu) = \nonumber \\
	&= \ln{\left[ \exp{\left( 5.960 - \dfrac{\sqrt{3}}{\pi} \ln\left[ Z\left(\dfrac{\nu}{\text{GHz}}\right)\left(\dfrac{T_{\rm e}}{10^4\,\text{K}}\right)^{-3/2}\right] \right)} + e \right]},
\end{align}
where $e$ is Napier's constant and we set the charge number to $Z=1$ (hydrogen plasma).
This modelling for the free-free emission is found to be more accurate than the conventional 
one~\citep{wilson09} over a broader 
frequency range, and was also employed in~\citet{planck_ir_xv}. We want to fit for the value 
of the emission measure EM, which is defined 
as the squared electron density integrated along the line of sight and is generally adopted 
as an indicator of the strength of the free-free emission. Notice that it would not be possible to fit 
simultaneously for the emission measure and the electron temperature, given the strong degeneracy 
between the two parameters that is clear from equation~\eqref{eq:tau_ff}. For the electron 
temperature we then 
adopt the values extracted from the public \textit{Planck} COMMANDER free-free maps; more 
specifically, for each region we take the average of the $T_{\rm e}$ values found in the same 
aperture used to extract the flux density. We obtain the values $7000\,\text{K}$ for both W49 and W51, 
and $6861\,\text{K}$ for IC443. 

The thermal emission from ISM dust grains is clearly evident in the FIR intensity SEDs for all 
regions. As customary, we model this emission as a modified blackbody, where in the optically 
thin regime the dust blackbody spectrum $B_{\nu}(T_{\rm dust})$ is weighted by the dust optical 
depth $\tau_{\text{dust}}$:
\begin{align}
\label{eq:tdust}
	S_{\rm dust}(\nu) &= \tau_{\text{dust}}(\nu)  B_{\nu}(T_{\rm dust})\Omega = \nonumber \\
	&= \tau_{\rm 250}\left(\frac{\nu}{1200\,\text{GHz}}\right)^{\beta_{\rm dust}}  B_{\nu}(T_{\rm dust})\Omega.
\end{align}
In the second equality we expressed the frequency scaling for the dust opacity as a power law around 
a pivot frequency of 1200\,GHz (or $250\,\mu\text{m}$). In this case we fit for the dust temperature 
$T_{\rm dust}$, emissivity $\beta_{\rm dust}$ and optical depth $\tau_{\rm 250}$ at $250\,\mu\text{m}$.

The AME emission is modelled phenomenologically as a parabola in the log-log plane~\citep{stevenson14}:
\begin{equation}
	\label{eq:ame}
	S_{\rm AME}(\nu) = A_{\rm AME}\cdot\exp{\left\{-\frac{1}{2}\left[ \frac{\ln({\nu / \nu_{\rm AME}})}{W_{\rm AME}} \right]^2\right\}}.
\end{equation}
In this case we fit for the peak frequency $\nu_{\rm AME}$, the peak amplitude $A_{\rm AME}$ and 
the width of the parabola $W_{\rm AME}$. 
A similar form for the AME fitting function, although 
with a different meaning of the parameters, was proposed by~\citet{bonaldi07}, and employed in the 
previous QUIJOTE publications. Such a model, however, has the drawback of coupling the peak frequency 
and the width of the parabola (meaning that variations in $\nu_{\rm AME}$ would alter the value of 
$W_{\rm AME}$). The functional form from equation~\eqref{eq:ame}, instead, disentagles the parameters 
$\nu_{\rm AME}$ and $W_{\rm AME}$, which can now be varied independently. 

Finally, we fit for a CMB contribution, expressed as the differential of the flux 
density with respect to the
temperature times the CMB temperature fluctutation $\Delta T_{\rm CMB}$:
\begin{equation}
	S_{\rm CMB}(\nu) = \dfrac{h^2 \nu^4}{2k_{\rm B}T_{\rm CMB}^2 c^2} \sinh^{-2}{\left(\dfrac{h\nu}{2k_{\rm B}T_{\rm CMB}}\right)} \,\Omega \,\Delta T_{\rm CMB},
\end{equation}
where $h$ is the Planck's constant and we adopt the reference CMB temperature 
$T_{\rm CMB}=2.7255\,\text{K}$~\citep{fixsen09}. In this case we fit for the value of $\Delta T_{\rm CMB}$.

The SED modelling is done with a  Levenberg-Marquardt least-square fit~\citep{markwardt09} between our data points and the sum of the chosen foreground models, implemented 
via the Interactive Data Language (IDL) {\sc MPFITFUN} function\footnote{\url{https://www.l3harrisgeospatial.com/docs/mpfitfun.html}.}. 
As anticipated in Section~\ref{sec:seds}, the model is used to 
compute colour correction coefficients for all frequencies above 10\,GHz; these corrections are 
applied to the original flux densities and to their original uncertainties,
and the fit is repeated, until convergence. The same procedure is adopted for both 
the intensity and polarisation SEDs. Before choosing the final model that 
best reproduces the SED of each source, we explore all different foreground combinations; this iterative correction 
process is applied consistently in each case. 

Notice that the $\chi^2$ minimization performed in our fit assumes that all measurements are 
statistically independent. This is not entirely true for adjacent QUIJOTE frequencies belonging 
to the same horn, whose noise is correlated~\citep[see][for a more complete discussion]{rubino_martin22}. 
We check for the effect this correlation can have on our source modelling in the following way.  
We combine the 11 and 13\,GHz flux densities and errors into an effective 12\,GHz measurement, using the correlation 
coefficients reported in~\citet{rubino_martin22} when computing the effective uncertainty; the same 
procedure is repeated for the 17 and 19\,GHz points, yielding and effective 18\,GHz flux 
density estimate. 
When repeating the fit using this set of reduced QUIJOTE data, we find that the resulting best-fit 
parameters are compatible within their uncertainties with the full QUIJOTE data case. We 
therefore conclude that the existing noise correlations between adjacent QUIJOTE frequencies are 
negligible as far as our source modelling is concerned.  

Information on the polarised emission can also be extracted from the frequency dependence of
the polarisation angle shown in Fig.~\ref{fig:polang}. Variations of $\gamma$ with 
frequency can be ascribed to Faraday rotation from the neighbouring ionised medium. 
The effect can be modelled, as a function of wavelength, as:
\begin{equation}
	\label{eq:rm}
	\gamma(\lambda) = \gamma_0 + \text{RM}\,\lambda^2,
\end{equation}
where $\gamma_0$ is the polarisation angle for $\lambda=0$ (or $\nu\rightarrow\infty$) 
and $\text{RM}$ is the rotation measure, which is proportional to the line-of-sight integral 
of the electron density times the parallel component of the magnetic field. We fit for 
 $\gamma_0$ and $\text{RM}$ over our measured $\gamma$ values, and report the best-fit 
 estimates in each panel in Fig.~\ref{fig:polang}. The resulting model for the polarisation 
 angle is overplotted as a dashed line to the data.

\subsection{Application to W49, W51 and IC443}
\label{ssec:apply_fit}

The SEDs in total intensity shown in Fig.~\ref{fig:ip_seds} can be modelled by 
a suitable combination of the four foregrounds mentioned in the beginning of this 
section. The contribution of thermal dust emission is clear in all the regions, 
producing the characteristic peak at FIR frequencies. The spectrum in the 
lowest frequency range shows a decrease which is due to either synchrotron or free-free 
emission, or a combination of the two. Any possible AME contribution would emerge 
in the microwave range as an excess signal with respect to the combination of synchrotron 
and the thermal emissions. The actual detection of AME towards these regions is then 
subject to a clear characterisation of the other foregrounds; this is particularly 
crucial for the low frequency points, where the free-free and synchrotron components 
are degenerate. For each region we will then repeat the fit by fixing the spectral 
index of the synchrotron power law to the value estimated from the polarised 
intensity SED, and fit only for its amplitude. We try 
different combinations of the four foreground emissions and finally quote the results 
that provide the best modelling of the measured SED; this is assessed evaluating both the 
resulting $\chi^2$ and the 
physical meaning of the best-fit parameter values. A summary of the fit results is reported 
in Tables~\ref{tab:w49_fit},~\ref{tab:w51_fit} and~\ref{tab:ic443_fit} for W49, W51 and 
IC443, respectively. The tables report the best-fit parameter values, the associated 
$\chi^2$ and the corresponding reduced chi-square defined as $\chi^2_{\rm red}=\chi^2/\text{dof}$, 
where the number of degree of freedom (dof) is computed as the number of fitted points 
minus the number of free parameters. 

The polarised SEDs plotted in Fig.~\ref{fig:ip_seds} show the joint contribution from a decaying 
component at low frequencies, and a raising component in the FIR range. The latter is to be 
ascribed to thermal dust emission, which is known to be polarised with fraction up to 20\% in some 
regions of the sky~\citep{planck_18_xi}; in our case the polarised flux densities 
suggest a low degree of 
polarisation, at the per cent level. The decaying part of the spectrum can be due either to free-free 
or to synchrotron. Although the former can yield a non-null residual polarisation fraction at the 
level $\lesssim 1\%$~\citep{trujillo_bueno02}, as commented in Section~\ref{sec:introduction} it is 
generally assumed a non-polarised foreground. The polarised signal observed towards our 
regions at low frequency is then most likely to be attributed to synchrotron emission. 

In principle, we could adopt the functional forms from equations~\eqref{eq:sync} and~\eqref{eq:tdust} 
and fit for a combination of the two foregrounds in polarisation. However, our FIR polarised flux 
densities only cover a portion of the rising dust spectrum, which is not effective in constraining the 
combination of the dust parameters. For this reason, we fix the dust temperature to 
the value obtained from the study of the corresponding SED in total intensity, and only fit for the 
dust emissivity and optical depth. Notice that we also considered a possible polarised 
AME contribution in our SED model, using the same functional form from equation~\eqref{eq:ame}. 
However, in all cases we found that the inclusion of a polarised AME component degrades the 
quality of the fit and yields non-physical values for the associated parameters. We conclude 
that there is no proof of any measurable AME polarised emission; as such, we decided not to 
include this foreground in the polarisation fit. The fit results for the three regions in 
polarisation are overplotted to the colour-corrected flux densities in the right panels of 
Fig.~\ref{fig:ip_seds}. In the fit we also kept the points that only provide upper limits on  
polarisation, by setting their flux density to zero and using their 1-$\sigma$ uncertainty.

\subsubsection{W49} 
\label{sssec:w49_fit}
For W49, we find that the SED in intensity is best modelled by a combination of all the 
aforementioned components with the exception of synchrotron; the resulting best-fit parameter 
values are listed in Table~\ref{tab:w49_fit}. 
 Our estimates for the free-free emission measure, the 
CMB temperature fluctuation and the thermal dust parameters are in agreement with those obtained 
in\footnote{For the value of the optical depth we refer to the \textit{erratum}~\citet{planck_erratum}.}~\citet{planck_ir_xv}.
At low frequencies, the contribution from the free-free emission alone 
provides the best fit to the SED; when including a synchrotron component, the resulting best-fit
spectral index is excessively steep ($\alpha_{\rm s}\simeq-1.3$) and its amplitude
$A_{\rm s}$ is compatible with zero, so that we decided not to include a synchrotron contribution. 
Such a result can be understood, given that the bulk of the emission 
 comes from the HII regions associated with W49A. The signal observed in polarisation 
towards W49 can be ascribed to the SNR W49B or, more likely, to the diffuse Galactic synchrotron emission; in any 
case, its contribution in total intensity is hindered by the brighter thermal \textit{bremmstrahlung} 
coming from the star-forming regions. Hence, for simplicity, we decided not to include it in 
our source modelling; although we know that a synchrotron component is present, its inclusion 
yields non-physical results for its amplitude and spectral index, while only marginally changing 
the parameter values for the other emissions. 

The situation is different when analysing the polarised SED, which clearly shows a decreasing 
spectrum at low frequencies that can be ascribed to synchrotron; we then fit the spectrum with 
a combination of synchrotron and thermal dust, fixing $T_{\rm dust}$ to the value obtained in 
total intensity. The resulting spectral index is $\alpha_{\rm s}= -0.67\pm0.10$, which is 
consistent with the expected range for the synchrotron emission.
It is then possible to try and fit for a synchrotron component in total intensity, fixing the 
spectral index to the value obtained in polarisation and only leaving the amplitude $A_{\rm s}$ as 
a free parameter. This time the synchrotron amplitude is detected at $2\,\sigma$, while the best-fit
values for the free-free emission measure and the CMB temperature fluctuation have respectively a 
worse and better agreement with~\citet{planck_ir_xv}; this can be expected as this reference does 
not include a synchrotron component in the source modelling, so that their free-free result accounts for the whole 
amplitude of the emission at low frequencies. 

\input{tables/w49_fitres.tex}  

Still, the joint contribution of free-free, CMB and thermal dust is not enough to 
account for the microwave flux densities in total intensity, with QUIJOTE 
points clearly suggesting a downturn of the AME excess in this 
frequency range. We found that the inclusion of our AME functional form yields the best modelling with 
$\chi^2_{\rm red}=0.69$ without synchrotron and $\chi^2_{\rm red}=0.52$ when fixing $\alpha_{\rm s}$ 
(by comparison, a fit without AME would result in $\chi^2_{\rm red}=3.60$ and $\chi^2_{\rm red}=3.78$
respectively, with a flip in sign for the CMB temperature fluctuation).
In the end, we can choose the modelling that includes the synchrotron with fixed spectral index as 
our reference model for W49, as it provides the best $\chi^2_{\rm red}$ value and yields physically reasonable 
values for all resulting parameters. 
The AME peak frequency is $\nu_{\rm AME}=20.0\pm1.4\,\text{GHz}$, where it is 
detected with a significance of $\sim 4.7\sigma$ and accounts for $\sim 55\%$ of the observed 
intensity emission towards W49. 

W49 does not appear in QUIJOTE maps as a strongly polarised source. In Fig.~\ref{fig:ws_qjt_maps} it is 
visible in $Q$ as a local increase in the intensity of the polarised background, while there are 
only hints of positive $U$ towards the source. The flux densities quoted in 
Table~\ref{tab:w49_fluxes} confirm 
this picture with most of the signal coming from positive $Q$. Since this polarised state is typical 
of the diffuse Galactic plane emission, the polarisation observed towards W49 may be produced by 
local intensification of the Galactic foreground, rather than by synchrotron emission proceeding 
specifically from the SNR W49B. The DRAO maps in Fig.~\ref{fig:ws_smt_maps} show a different 
configuration, in which the positive $Q$ towards the centre of the region is smeared, while $U$ 
shows a clear positive signal; this can be a result of Faraday rotation of the polarised direction 
showing up at the 21\,cm wavelength. This 
observation is corroborated by the fact that the polarised intensity spectrum in Fig.~\ref{fig:ip_seds} 
flattens at the lowest frequencies, suggesting a loss of polarised flux density (Faraday 
depolarisation). The $Q$ and $U$ flux densities in Table~\ref{tab:w49_fluxes} show that this 
effect already begins to be important at the Urumqi 5\,GHz frequency, although their high uncertainties 
only provide an upper limit for $P$, as shown in the SED plot. A more direct representation of 
this effect is provided by the spectral distribution of the polarisation angle in Fig.~\ref{fig:polang}.
While at $\nu\gtrsim10\,\text{GHz}$ the angle is constantly close to zero, which means a mainly positive 
$Q$ state, the DRAO and Urumqi points show a deviation that can be fitted by our Faraday rotation 
model with a rotation measure $RM=(-14.9\pm5.1)\,\text{rad}/\text{m}^2$. We stress that, although we 
measure a Faraday rotation effect, the associated synchrotron emission likely does not proceed from 
the SNR W49B, but rather from the diffuse Galactic polarised emission; as a further proof of this, 
we notice that the peak of the positive U emission in the DRAO map is shifted with respect to the 
position of W49. Nonetheless, as this polarised signal is captured by our aperture, it is worth 
showing the associated modelling in $P$ and $\gamma$ as part of our study of QUIJOTE data towards 
this region.

\subsubsection{W51} 
\label{sssec:w51_fit}
For W51 the situation is very similar to W49: the local massive molecular clouds and 
associated HII regions are the main source of radio emission at low frequencies, where they 
mask the non-thermal emission from the SNR W51C. The SED is best fitted by a combination 
of free-free, AME, CMB and thermal dust, with the estimates on the parameters reported in 
Table~\ref{tab:w51_fit}; the CMB in this case shows a negative contribution, which however is 
compatible with zero. 
Our dust parameters are in reasonably good agreement with the ones reported 
in~\citet{demetroullas15}, considering that their fit includes a synchrotron component and 
does not include AME or CMB.
In our fit, the inclusion of AME determines a considerably better agreement between the model and 
the data points, improving the $\chi^2_{\rm red}$ value from 2.06 to 1.23. The inclusion of 
a synchrotron component, instead, would make the free-free emission measure compatible with zero and yield 
non-physical values for the AME parameters; when keeping the synchrotron and removing the AME component 
we find again a free-free amplitude compatible with zero and a very flat synchrotron spectrum 
($\alpha_{\rm s}\simeq-0.1$). We therefore decided not to include the synchrotron component in our intensity fit. 

However, unlike W49, W51 clearly stands out against the Galactic plane as a bright polarised source, with 
both negative $Q$ and $U$. Among the three regions considered in this work, it is the one that provides 
the highest polarised flux densities; still, this increase in the polarised flux density is 
associated with an increase in the total intensity towards the source, so that the resulting polarisation fractions 
in Table~\ref{tab:w51_fluxes} are at the same level as for the other regions. 
Also, for W51 the effect of Faraday depolarisation appears even stronger than in the case of W49: 
this can be seen already from the Urumqi and DRAO maps in Fig.~\ref{fig:ws_smt_maps}. The screening 
effect is likely due to 
intervening ionised gas along the line of sight towards W51C, possibly associated with the HII 
regions from the neighbouring W51A. As reported in Table~\ref{tab:w51_fluxes}, at the Urumqi frequency 
the $U$ flux density is compatible with zero, while the polarised SED in 
Fig.~\ref{fig:ip_seds} shows that the DRAO point flux density is at the same level as the 
lowest QUIJOTE frequencies.
The spectral distribution of the polarisation angle shown in Fig.~\ref{fig:polang} once again confirms
this picture, with a rather constant angle at microwave frequencies and a sharp decrease at 
$\nu\lesssim10\,\text{GHz}$. The fit of our Faraday rotation model yields 
$RM=(-17.6\pm3.1)\,\text{rad}/\text{m}^2$, 
which is higher in modulus than the estimate obtained for W49, confirming the effect is more relevant.
Note that in this case we did not include the \textit{Planck}-HFI points in the fit, as in the 
plot they show a break with the microwave trend; this is probably due to polarised dust contribution, either from the diffuse Galactic background or associated with the HII regions in W51. The effect is also quite evident from the bottom row panels in Fig.~\ref{fig:ws_smt_maps}. 

\input{tables/w51_fitres.tex}  

Once again, the polarised SED is best fitted by a combination of a synchrotron and a thermal dust component 
alone. Including all our available points we obtain the values $\alpha_{\rm s}=-0.37\pm0.05$ and 
$S_{\rm sync}^{\rm 1\,\text{GHz}}=12.7\pm2.0$ with $\chi^2_{\rm red}=2.61$. \citet{demetroullas15} 
obtained the estimate $\alpha_{\rm s}=-0.58\pm0.06$ when using CBI 31\,GHz and Effelsberg 2.7\,GHz 
data. Our flatter spectrum can be a result of Faraday depolarisation at the lowest frequencies, 
as explained. In fact, if we remove the DRAO point from the fit and only keep frequency points 
from 4.8\,GHz upwards (the Urumqi point only providing an upper limit), we obtain the values reported 
in Table~\ref{tab:w51_fit}, with the compatible estimate $\alpha_{\rm s}=-0.51\pm0.07$. 
We conclude that the Faraday depolarisation severely affects the DRAO point, in such a way that its 
inclusion in the fit would result in a biased estimation of the synchrotron spectral index. Hence, we keep 
the fit results obtained removing the DRAO point as our fiducial estimates. 

Going back to the intensity SED, we still tried to include a synchrotron component 
by fixing its spectral index to the value $\alpha_{\rm s}=-0.51$ fitted in polarisation. 
Although the goodness of fit is maintained ($\chi^2_{\rm red}=1.13$) this modification 
in the modelling significantly alters the values of the emission measure (which is now compatible with zero), 
the CMB (again compatible with zero but positive)
and the AME parameters, the latter showing an unphysically large value for $W_{\rm AME}$, and a very low value for the 
peak frequency. We also tried and fixed the synchrotron amplitude too, by assuming a polarisation fraction 
of 20\%; from the polarisation best-fit modelling, this results in $A_{\rm s}=99\,\text{Jy}$. When fixing both the 
synchrotron amplitude and spectral index, compared to the case of no synchrotron, we obtain compatible 
best-fit parameters for the thermal dust, a slightly larger value for the emission measure 
$\text{EM}=(2246\pm250)\,\text{pc}\,\text{cm}^{-6}$, a sign-flipped CMB contribution (still compatible with 
zero) $\Delta T_{\rm CMB} = (6\pm 121)\,\mu K$
and AME parameters equal to $A_{\rm AME}=(164\pm36)\,\text{Jy}$, 
$W_{\rm AME}=1.2\pm0.4$ and $\nu_{\rm AME}=(16.8\pm3.4)\,\text{GHz}$, with $\chi^2_{\rm red}=1.15$. 
Although the AME width is still rather large, this example shows that the observed SED in intensity is in 
fact compatible with the inclusion of a synchrotron component at reasonable levels; however, we do not adopt this 
fit as our fiducial model for W51 due to the rather arbitrary value we assigned to the polarisation fraction. 
In conclusion, the intensity SED suggests that although a synchrotron component is 
present in the source emission, the observed low-frequency data are best modelled by  
a free-free component alone, which is by far the dominant foreground emission mechanism in this frequency
range. The inclusion of the AME component, on the contrary, always improves the quality of this fit,
changing the $\chi^2_{\rm red}$ from $2.17$ to $1.13$ (from $3.05$ to $1.15$ when the synchrotron amplitude is
also fixed). For all these considerations, we fix the combination of free-free, AME, CMB and thermal dust, 
with no synchrotron, as the best model for W51 in total intensity. Towards this region AME is detected at the 
peak frequency $\nu_{\rm AME}=(17.7\pm3.6)\,\text{GHz}$ with a significance of $\sim 4.0\,\sigma$, contributing 
to the observed intensity emission with a $\sim21\%$ in flux density. Its relative significance, then, is 
lower compared to W49.

\subsubsection{IC443} 
\label{sssec:ic443_fit}
In total intensity IC443 shows a different combination of emission mechanisms compared 
to the other two sources. In this case, we find that the best-fit model consists of 
synchrotron (with spectral index $\alpha_{\rm s}=-0.29\pm0.03$), free-free, CMB and 
thermal dust, with the parameters reported in the second column of 
Table~\ref{tab:ic443_fit} and a final $\chi^2_{\rm red}=0.73$. Any combination that does not 
include a synchrotron component would result in values of $\chi^2_{\rm red}>2$, non-physical 
values for most parameters and an overall bad agreement between the model and the points at 
low frequency. This proves that the bulk of the IC443 emission is non-thermal and proceeds 
from the local SNR. On the contrary, a model without a 
free-free component could be valid as well, yielding $\chi^2_{\rm red}=0.69$ and  
values for the other parameters compatible with those reported in the second column of 
Table~\ref{tab:ic443_fit}. Thermal \textit{bremmstrahlung} radiation from IC443 was 
initially claimed in~\citet{onic12}, although discarded by subsequent work; it is clear that 
free-free is a subdominant component in this region, but its presence cannot be discarded 
\textit{a priori}. Since it is detected in our fit with reasonable values for the other 
parameters, it is worth including it in the source modelling. 
On the contrary, the inclusion of AME as an additional component would yield 
$\chi^2_{\rm red}=0.80$, with quite low values for the AME amplitude 
($A_{\rm AME}=4.4\pm3.8\,\text{Jy}$), width ($W_{\rm AME}=0.21\pm0.19$) and peak frequency 
($\nu_{\rm AME}=16.3\pm2.7$); similar considerations hold when the CMB contribution is removed 
from the fit. We conclude that the measured SED in IC443 does not provide any 
compelling evidence for the presence of AME towards the region. 
We stress that dust emission towards IC443 is well-reconstructed, but its amplitude is at least 
one order of magnitude lower compared to W49 and W51. This suggests a limited presence of dust 
in the region; as AME is generally observed in association with dust, it is not likely to be 
observed towards IC443. 

IC443 is detected in polarisation across our microwave frequencies as a positive $Q$ and $U$ 
source. The low dust amplitude in total intensity also results in very low flux densities 
in polarisation; 
even at lower frequencies, the flux densities in $Q$ and $U$ are often consistent with zero. 
In this case the 
debiasing employed to compute $P$ flux densities is important to prevent overestimating the polarised 
intensity. Still, although the microwave polarised flux densities are 
generally below 1\,Jy, in this case 
the position of the source outside the Galactic plane allows the detection of its polarised emission with 
less confusion with the background. Most likely, this polarised signal is synchrotron emission from 
the SNR, and it is reasonable to fit for the corresponding power law on the polarised SED. 
At 1.4\,GHz the source still yields a positive $U$ flux density, and a value of $Q$ consistent with zero; 
the DRAO maps in Fig.~\ref{fig:ic443_smt_maps} show hints of positive polarisation signal towards 
the centre of 
the aperture, although being in general subdominant with respect to other background structures. 
At 5\,GHz the source yields a strong positive $Q$ flux density, and a change of sign in $U$. These 
considerations may indicate that the two lowest 
frequency points are once again affected by Faraday rotation, although their error bars are too large 
to firmly establish it. In fact, looking at the SEDs plot in 
Fig.~\ref{fig:ip_seds}, there seems not to be a net flux density loss below 10\,GHz, and 
the spectrum decays as expected for a synchrotron emission. As a further confirmation, we can 
see from Fig.~\ref{fig:polang} that in this case the variation of the polarisation angle at 
low frequency is very mild, and as a result our Faraday rotation fit yields a value of 
$RM=(-5.7\pm10.3)\,\text{rad}/\text{m}^2$ for the rotation measure, 
which is compatible with zero and considerably lower in modulus than the one measured for W49 or W51.
We thus conclude that, although we observe hints of Faraday rotation towards IC443, the effect 
is negligible as far as our analysis is concerned. We therefore proceed with a fit for the 
polarised SED using all available points, and modelling again the emission as a combination 
of synchrotron and dust; the resulting parameters are reported in the fourth column of 
Table~\ref{tab:ic443_fit}, with a synchrotron spectral index of $\alpha=-0.37\pm0.13$ and 
$\chi^2_{\rm red}=0.90$. 

Going back to the SED in total intensity, when fixing the synchrotron 
spectral index to the value obtained in polarisation, we find the best-fit 
is obtained with the same model adopted when the synchrotron spectral index 
was left free, i.e. a combination of synchrotron, free-free, CMB and thermal dust. 
Compared to the case of free spectral index, the synchrotron amplitude decreases
while the measurement of the free-free emission measure becomes more significant; 
the parameter values associated with CMB and thermal dust, instead, are only marginally 
affected (third column of Table~\ref{tab:ic443_fit}). Once again, the 
inclusion of AME would yield non-physical values for the associated parameters, especially 
for its peak frequency ($\nu_{\rm AME}\simeq 6\,\text{GHz}$). 
The fact that the free-free amplitude is higher in this case
can be the result of a steepening of the synchrotron spectrum at frequencies above a few GHz. 
In that case, because the synchotron is fitted in polarisation starting from $\sim5\,\text{GHz}$ 
(the DRAO point really only provides an upper limit), we would expect the value of 
$\alpha_{\rm s}$ to be larger in modulus compared to the result fitted in total intensity. 
When forcing such a steep spectrum in the intensity SED, the fit better accommodates 
the lowest frequency points by lowering the synchrotron amplitude and including a flatter 
spectral contribution as is the one from the free-free emission. So, our detection of 
thermal emission in IC443 may simply be a result of an inaccurate modelling of the synchrotron
spectrum across the frequency range we explore. 

In order to better test this hypotesis, we also try and fit for a curved synchrotron model, 
which differs from the parametrisation in equation~\eqref{eq:sync} for the inclusion of a cut-off 
frequency $\nu_{\rm 0,s}$:
\begin{equation}
\label{eq:curv_sync}
	S_{\rm sync}^{\rm curv}(\nu) = A_{\rm s} \left( \frac{\nu}{\nu_{\rm p}} \right)^{\alpha_{\rm s}} \,\exp{\left(-\dfrac{\nu}{\nu_{\rm 0,s}}\right)}.
\end{equation}
With this model, we find that the best fit to the IC443 intensity spectrum is provided by 
a combination of synchrotron, CMB and thermal dust, with the best value of $\chi^2_{\rm red}=0.61$ 
obtained for this region and a spectral index which is lower in modulus 
than one obtained for a simple power-law synchrotron model (last column of Table~\ref{tab:ic443_fit}). 
In this case the CMB component flips sign compared to the case of the simpler synchrotron model, although
its amplitude is compatible with zero, confirming that it is a subdominant component in the SED.
The inclusion of AME generally yields unphysical values for its parameters, 
and the inclusion of free-free results in too low a value for the synchrotron cut-off frequency 
(around 60\,GHz). We then decided to fit IC443 as a combination of curved 
synchrotron, CMB and thermal dust; this is the model that is overplotted to the data points in 
Fig.~\ref{fig:ip_seds}, and the resulting parameters are quoted in the fifth column of 
Table~\ref{tab:ic443_fit}.

\input{tables/ic443_fitres.tex}  

Previous works already tackled the modelling of IC443 spectrum from radio 
to FIR frequencies. As already mentioned,~\citet{onic12} claimed the detection of 
free-free towards the region, accounting up to 57\% of the observed emission at 1\,GHz. Our model with a 
free spectral index results in a free-free amplitude of $\sim10\,\text{Jy}$, or $\sim10\%$ of the total,
while the fixed spectral index model yields $\sim36\,\text{Jy}$, or $\sim33\%$ of the total; 
both these estimates are in agreement with the quoted result. We already stressed, 
however, how our finding could be simply the result of forcing too steep a synchrotron 
spectrum. In~\citet{planck_ir_xxxi} the existence of a thermal component was ruled out, and 
IC443 was modelled as a simple combination of synchrotron and thermal dust; for the latter, 
they quote $\beta_{\rm dust}=1.5$, in agreement with our finding, and $T_{\rm dust}=16\,\text{K}$, 
which is $\sim 2\sigma$ lower than our estimate (although the reference does not 
quote error bars, which makes an acutal comparison unreliable). As per the synchrotron, they 
favour $\alpha_{\rm s}=-0.36$, compatible with our estimate in polarisation, although they 
mention that for frequencies above 40\,GHz the spectrum shows a steepening to 
$\alpha_{\rm s}=-1.5$. This observation corroborates our use of a curved synchrotron model. 
In~\citet{onic17} the IC443 SED from 400\,MHz to 143\,GHz was fitted using a combination of 
synchrotron, thermal dust and AME (not favouring any free-free contribution). Their work 
also stressed the presence of a break in 
the synchrotron power law visible above $\sim5\,\text{GHz}$, and they also considered 
a curved synchrotron model as in our equation~\ref{eq:curv_sync}. Different implementations pointed towards 
a cut-off frequency in the range $112\,\text{GHz}<\nu_{\rm 0,s}<152\,\text{GHz}$ and spectral index 
in the range $-0.39<\alpha_{\rm s}<-0.35$; their estimates for the dust temperature are in agreement with 
\textit{Planck} results, while their emissivity results are slightly higher, in the 
range $1.97<\beta_{\rm dust}<2.345$. Most importantly, they detected an excess of 
emission at 30\,GHz
which they interpreted as AME. A very similar analysis was conducted in~\citet{loru19}, measuring 
a curved synchrotron with $\alpha_{\rm s}=-0.38$ and $\nu_{\rm 0,s}=(148\pm23)\,\text{GHz}$, 
and claiming a detection of AME peaking at $\nu_{\rm 0,AME}=(28\pm3)\,\text{GHz}$. The synchrotron
cut-off frequency obtained in our analysis is consistent with these results. 

\begin{figure*} 
\centering
\includegraphics[trim= 0mm 0mm 0mm 0mm, scale=0.34]{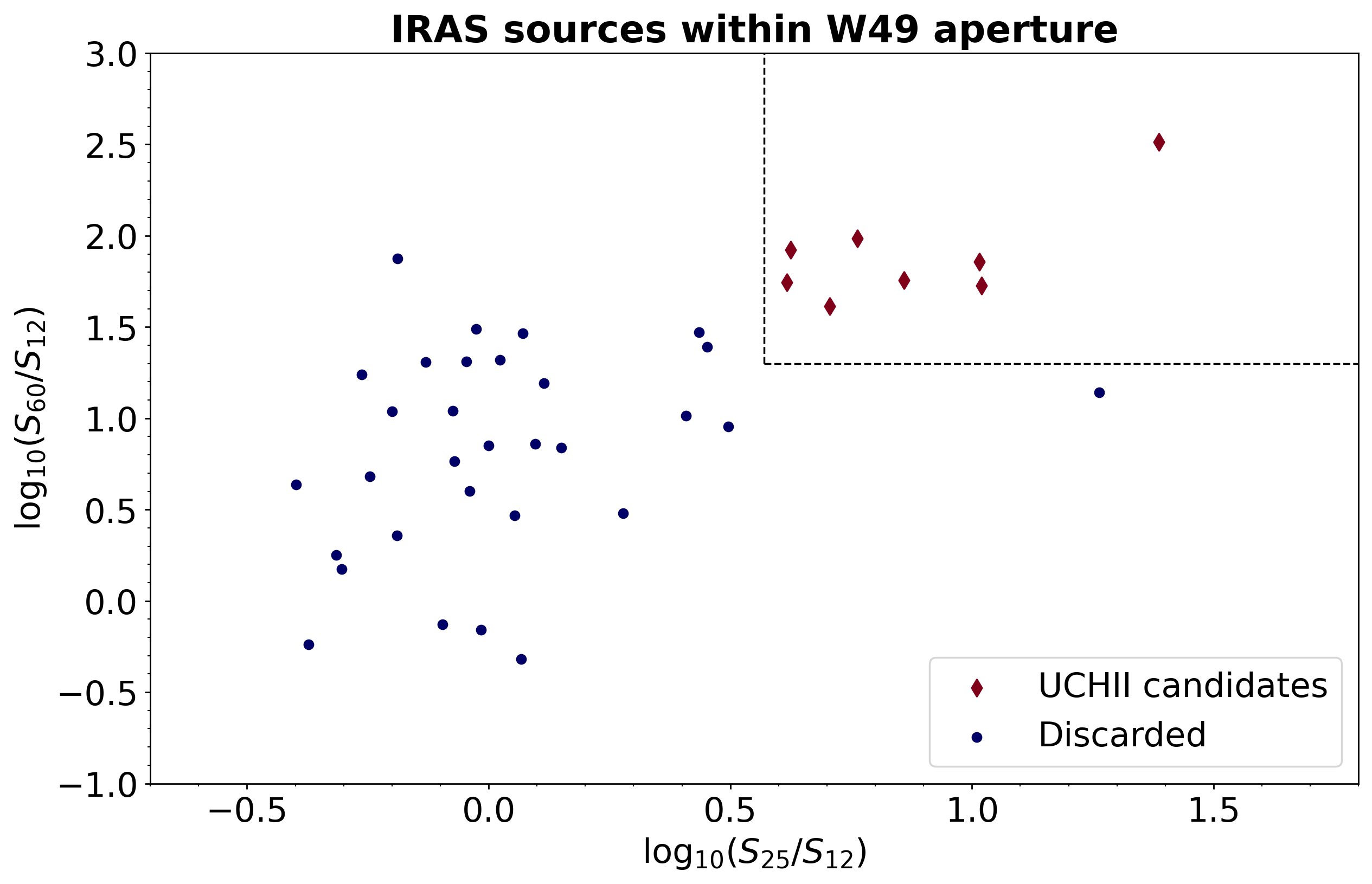}
\includegraphics[trim= 0mm 0mm 0mm 0mm, scale=0.34]{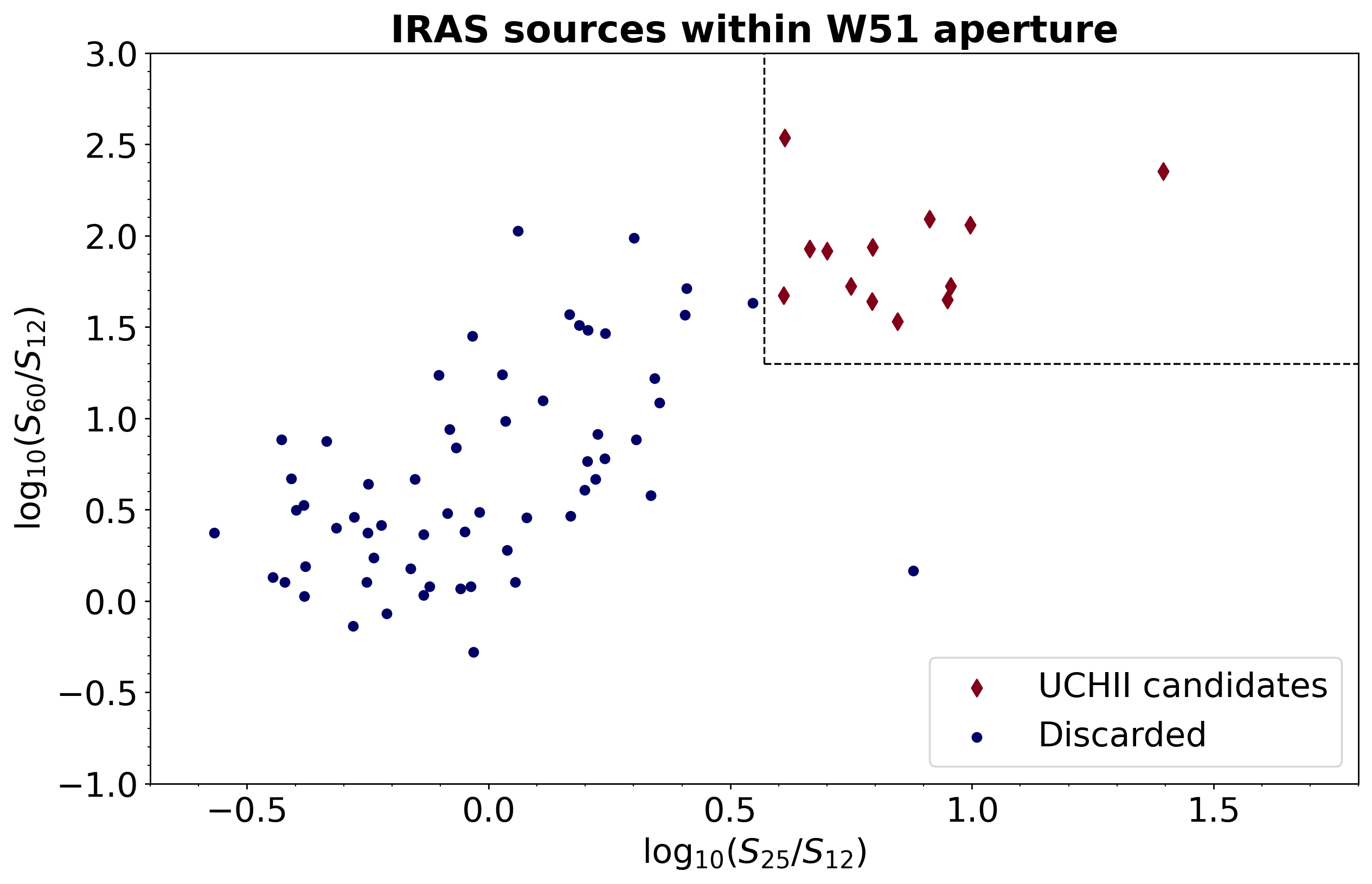} 
	\caption{Extraction of candidate UCHII regions using \textit{IRAS} data for both W49 (left panel) 
	and W51 (right panel). Points represent \textit{IRAS} galactic IR sources entering the aperture 
	employed for the photometric study of each region, displayed in a colour-colour plot based on 
	\textit{IRAS} measured flux densities. UCHII regions are expected to yield ratios 
	$\log_{10}{(S_{60}/S_{12})}\gtrsim1.30$ and $\log_{10}{(S_{25}/S_{12})}\gtrsim0.57$; the group 
	of sources satisfying these condition is shown with red diamonds, with the dashed lines marking the 
	flux density boundaries. In this way we select 8 sources for W49 and 13 sources for W51, and use 
	them to estimate the contamination from UCHII to our AME detections. See Section~\ref{sssec:uchii} 
	for details.}
\label{fig:uchii}
\end{figure*} 

Although these previous works would agree in establishing the presence of AME towards IC443, 
there are some issues that should be pointed out. First of all, their analyses rely on the 
combination of flux density estimates coming from previous results in the literature, obtained 
using different types of radio observations, different angular resolutions and different analysis techniques.
Although~\citet{onic17} recognised the inherent risks of such an approach, which makes their 
study somehow qualitative, they still firmly established the need for an extra emission at 
microwave frequencies to account for the observed excess. Besides, both~\citet{onic17} 
and~\citet{loru19} adopted the microwave flux densities from~\citet{planck_ir_xxxi}. Those flux
densities were measured with aperture photometry, adopting a variable size of the aperture as a function 
of the considered frequency. The analysis in~\citet{planck_ir_xxxi} introduced a correction factor, 
equal to 1.4 for IC443, to account for the loss of flux density that is not captured inside their aperture 
due to the beam convolution. The authors did not applied this correction
to well-resolved sources, defined as those with diameter 
50\% larger than the beam FWHM. As the \textit{Planck} beam FWHM decreases with increasing 
frequency, this criterion implies that for IC443 the correction is applied at 30 and 44\,GHz, 
but not from 70\,GHz onwards. Their analysis adapted the size of the aperture at each frequency, 
in such a way that the same fraction of flux density should be retained inside the aperture; hence,
these correction factors should be applied to all their frequency points, and not only to 30 and 44 GHz. 
This approach produces an artificial dip of the SED at 70\,GHz, 
which the authors interpreted as the downturn of AME. We can also argue that this strategy for the 
flux density extraction is not appropriate in cases of significant background fluctuations, as it is 
the case for IC443; in fact, none of those works adopted any masking around IC443, so that the aperture photometry 
estimation of the flux density emission and/or the background level can be easily biased by the 
contribution from the neighbouring sources Sh2-249 and Sh2-252. The analysis we have presented 
in this paper, however, not only took care of the mask, but also measured the flux densities adopting 
a common angular resolution and a consistent methodology (with fixed apertures and background annuli) 
that is applied throughout the considered frequency range. In conclusion, we do believe that the 
available data set points towards an absence of AME emission towards IC443; at the very least, we can 
affirm that if an AME component is present, it is dominated by the uncertainties in our flux density measurements, and is certainly much fainter than the emission detected towards W49 and W51. 
Although we still observe a dip in the IC443 spectrum around 
70\,GHz, it can be more reasonably fitted with a curved synchrotron component, and is not  
sufficently compelling to justify the inclusion of AME at $\sim30\,\text{GHz}$.

\subsection{Further considerations on AME}
\label{ssec:ame_discussion}
We conclude our analysis by commenting some further aspects related to the detection of AME in 
W49 and W51. First of all, we have to stress that, because we are working with maps smoothed at a 1-degree 
angular resolution and we use apertures of typical radius of $\sim 1\,\text{deg}$, the measured flux 
densities may have a significant contribution from the region neighbouring each source.
Therefore, although we detect AME towards W49 and W51, it is hard to assess whether the 
measured signal proceeds entirely from the sources or has an important contribution from the diffuse Galactic 
emission in their surroundings. Although in the rest of this section we will still discuss about 
AME emission from W49 and W51, it is important to acknowledge this possibility. Further discussion 
about this matter can be found in Appendix~\ref{app:ame}.

\subsubsection{Contribution from ultra-compact HII regions} 
\label{sssec:uchii}

UCHII regions are produced by recently formed massive stars which are still embedded in their 
molecular clouds; those with $\text{EM}\gtrsim10^7\,\text{cm}^{-6}\text{pc}$ can have a significant optically 
thick free-free emission at frequencies up to 10\,GHz or higher. This emission can produce 
an excess in the measured flux density at QUIJOTE frequencies, thus mimicking the effect of AME on 
the SEDs of W49 and W51. The goal of this section is to assess to what extent the  
contribution from UCHII can affect the significance of our AME detection. 

At first we follow the procedure described 
in~\citet{planck_ir_xv} and also adopted by~\citet{poidevin22}, to which we redirect for 
a more detailed explanation and further references. This method is based on the extraction of 
candidate UCHII sources 
 entering our chosen apertures and on the estimation of their total flux density at AME frequencies. 
For this we employ data from the \textit{IRAS} Point Source 
Catalogue\footnote{\url{https://heasarc.gsfc.nasa.gov/W3Browse/iras/iraspsc.html}.}, 
which provides flux density measurements for point sources at 12, 25, 60, and 100$\,\mu\text{m}$.
For both W49 and W51 we query the catalogue to extract the IR sources contained in the 
aperture employed for our photometric flux density measurements; in addition, we discard sources 
which are extragalactic or upper limits in either $25\,\mu\text{m}$ or $60\,\mu\text{m}$.
We then place the retrieved IR sources on 
a $\log_{10}{(S_{25}/S_{12})}$--$\log_{10}{(S_{60}/S_{12})}$ plane as shown in Fig.~\ref{fig:uchii}, 
where $S_{x}$ denotes the \textit{IRAS} measured flux density at wavelength $x$. UCHII sources are 
expected to have $\log_{10}{(S_{60}/S_{12})}\gtrsim1.30$ and $\log_{10}{(S_{25}/S_{12})}\gtrsim0.57$;
as marked in Fig.~\ref{fig:uchii}, this leaves 8 IR sources in W49 and 13 sources in W51. 
\textit{IRAS} data allow to evaluate the associated total flux density 
at 100\,$\mu\text{m}$. For an UCHII 
region, the ratio between this flux density and the flux density at 15\,GHz is found to lie in the range 
1000-400000, which allows us to place both an upper and a lower limit to the expected UCHII 
contamination at QUIJOTE frequencies. The results are reported in Table~\ref{tab:uchii}, 
where they are compared with the residual AME amplitudes at the same frequency.  
We see that in the worst case scenario the contribution from UCHII sources can account for  
$\sim55\%$ of the AME flux density towards W49, and for $\sim32\%$ of the AME flux density 
towards W51. We stress that this not only is an upper limit, but according to~\citet{planck_ir_xv} it is 
also likely overestimated by a factor of a few. The corresponding lower limit estimate is 
completely negligible compared to the measured AME flux density, which confirms our detection of 
AME cannot be ruled out on the basis of this simple extrapolation. 

In order to get an independent estimate of the possible contamination from UCHII regions, we consider  
data from the CORNISH\footnote{Co-Ordinated Radio 'N' Infrared Survey for High-mass star formation, 
\url{https://cornish.leeds.ac.uk/public/index.php}.} continuum survey of the Galactic plane at 5\,GHz 
with a 1.5\,arcsec resolution~\citep{purcell13}. CORNISH legacy data include a catalogue of 
identified UCHII regions, which we query to select the ones entering our apertures. We find 
24 sources within the W49 aperture and 8 sources within the W51 aperture, contributing with the 
5\,GHz flux densities quoted in Table~\ref{tab:uchii}. Now, there is no standard way to extrapolate 
these flux densities to the peak frequency of our AME detection. Nonetheless, we can refer 
to the SEDs plotted in~\citet{kurtz94} for several different UCHII regions, which were built 
using VLA integrated flux density measurements at 8.3 and 15\,GHz in combination 
with \textit{IRAS} flux densities and 
other ancillary data. The plots do not show any appreciable variation of the spectrum for 
frequencies $\lesssim100\,\text{GHz}$. Hence, to a first approximation we can extrapolate 
our UCHII flux density measurements at 5\,GHz to 15\,GHz for a comparison with the corresponding AME 
residuals. We find that the contribution from UCHII amounts to $\sim10\%$ of 
the AME amplitude in W49 and to $\sim5\%$ in W51. We conclude that, although a non-negligible contribution from UCHII is entering 
our measured flux densities, it is not enough to account for the local SED excess amplitude and rule out 
the presence of AME. Novel, tailored observations to these regions are required to better 
constrain the UCHII contribution. 

\input{tables/uchii.tex}  
\input{tables/polame.tex} 

\subsubsection{Constraints on AME polarisation} 
\label{sssec:polame}

One of the driving goals of this analysis was the search for a possible polarised AME emission, 
which has not been detected to date and could shed light on the mechanism responsible for this 
foreground (Section~\ref{sec:introduction}). We have already commented that for both W49 and W51 
the polarised SED is suitably modelled by a combination of synchrotron emission and dust emission 
alone. Although AME is not directly detected in our polarisation fits, we can proceed in a way similar to 
what was done in~\citet{genova_santos17} and~\citet{poidevin19} to quote upper limits for the 
AME polarisation fraction. 

We proceed as follows. We consider the frequency points where AME is most significant and less 
affected by other foregrounds, i.e. those in the frequency range from 17\,GHz to 61\,GHz. In 
this frequency range we generate a set of 1000 AME residual SEDs in intensity, computed using 
equation~\eqref{eq:ame} and adopting random AME parameters; the latter are sampled from a Gaussian 
distribution centred on the best-fit estimates from Tables~\ref{tab:w49_fit} and~\ref{tab:w51_fit}, 
and scattered according to their uncertainties. The dispersion of these realisations at the chosen 
frequency points allows to evaluate the uncertainty for the local AME intensity $I_{\rm AME}$. 
The resulting flux densities are quoted in Table~\ref{tab:polame}, together with 
the polarised $P$ flux densities at the same frequencies\footnote{The $P$ values in Table~\ref{tab:polame} 
have been colour-corrected using the CC coefficients reported in the last column of Tables~\ref{tab:w49_fluxes} 
and~\ref{tab:w51_fluxes}; hence, they are slightly different from the polarised flux densities already 
quoted in the photometric analysis section.}. 
We then compute the associated polarisation fractions as $P/I_{\rm AME}$, 
which are reported in the fourth column of Table~\ref{tab:polame} in percent units.

Clearly, we know that at these frequencies an important polarised contribution comes from both 
the synchrotron and the dust emission. These fractions should therefore be interpreted as upper 
limits on the AME polarisation fraction. In fact, while the AME intensity peaks around 
20\,GHz, the polarised flux densities steadily decrease over the considered frequency range, and 
are most likely associated with synchrotron. For this reason, in the last column of 
Table~\ref{tab:polame} we report explicitly the upper limits on the AME polarisation fraction 
$\Pi_{\rm AME}$, which are computed as the 95\% confidence level upper boundaries of the $P/I_{\rm AME}$ values. 
We can quote the tighest constraints 
$\Pi_{\rm AME} \lesssim 1.2\%$ for W49 at 44.1\,GHz and $\Pi_{\rm AME} \lesssim 6.0\%$ for W51 at 22.8\,GHz.
The constraint for W51 is quite loose and does not allow to provide any strong claim on the nature of 
AME; the constraint for W49 is instead tighter, and favours the spinning dust model 
as the most plausible physical mechanism for AME. Follow up studies of W49 and W51 
with higher resolution are required to better characterise 
their polarised emission, as the contamination from the observed polarised synchrotron is still the 
major factor hampering the measurement of a real AME polarised upper limit towards these region. 

We conclude this section acknowledging that our polarisation fractions  
may be underestimated, as the integration of the Stokes parameters inside the apertures may 
result in depolarisation. To check for this effect, we can try and evaluate the level of possible aperture 
depolarisation effects on the dust flux densities; we consider the \textit{Planck} 353 GHz map, 
where the dust polarisation flux densities are the highest.  
We then compute the (debiased) polarised intensity $P$ for individual pixels in the region surrounding W49 and 
W51, and obtain a new estimate on the polarisation of these sources by applying aperture photometry 
directly on the $P$-map. For clarity, we shall label these new polarised flux densities as $P^{\rm (pix)}$, 
to stress that they are obtained integrating the values of individual pixels, while we refer to the 
flux densities we employed up to now as $P^{\rm (ap)}$, as they are computed from the values of $Q$ 
and $U$ integrated over the apertures. Using the maps smoothed to a 1-degree resolution, and the same 
apertures as quoted in Table~\ref{tab:sources_info}, we find $P^{\rm (pix)}=181\,\text{Jy}$ for W49 and 
$P^{\rm (pix)}=176\,\text{Jy}$ for W51, yielding respectively the polarisation fractions 
$\Pi^{\rm (pix)}=1.9\%$ and $\Pi^{\rm (pix)}=1.1\%$. These numbers are comparable with the estimates 
of $\Pi^{\rm (ap)}$ reported in Tables~\ref{tab:w49_fluxes} and~\ref{tab:w51_fluxes}. 
Even when considering the \textit{Planck}
353 GHz maps at their original resolution, we obtain $\Pi^{\rm (ap)}=1.0\%$ and $\Pi^{\rm (pix)}=1.0\%$ 
for W49, and $\Pi^{\rm (ap)}=1.2\%$ and $\Pi^{\rm (pix)}=1.1\%$ for W51 (in this case we are employing 
smaller apertures of $r_{\rm ap}=20\,\text{arcmin}$ for W49 and $r_{\rm ap}=30\,\text{arcmin}$ for W51, 
which are large enough to include the full extent of each source). Hence, integrating over our chosen 
apertures does not provide any evident depolarisation effect for the dust emission. If we make 
the reasonable assumption that the polarisation of AME, if present, would have the same orientation 
as the polarisation of dust, we conclude that beam depolarisation is not a major issue in our analysis. 

Finally, we mention that also the mean orientation of the local magnetic field with respect to the plane of the sky affects the intrinsic level of dust polarisation~\citep{planck_18_xii}, and is most likely the main reason for the low measured dust polarisation fractions towards W49 and W51. This effect may also result in a low AME polarisation in these particular sources. We then stress that the upper limits we quote in Table~\ref{tab:polame} could be a result of the geometrical configuration of the sources and the morphology of the local magnetic field, and cannot be used alone to infer a general property of AME polarisation. A general, physical modelling of this foreground emission goes beyond the scope of this work. The results from Table~\ref{tab:polame} refer specifically to measurements obtained towards W49 and W51 at the angular scales set by our apertures ($\sim 1$ deg). The main conclusion of our study still remains that AME polarisation is not detected towards W49 and W51 at this angular resolution, when considering the data set employed in this paper.


\section{Conclusions}
\label{sec:conclusions}

We have presented novel microwave intensity and polarisation data for the Galactic 
regions W49, W51 and IC443 obtained with the QUIJOTE experiment; data were acquired with the 
MFI instrument at 11, 13, 17 and 19\,GHz. This study
is part of the QUIJOTE survey of astrophysically relevant Galactic regions, with 
the aim of characterising the local foreground emission. In this case, the main 
goal was the assessment of the level of AME towards SNRs, and the estimate 
of the spectral index of synchrotron emission. The selected regions allow to compare different 
local environments: whereas IC443 is a relatively isolated SNR, W49 and W51 host SNRs
in association with molecular clouds. 

We described the observations performed with QUIJOTE and presented the resulting 
maps in the Stokes parameters $I$, $Q$ and $U$ towards the three regions. A null-test on 
the map allowed to confirm the good quality of these data products, as the computed 
sensitivity in polarisation is consistent with the instrument specifications.  
The sources are well detected in total intensity; hints of polarised emission are visible
towards W49, with a much more significant detection towards W51 and IC443.
The inclusion 
of dedicated raster scan observations allow to reduce the noise level in polarisation by 
a factor $\sim2$ in the source areas compared to the wide survey maps only.

We combined the QUIJOTE maps with a set of ancillary data to study the source properties in 
the frequency range $[0.4,3000]\,\text{GHz}$. Flux densities for the Stokes parameters $I$, $Q$ and $U$
were measured on the maps with the aperture photometry technique, choosing aperture sizes 
suitable to each source angular extension and estimating the uncertainties as the rms 
of the flux densities from random apertures surrounding each source position. The polarisation $P$ and
polarisation fraction $\Pi$ were obtained combining the $Q$ and $U$ values with a proper 
debiasing which ensures the final estimates are not overestimated. We also computed the polarisation
angle $\gamma$ as a function of frequency. 

The SEDs in total intensity were modelled with a multi-component fit including different combinations
of synchrotron, free-free, AME, CMB and thermal dust emission. We performed a similar fit in the polarised SEDs, 
where we considered possible contributions 
from synchrotron, AME and thermal dust. The highest frequency available for polarisation data, 
353\,GHz, does not include the peak of the dust emission, implying the dust component would be 
poorly fitted; for this reason we fixed the dust temperature to the value obtained from the 
intensity SED prior to the fit.

For all three sources, the thermal dust component is well reconstructed in total intensity, 
with general agreement with the literature, while the synchrotron provides a good fit for
the polarised SED, most likely proceeding from the local SNRs (or from diffuse Galactic emission 
in the case of W49). 
Still, we found that the total intensity SEDs for the sources W49 and W51 can be conveniently 
modelled without the inclusion of a synchrotron emission, confirming that the 
bulk of the local low-frequency emission is thermal; in W49, however, it is possible to accommodate a 
synchrotron component when fixing its spectral index to the value fitted on the polarised SED. 
We detected AME in these 
two regions, the QUIJOTE points proving crucial in confirming the downturn of AME emission at 
frequencies below 20\,GHz. We employed \textit{IRAS} and VLA data 
to assess the contribution of UCHII regions to the observed excess in the emission at microwave 
frequencies; this contamination is the main reason why W49 was discarded as a reliable AME source 
in~\citet{planck_ir_xv}. We found that, although a measurable UCHII contribution is definitely 
present in our SEDs, its amplitude is not enough to account for the observed spectral bump at 
AME frequencies. We concluded that W49 and W51 host AME; this is the first time such claim 
is made for the W51 region~\citep[this result is also confirmed in][]{poidevin22}. However, we also acknowledged that, because our analysis has been conducted 
at a 1\,deg resolution, it is not possible to assess whether AME proceeds indeed from the sources 
themselves, or enters our apertures from the neighbouring background. Still, our analysis 
confirms the presence of AME in total intensity towards these regions; tailored, higher resolution observations  
are required in order to better constrain its spatial origin. 

We found that no AME contribution is required to improve the modelling of the 
polarised SEDs, thus confirming that polarised AME emission is not detected.
We then derived conservative upper limits for the AME polarisation fraction towards 
the two regions, finding $\Pi_{\rm AME} \lesssim 1.2\%$ for W49 at 44.1\,GHz 
and $\Pi_{\rm AME} \lesssim 6.0\%$ for W51 at 22.8\,GHz. The upper limit for W49 is relatively tight 
and suggests a very low or possibly null AME polarisation; the upper limit for W51 is instead quite 
loose and non-informative in this sense. In reality, due to the confusion with the 
intervening polarised synchrotron emission, we expect both upper limits to be overestimated, 
thus favouring models of AME with a low degree of polarisation (SDE or amorphous dust models). 
We also stressed that aperture depolarisation may result in an actual underestimation of the 
AME polarisation fractions, although a test performed for the polarised dust emission on 
\textit{Planck} 353 GHz maps suggests this issue should not be too relevant. The main conclusion 
is that AME polarisation is not detected in this study.

In the case of IC443, instead, we found that the intensity spectrum 
is best fitted by a combination of synchrotron, CMB and thermal dust, without any need 
for a free-free or AME contribution; the best modelling is achieved by introducing a 
 break in the synchrotron spectrum. This result seems to confirm that AME is not naturally 
associated with SNRs, but more likely with dust-richer environments like molecular clouds;
however, it is in tension with previous works that claimed an AME contribution is required 
to model the microwave spectrum of IC443. We discussed how such a claim could be the result of 
a systematic bias in the flux density measurements adopted in those works.


\input{acknowledgements.tex}
\section*{Data Availability Statement}
The QUIJOTE raster scan data at the core of this study are proprietary of the 
QUIJOTE Collaboration, however they are available on reasonable request to the corresponding 
author. The QUIJOTE nominal mode maps, which are also used in 
this analysis, are expected to be made publicly available in 
the first QUIJOTE data release. Other ancillary data employed in this work are publicly 
available and can be accessed online as detailed in the paper text.

\bibliographystyle{mnras}
\bibliography{bibliography} 

\appendix
\section{Flux contribution from neighbouring areas}
\label{app:ame}

As already mentioned in the beginning of Sec.~\ref{ssec:ame_discussion}, given the common resolution 
of 1 degree that we adopt throughout the frequency range we explore, the AME signal we detect towards 
W49 and W51 may have a significant contribution coming from the Galactic emission that surrounds them. 
In this section we shall refer to the area surrounding each source, with the exclusion of the source 
itself, as the ``region''.
Although it is not possible to extract further information about the actual spatial distribution 
of the measured flux densities, we can still get a rough estimate of the level of the region contribution 
in our chosen apertures. 

We repeat the measurement of the flux densities of W49 and W51, using this time the maps prior to 
the convolution with a 1-degree beam. We focus in particular on the maps with the highest original 
resolution, where the sources are better resolved; we set the threshold 
$\theta_{\rm FWHM}<10\,\text{arcmin}$, which selects the Urumqi map at low frequencies and the 
\textit{Planck}-HFI maps at higher frequencies. This time we can choose a smaller 
aperture, in order to only encompass the source angular extent; we select an internal radius of 
$r_{\rm ap}=20\,\text{arcmin}$ for W49 and $r_{\rm ap}=30\,\text{arcmin}$ for W51. These values 
are chosen to ensure each source is completely encompassed by the aperture, and they are slightly larger than 
the nominal source diameters because even these smaller beams produce a broadening of the source signal
to some extent. The background annulus is chosen in each case with a separation of $20\,\text{arcmin}$ 
from the aperture, and with a size of $20\,\text{arcmin}$ (i.e., $r_{\rm min}=r_{\rm ap}+20\,\text{arcmin}$ 
and $r_{\rm max}=r_{\rm min}+20\,\text{arcmin}$). The comparison of 
the flux densities computed this way and the ones computed on the 1-degree smoothed
maps with the nominal 
apertures shown in Table~\ref{tab:sources_info}, enables us to obtain information on the typical region 
contribution in our measurements. 

\begin{figure} 
\centering
\includegraphics[trim= 10mm 0mm 0mm 0mm, scale=0.38]{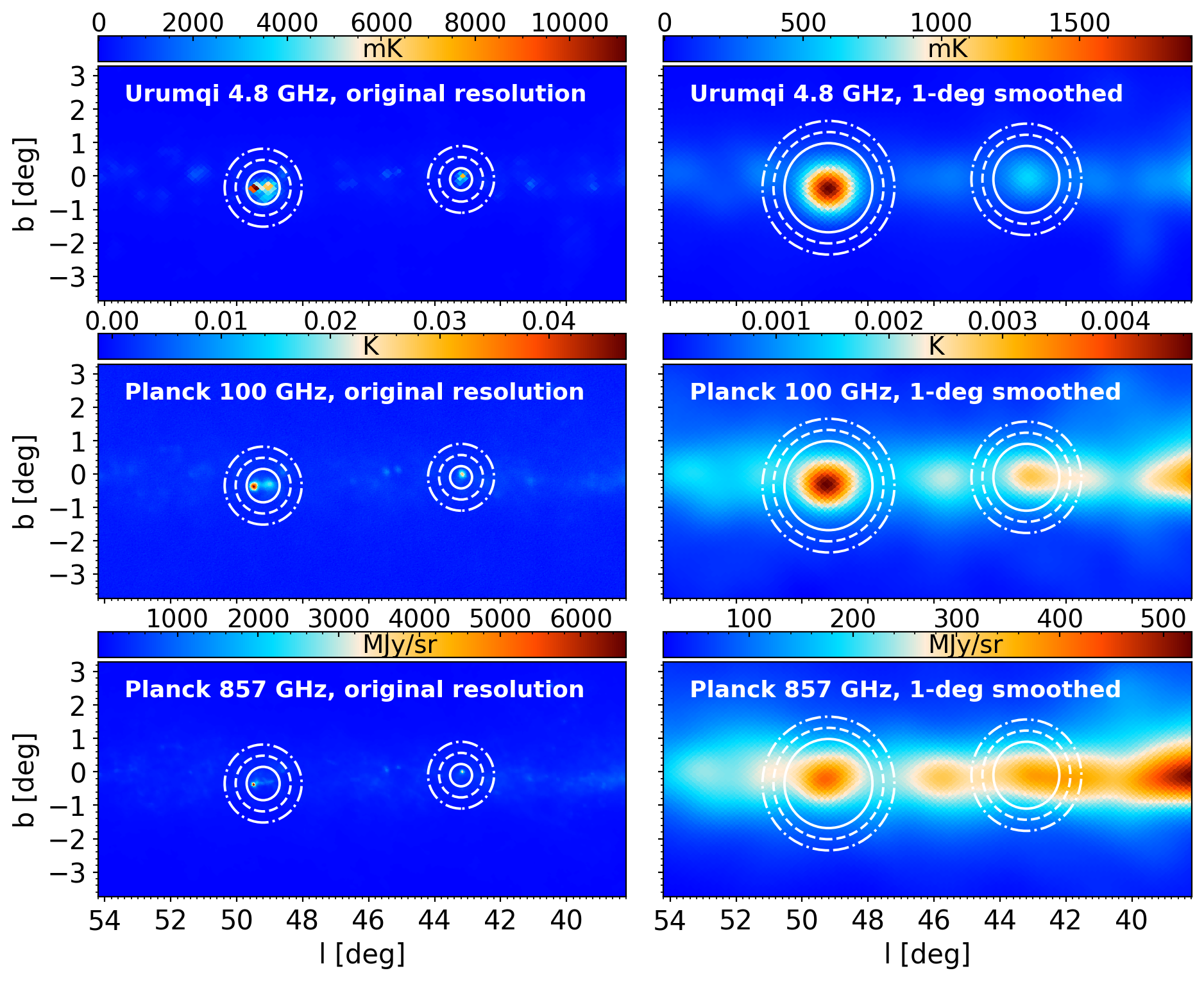}
	\caption{A comparison between the maps at their original resolution (left) and smoothed to 
	a 1-degree beam (right) for three sample frequencies, in the area encompassing W49 and W51. The 
	apertures employed to compute the flux densities reported in Table~\ref{tab:reg_contrib} are 
	also shown.}
\label{fig:reg_ame}
\end{figure}

The results are summarised in Table~\ref{tab:reg_contrib}, where for each of the selected maps we 
report its original FWHM value, the flux densities $S_{\rm or}$ computed on the map at its parent angular resolution, and 
the flux densities $S_{\rm 1d}$ computed on the 1-degree smoothed map. 
The difference $S_{\rm 1d}-S_{\rm or}$ is our estimate of the flux proceeding from the region that 
makes it into our apertures; it is reported in Table~\ref{tab:reg_contrib} as a fractional contribution 
with respect to the 1-degree flux $S_{\rm 1d}$. In Fig.~\ref{fig:reg_ame} we also show, for three sample 
frequencies, the comparison between the maps at their original resolution and the smoothed maps, together 
with the circles employed for the photometric analysis in each case. 

We notice that the fractional region flux density in our apertures is in general larger than 
50\%. Now, because this is the total flux, which results from the combination of different emission 
mechanisms, we cannot say that this is the fraction of background AME that enters the apertures. 
In fact, the ratios tend to be larger at higher frequencies, where we are practically dominated by 
thermal dust emission and we expect indeed a significant contribution from the diffuse Galactic 
dust~\citep{planck_15_xxv}. 
On the contrary, at the Urumqi frequency, 
in these regions we are dominated by free-free emission, which is also a relevant component of the 
diffuse emission. 

These results show that it is indeed likely that a contribution of the AME signal does not come from 
our sources; still, it is not possible to rule out the presence of AME in the molecular clouds 
associated with W49 and W51. As already mentioned, we are limited by the coarser resolution 
of most of the surveys we consider, and it is not possible to circumvent this issue. We stress
in particular that similar analyses we have cited in the text are affected by the same type 
of uncertainty~\citep[e.g., ][]{planck_ir_xv, demetroullas15, planck_ir_xxxi}. In our case, we 
ensured we acknowledge this issue when presenting our results. 

\input{tables/reg_contrib.tex}

\bsp	
\label{lastpage}
\end{document}

%% file: authors_and_institutions.tex
\author[D. Tramonte et al.]{
D. Tramonte,$^{1,2,3,4}$\thanks{E-mail: tramonte@pmo.ac.cn}
R.~T. G\'enova-Santos,$^{3,4}$
J.~A. Rubi{\~n}o-Mart\'in,$^{3,4}$
P. Vielva,$^{5}$
\newauthor
F. Poidevin,$^{3,4}$
C.~H. L\'opez-Caraballo,$^{3,4}$
M.~W. Peel,$^{3,4}$
M. Ashdown,$^{6,7}$
E. Artal,$^{8}$
\newauthor
R.~B. Barreiro,$^{5}$
F.~J. Casas,$^{5}$
E. de la Hoz,$^{5,9}$
M. Fern\'andez-Torreiro,$^{3,4}$
F. Guidi,$^{10,3,4}$
\newauthor
D. Herranz,$^{5}$
R.~J. Hoyland,$^{3,4}$
A.~N. Lasenby,$^{6,7}$
E. Martinez-Gonzalez,$^{5}$
L. Piccirillo,$^{11}$
\newauthor
R. Rebolo,$^{3,4,12}$
B. Ruiz-Granados,$^{3,4,13}$
F. Vansyngel$^{3,4}$
and R.~A. {Watson}$^{11}$ 
\\
\\
$^{1}$Purple Mountain Observatory, CAS, No.10 Yuanhua Road, Qixia District, Nanjing 210034, China\\
$^{2}$NAOC-UKZN Computational Astrophysics Center (NUCAC), University of Kwazulu-Natal, Durban 4000, South Africa\\
$^{3}$Instituto de Astrof\'{\i}sica de Canarias, E-38200 La Laguna, Tenerife, Spain\\
$^{4}$Departamento de Astrof\'{\i}sica, Universidad de La Laguna, E-38206 La Laguna, Tenerife, Spain\\
$^{5}$Instituto de F\'{\i}sica de Cantabria (IFCA), CSIC-Univ. de Cantabria, Avda. los Castros, s/n, E-39005 Santander, Spain\\
$^{6}$Astrophysics Group, Cavendish Laboratory, University of Cambridge, J J Thomson Avenue, Cambridge CB3 0HE, UK\\
$^{7}$Kavli Institute for Cosmology, University of Cambridge, Madingley Road, Cambridge CB3 0HA, UK\\
$^{8}$Departamento de Ingenier\'{\i}a de COMunicaciones (DICOM), Universidad de Cantabria, Plaza de la Ciencia s/n, 39005 Santander, Spain\\
$^{9}$Dpto. de F\'{\i}sica Moderna, Universidad de Cantabria, Avda. los Castros s/n, E-39005 Santander, Spain \\
$^{10}$Institut d'Astrophysique de Paris, UMR 7095, CNRS \& Sorbonne Universit\'e, 98 bis boulevard Arago, 75014 Paris, France\\
$^{11}$Jodrell Bank Centre for Astrophysics, Alan Turing Building, Department of Physics and Astronomy, School of Nature Sciences, 
University of Manchester,\\ Oxford Road, Manchester M13 9PL, U.K\\
$^{12}$Consejo Superior de Investigaciones Cientificas, E-28006 Madrid, Spain\\
$^{13}$Departamento de F\'{\i}sica. Facultad de Ciencias. Universidad de C\'ordoba.  Campus de Rabanales, Edif. C2. Planta Baja.  E-14071 C\'ordoba, Spain
}

%% file: tables/nulltest_table.tex
\begin{table*}
\centering
\caption{Results of the null test performed on QUIJOTE maps. For each frequency and Stokes parameter we compare the 1-degree beam rms obtained in the original map (Map) and in the null test map (NT). The last column reports the resulting MFI instantaneous sensitivity in polarisation.}
\label{tab:nulltest}
\setlength{\tabcolsep}{1.5em}
\begin{tabular}{ccccccccccc}
\hline
\hline
 Frequency & \multicolumn{2}{c}{$\sigma_{I}$ [$\mu$K$\,\text{beam}^{-1}$] } & & \multicolumn{2}{c}{$\sigma_{Q}$ [$\mu$K$\,\text{beam}^{-1}$]} & &  \multicolumn{2}{c}{$\sigma_{U}$  [$\mu$K$\,\text{beam}^{-1}$]} & & $\sigma_{Q,U}$ [mK$\,\text{s}^{1/2}$] \\ 
\\[-1em]
[GHz]  & Map & NT & & Map & NT & & Map & NT & & NT \\
\\[-1em]
\hline
\hline

\multicolumn{11}{c}{W49/W51}
\\[-1em]
\\[-1em]
\\[-1em]
\\[-1em]
\\[-1em]
\hline
11.0   &  617.4   &   53.8   &     &   25.8   &   22.5   &     &   21.9   &   21.3     &   &       1.4  \\
13.0   &  489.8   &   34.6   &     &   18.9   &   16.2   &     &   15.9   &   15.1     &   &       1.2  \\
17.0   &  300.7   &   57.5   &     &   14.1   &   13.1   &     &   13.8   &   13.2     &   &       1.4  \\
19.0   &  252.9   &   71.7   &     &   15.9   &   12.8   &     &   15.3   &   14.6     &   &       1.4  \\

\hline
\multicolumn{11}{c}{IC443}
\\[-1em]
\\[-1em]
\\[-1em]
\\[-1em]
\\[-1em]
\hline
11.0   &   71.1   &   59.8   &     &   21.2   &   19.6   &     &   21.1   &   20.0     &   &       1.1  \\
13.0   &   52.2   &   44.7   &     &   20.1   &   18.4   &     &   18.4   &   19.0     &   &       1.0  \\
17.0   &   80.9   &   62.4   &     &   16.0   &   15.3   &     &   19.2   &   15.8     &   &       1.2  \\
19.0   &  100.1   &   79.3   &     &   18.8   &   17.6   &     &   19.2   &   17.1     &   &       1.4  \\

\hline
\hline
\end{tabular}
\end{table*}

%% file: tables/w49_fluxtable.tex
\begin{table*}
\centering
\caption{Summary table for the intensity and (when available) polarised flux densities obtained for W49 at different frequencies. Values for the Stokes parameters $I$, $Q$ and $U$ are extracted from the QUIJOTE and ancillary maps degraded at a 1 degree resolution, while the polarised intensity $P$, the polarised fraction $\Pi$ and the polarisation angle $\gamma$ are computed as detailed in Section~\ref{sec:seds}. The last column reports the colour correction coefficients for intensity and polarisation, computed following the methodology described in Section~\ref{ssec:fit_method}; when polarisation data are not available we quote the colour correction in total intensity only.}
\label{tab:w49_fluxes}
\setlength{\tabcolsep}{1.em}
\renewcommand{\arraystretch}{1.1}
\begin{tabular}{cc|ccccccc}
\hline
\hline
 \multicolumn{9}{c}{W49}   \\ 
\hline
 Survey & Freq. &   $I$ & $Q$ & $U$ & $P$ & $\Pi$ & $\gamma$ & CC \\ 
\\[-1em]
  & [GHz] &   [Jy] & [Jy] & [Jy] & [Jy]  & \% & [deg] & $(I,P)$ \\ 
\hline
\hline

Haslam & 0.4 & 208$\pm$35 &  -  &  -  &  -  &  -  &  -  & - \\
Berkhuijsen & 0.8 & 167$\pm$23 &  -  &  -  &  -  &  -  &  -  & - \\
Reich & 1.4 & 134$\pm$16 &  -  &  -  &  -  &  -  &  -  & - \\
DRAO & 1.4 &  -  & 0.7$\pm$2.1 & 4.5$\pm$1.3 & 4.2$\pm$1.7 & 3.2$^{+1.3}_{-1.4}$ & -40.4$\pm$12.9 & - \\
Jonas & 2.3 & 105$\pm$13 &  -  &  -  &  -  &  -  &  -  & - \\
Urumqi & 4.8 & 150$\pm$16 & 2.6$\pm$7.8 & 3.6$\pm$7.8 & $\lesssim$17.0 & $\lesssim$11.2 & -26.9$\pm$50.6 & - \\
QUIJOTE & 11.1 & 143$\pm$8 & 1.9$\pm$0.6 & 0.6$\pm$0.5 & 1.9$\pm$0.6 & 1.3$^{+0.4}_{-0.5}$ & -8.4$\pm$7.7 & 0.986, 0.974 \\
QUIJOTE & 12.9 & 148$\pm$8 & 1.3$\pm$0.6 & 0.2$\pm$0.3 & 1.2$\pm$0.4 & 0.8$\pm$0.3 & -4.5$\pm$6.1 & 1.001, 0.998 \\
QUIJOTE & 16.8 & 158$\pm$9 & 1.2$\pm$0.3 & 0.2$\pm$0.2 & 1.2$\pm$0.3 & 0.7$\pm$0.2 & -3.8$\pm$5.4 & 1.007, 1.019 \\
QUIJOTE & 18.8 & 160$\pm$9 & 1.2$\pm$0.3 & 0.2$\pm$0.2 & 1.2$\pm$0.3 & 0.8$\pm$0.2 & -4.7$\pm$5.3 & 1.008, 1.012 \\
\textit{WMAP} & 22.8 & 161$\pm$6 & 1.0$\pm$0.3 & 0.3$\pm$0.2 & 1.1$\pm$0.2 & 0.7$\pm$0.1 & -7.8$\pm$4.8 & 0.973, 0.960 \\
\textit{Planck}-LFI & 28.4 & 149$\pm$6 & 1.0$\pm$0.2 & 0.1$\pm$0.2 & 1.0$\pm$0.2 & 0.7$\pm$0.1 & -2.9$\pm$4.1 & 1.006, 1.002 \\
\textit{WMAP} & 33.0 & 144$\pm$5 & 0.8$\pm$0.3 & 0.1$\pm$0.2 & 0.8$\pm$0.3 & 0.5$\pm$0.2 & -1.8$\pm$8.2 & 0.982, 0.978 \\
\textit{WMAP} & 40.6 & 134$\pm$4 & 0.5$\pm$0.4 & 0.3$\pm$0.3 & 0.4$^{+0.2}_{-0.4}$ & $\lesssim$0.9 & -16.7$\pm$16.1 & 0.994, 0.992 \\
\textit{Planck}-LFI & 44.1 & 132$\pm$5 & 0.4$\pm$0.2 & 0.1$\pm$0.2 & 0.3$\pm$0.2 & 0.2$^{+0.1}_{-0.2}$ & -7.3$\pm$15.2 & 0.992, 0.991 \\
\textit{WMAP} & 60.8 & 130$\pm$5 & 0.0$\pm$0.4 & 0.4$\pm$0.6 & $\lesssim$1.2 & $\lesssim$0.9 & -42.4$\pm$33.8 & 0.977, 0.985 \\
\textit{Planck}-LFI & 70.4 & 137$\pm$6 & 1.0$\pm$0.3 & 0.5$\pm$0.3 & 1.1$\pm$0.3 & 0.7$^{+0.3}_{-0.2}$ & -14.8$\pm$8.3 & 0.984, 0.985 \\
\textit{WMAP} & 93.5 & 189$\pm$9 & 1.5$\pm$2.0 & 0.0$\pm$1.6 & $\lesssim$4.3 & $\lesssim$2.3 & -0.7$\pm$31.3 & 0.993, 1.000 \\
\textit{Planck}-HFI & 100.0 & 220$\pm$11 & 3.4$\pm$0.5 & -0.1$\pm$0.3 & 3.4$\pm$0.4 & 1.5$\pm$0.2 & 1.1$\pm$2.4 & 0.965, 0.945 \\
\textit{Planck}-HFI & 143.0 & 493$\pm$24 & 9.6$\pm$1.5 & -0.2$\pm$0.9 & 9.6$\pm$1.2 & 1.9$\pm$0.3 & 0.6$\pm$2.7 & 0.984, 0.979 \\
\textit{Planck}-HFI & 217.0 & 2021$\pm$109 & 43.2$\pm$6.5 & 5.7$\pm$4.2 & 43.2$\pm$5.2 & 2.1$\pm$0.3 & -3.7$\pm$2.8 & 0.898, 0.896 \\
\textit{Planck}-HFI & 353.0 & 9666$\pm$573 & 190.2$\pm$31.7 & 43.9$\pm$18.6 & 193.7$^{+24.2}_{-24.3}$ & 2.0$\pm$0.3 & -6.5$\pm$2.9 & 0.896, 0.900 \\
\textit{Planck}-HFI & 545.0 & (3.58$\pm$0.28) $\times 10^{4}$ &  -  &  -  &  -  &  -  &  -  & 0.887 \\
\textit{Planck}-HFI & 857.0 & (1.18$\pm$0.09) $\times 10^{5}$ &  -  &  -  &  -  &  -  &  -  & 0.961 \\
DIRBE & 1249.1 & (2.66$\pm$0.32) $\times 10^{5}$ &  -  &  -  &  -  &  -  &  -  & 0.996 \\
DIRBE & 2141.4 & (4.14$\pm$0.50) $\times 10^{5}$ &  -  &  -  &  -  &  -  &  -  & 1.075 \\
DIRBE & 2997.9 & (2.13$\pm$0.26) $\times 10^{5}$ &  -  &  -  &  -  &  -  &  -  & 1.086 \\

\hline
\hline
\end{tabular}
\end{table*}

%% file: tables/w51_fluxtable.tex
\begin{table*}
\centering
\caption{Same as in Table~\ref{tab:w49_fluxes}, but for the W51 region.}
\label{tab:w51_fluxes}
\setlength{\tabcolsep}{1.em}
\renewcommand{\arraystretch}{1.1}
\begin{tabular}{cc|ccccccc}
\hline
\hline
 \multicolumn{9}{c}{W51}   \\ 
\hline
 Survey & Freq. &   $I$ & $Q$ & $U$ & $P$ & $\Pi$ & $\gamma$ & CC \\ 
\\[-1em]
  & [GHz] &   [Jy] & [Jy] & [Jy] & [Jy]  & \% & [deg] & $(I,P)$ \\ 
\hline
\hline

Haslam & 0.4 & 684$\pm$83 &  -  &  -  &  -  &  -  &  -  & - \\
Berkhuijsen & 0.8 & 653$\pm$71 &  -  &  -  &  -  &  -  &  -  & - \\
Reich & 1.4 & 411$\pm$43 &  -  &  -  &  -  &  -  &  -  & - \\
DRAO & 1.4 &  -  & 4.3$\pm$2.1 & -6.3$\pm$1.9 & 7.4$^{+2.0}_{-2.1}$ & 1.8$^{+0.6}_{-0.5}$ & 27.8$\pm$7.7 & - \\
Jonas & 2.3 & 591$\pm$61 &  -  &  -  &  -  &  -  &  -  & - \\
Urumqi & 4.8 & 657$\pm$66 & -3.0$\pm$8.4 & 2.7$\pm$8.4 & $\lesssim$17.8 & $\lesssim$2.7 & -68.9$\pm$58.8 & - \\
QUIJOTE & 11.1 & 508$\pm$25 & -4.4$\pm$0.8 & -0.7$\pm$0.7 & 4.4$\pm$0.7 & 0.8$^{+0.2}_{-0.1}$ & 85.3$\pm$4.4 & 0.982, 0.975 \\
QUIJOTE & 12.9 & 499$\pm$25 & -5.6$\pm$0.6 & -4.2$\pm$0.8 & 6.9$\pm$0.7 & 1.4$\pm$0.2 & 71.5$\pm$2.9 & 1.001, 0.998 \\
QUIJOTE & 16.8 & 502$\pm$25 & -2.6$\pm$0.4 & -3.4$\pm$0.5 & 4.3$\pm$0.4 & 0.8$\pm$0.1 & 64.0$\pm$2.8 & 1.008, 1.018 \\
QUIJOTE & 18.8 & 504$\pm$25 & -3.2$\pm$0.5 & -4.0$\pm$0.4 & 5.1$\pm$0.5 & 1.0$\pm$0.1 & 64.3$\pm$2.8 & 1.008, 1.011 \\
\textit{WMAP} & 22.8 & 498$\pm$15 & -3.2$\pm$0.2 & -2.4$\pm$0.4 & 4.0$\pm$0.3 & 0.8$\pm$0.1 & 71.3$\pm$2.4 & 0.971, 0.963 \\
\textit{Planck}-LFI & 28.4 & 470$\pm$14 & -3.4$\pm$0.3 & -1.8$\pm$0.3 & 3.8$\pm$0.3 & 0.8$\pm$0.1 & 76.4$\pm$2.5 & 1.006, 1.004 \\
\textit{WMAP} & 33.0 & 466$\pm$14 & -3.1$\pm$0.4 & -1.6$\pm$0.4 & 3.5$\pm$0.4 & 0.7$\pm$0.1 & 76.2$\pm$3.2 & 0.982, 0.979 \\
\textit{WMAP} & 40.6 & 435$\pm$13 & -2.9$\pm$0.5 & -1.4$\pm$0.4 & 3.2$\pm$0.5 & 0.7$\pm$0.1 & 77.4$\pm$3.8 & 0.995, 0.993 \\
\textit{Planck}-LFI & 44.1 & 426$\pm$13 & -2.9$\pm$0.2 & -1.6$\pm$0.3 & 3.3$\pm$0.3 & 0.8$\pm$0.1 & 75.5$\pm$2.6 & 0.993, 0.991 \\
\textit{WMAP} & 60.8 & 405$\pm$14 & -2.1$\pm$0.9 & -1.4$\pm$0.5 & 2.4$\pm$0.7 & 0.6$\pm$0.2 & 72.5$\pm$7.6 & 0.976, 0.972 \\
\textit{Planck}-LFI & 70.4 & 406$\pm$14 & -1.4$\pm$0.6 & -1.2$\pm$0.5 & 1.7$\pm$0.5 & 0.4$\pm$0.1 & 70.2$\pm$8.1 & 0.985, 0.982 \\
\textit{WMAP} & 93.5 & 481$\pm$23 & -1.3$\pm$3.0 & -1.1$\pm$1.9 & $\lesssim$5.4 & $\lesssim$1.1 & 69.7$\pm$41.1 & 0.988, 0.985 \\
\textit{Planck}-HFI & 100.0 & 529$\pm$21 & 0.4$\pm$0.7 & -1.9$\pm$0.4 & 1.8$\pm$0.6 & 0.3$\pm$0.1 & 39.0$\pm$10.8 & 0.976, 0.970 \\
\textit{Planck}-HFI & 143.0 & 957$\pm$43 & 7.3$\pm$2.2 & 2.5$\pm$1.3 & 7.5$\pm$1.7 & 0.8$\pm$0.2 & -9.5$\pm$5.4 & 0.986, 0.970 \\
\textit{Planck}-HFI & 217.0 & 3545$\pm$197 & 39.3$\pm$9.2 & 15.1$\pm$4.7 & 41.6$\pm$6.6 & 1.2$\pm$0.2 & -10.5$\pm$3.7 & 0.900, 0.870 \\
\textit{Planck}-HFI & 353.0 & (1.63$\pm$0.10) $\times 10^{4}$ & 159.5$\pm$43.1 & 93.0$\pm$19.9 & 182.3$\pm$29.3 & 1.1$\pm$0.2 & -15.1$\pm$4.3 & 0.898, 0.870 \\
\textit{Planck}-HFI & 545.0 & (6.08$\pm$0.48) $\times 10^{4}$ &  -  &  -  &  -  &  -  &  -  & 0.887 \\
\textit{Planck}-HFI & 857.0 & (2.06$\pm$0.16) $\times 10^{5}$ &  -  &  -  &  -  &  -  &  -  & 0.958 \\
DIRBE & 1249.1 & (4.82$\pm$0.59) $\times 10^{5}$ &  -  &  -  &  -  &  -  &  -  & 0.982 \\
DIRBE & 2141.4 & (8.66$\pm$1.06) $\times 10^{5}$ &  -  &  -  &  -  &  -  &  -  & 1.081 \\
DIRBE & 2997.9 & (5.60$\pm$0.68) $\times 10^{5}$ &  -  &  -  &  -  &  -  &  -  & 1.072 \\

\hline
\hline
\end{tabular}
\end{table*}

%% file: tables/ic443_fluxtable.tex
\begin{table*}
\centering
\caption{Same as in Table~\ref{tab:w49_fluxes}, but for the IC443 region.}
\label{tab:ic443_fluxes}
\setlength{\tabcolsep}{1.em}
\renewcommand{\arraystretch}{1.1}
\begin{tabular}{cc|ccccccc}
\hline
\hline
 \multicolumn{9}{c}{IC443}   \\ 
\hline
 Survey & Freq. &   $I$ & $Q$ & $U$ & $P$ & $\Pi$ & $\gamma$ & CC \\ 
\\[-1em]
  & [GHz] &   [Jy] & [Jy] & [Jy] & [Jy]  & \% & [deg] & $(I,P)$ \\ 
\hline
\hline

Haslam & 0.4 & 130$\pm$14 &  -  &  -  &  -  &  -  &  -  & - \\
Berkhuijsen & 0.8 & 132$\pm$13 &  -  &  -  &  -  &  -  &  -  & - \\
Reich & 1.4 & 81$\pm$8 &  -  &  -  &  -  &  -  &  -  & - \\
DRAO & 1.4 &  -  & 0.2$\pm$3.3 & 3.6$\pm$3.5 & $\lesssim$8.9 & $\lesssim$11.0 & -43.4$\pm$26.4 & - \\
Urumqi & 4.8 & 88$\pm$8 & 1.6$\pm$1.0 & -0.2$\pm$1.0 & 0.9$^{+0.8}_{-0.9}$ & $\lesssim$4.0 & 3.1$\pm$18.2 & - \\
QUIJOTE & 11.1 & 57$\pm$3 & 1.0$\pm$0.2 & 0.6$\pm$0.1 & 1.1$\pm$0.1 & 1.9$\pm$0.3 & -15.8$\pm$3.5 & 0.977, 0.977 \\
QUIJOTE & 12.9 & 54$\pm$3 & 0.4$\pm$0.1 & 1.1$\pm$0.1 & 1.2$\pm$0.1 & 2.1$\pm$0.3 & -36.2$\pm$3.6 & 1.000, 0.999 \\
QUIJOTE & 16.8 & 52$\pm$3 & 0.3$\pm$0.2 & 0.6$\pm$0.1 & 0.7$\pm$0.2 & 1.3$^{+0.3}_{-0.4}$ & -31.2$\pm$7.8 & 1.008, 1.017 \\
QUIJOTE & 18.8 & 51$\pm$3 & 0.1$\pm$0.3 & 0.4$\pm$0.4 & $\lesssim$0.9 & $\lesssim$1.8 & -40.4$\pm$25.6 & 1.008, 1.011 \\
\textit{WMAP} & 22.8 & 47$\pm$2 & 0.4$\pm$0.2 & 0.7$\pm$0.1 & 0.8$\pm$0.1 & 1.7$\pm$0.3 & -29.6$\pm$4.9 & 0.966, 0.965 \\
\textit{Planck}-LFI & 28.4 & 42$\pm$2 & 0.8$\pm$0.1 & 0.4$\pm$0.1 & 0.9$\pm$0.1 & 2.1$\pm$0.3 & -13.8$\pm$3.5 & 1.005, 1.005 \\
\textit{WMAP} & 33.0 & 40$\pm$2 & 0.4$\pm$0.3 & 0.6$\pm$0.2 & 0.7$\pm$0.2 & 1.7$\pm$0.6 & -26.0$\pm$10.3 & 0.980, 0.981 \\
\textit{WMAP} & 40.6 & 37$\pm$2 & 0.3$\pm$0.3 & 0.5$\pm$0.5 & 0.4$^{+0.2}_{-0.4}$ & $\lesssim$3.5 & -31.8$\pm$15.1 & 0.994, 0.994 \\
\textit{Planck}-LFI & 44.1 & 35$\pm$2 & 0.5$\pm$0.3 & 0.6$\pm$0.3 & 0.6$^{+0.3}_{-0.4}$ & 1.8$^{+0.9}_{-1.0}$ & -24.5$\pm$12.5 & 0.992, 0.992 \\
\textit{WMAP} & 60.8 & 33$\pm$5 & 0.5$\pm$1.4 & 0.8$\pm$1.3 & $\lesssim$3.0 & $\lesssim$9.3 & -29.1$\pm$40.9 & 0.975, 0.973 \\
\textit{Planck}-LFI & 70.4 & 33$\pm$5 & -0.1$\pm$0.4 & 1.0$\pm$0.3 & 1.0$\pm$0.4 & 2.9$^{+1.2}_{-1.4}$ & -47.6$\pm$11.4 & 0.984, 0.984 \\
\textit{WMAP} & 93.5 & 43$\pm$11 & 0.6$\pm$2.2 & -0.4$\pm$2.7 & $\lesssim$5.0 & $\lesssim$11.6 & 15.3$\pm$102.5 & 0.989, 0.983 \\
\textit{Planck}-HFI & 100.0 & 48$\pm$11 & 0.3$\pm$0.4 & 0.5$\pm$0.3 & 0.3$\pm$0.3 & $\lesssim$2.8 & -28.9$\pm$20.6 & 0.975, 0.988 \\
\textit{Planck}-HFI & 143.0 & 80$\pm$23 & 0.5$\pm$0.6 & 1.0$\pm$0.7 & 0.9$^{+0.5}_{-0.7}$ & $\lesssim$3.4 & -32.0$\pm$15.4 & 0.991, 0.987 \\
\textit{Planck}-HFI & 217.0 & 282$\pm$84 & 3.7$\pm$3.4 & 2.9$\pm$2.6 & 2.8$^{+2.3}_{-2.8}$ & $\lesssim$4.2 & -19.3$\pm$17.8 & 0.910, 0.893 \\
\textit{Planck}-HFI & 353.0 & 1205$\pm$353 & 15.8$\pm$14.6 & 11.7$\pm$10.0 & 12.6$^{+8.9}_{-12.6}$ & $\lesssim$4.1 & -18.2$\pm$17.2 & 0.907, 0.884 \\
\textit{Planck}-HFI & 545.0 & 3934$\pm$1142 &  -  &  -  &  -  &  -  &  -  & 0.900 \\
\textit{Planck}-HFI & 857.0 & (1.09$\pm$0.31) $\times 10^{4}$ &  -  &  -  &  -  &  -  &  -  & 0.970 \\
DIRBE & 1249.1 & (2.00$\pm$0.31) $\times 10^{4}$ &  -  &  -  &  -  &  -  &  -  & 1.013 \\
DIRBE & 2141.4 & (2.54$\pm$0.40) $\times 10^{4}$ &  -  &  -  &  -  &  -  &  -  & 1.060 \\
DIRBE & 2997.9 & (1.21$\pm$0.20) $\times 10^{4}$ &  -  &  -  &  -  &  -  &  -  & 1.088 \\

\hline
\hline
\end{tabular}
\end{table*}

%% file: tables/w49_fitres.tex
\begin{table}
\centering
\caption{Results of the multi-component fits for the W49 region. The table reports the best-fit values for different parameters when fitting the total intensity $I$ and the polarised intensity $P$. The table also reports the results of the fit in intensity when fixing the synchrotron spectral index to the value fitted in polarisation. When fitting for $P$ the dust temperature is always fixed to the value obtained in total intensity.}
\label{tab:w49_fit}
\setlength{\tabcolsep}{0.8em}
\renewcommand{\arraystretch}{1.1}
\begin{tabular}{cccc}
\hline
\hline
 \multicolumn{4}{c}{W49}   \\ 
\hline
 Parameter & I & I (fixed $\alpha_{\rm s}$) & P \\ 
\hline
\hline
\\[-1em]
$A_{\rm s}$ [Jy] &  -  & 65.3$\pm$31.3 & 7.2$\pm$2.2\\
$\alpha_{\rm s}$ &  -  & -0.67 & -0.67$\pm$0.10\\
EM [cm$^{-6}$pc] & 1315$\pm$84 & 774$\pm$283 &  - \\
$A_{\rm AME}$ [Jy] & 55.0$\pm$7.2 & 85.9$\pm$17.6 &  - \\
$W_{\rm AME}$ & 0.6$\pm$0.1 & 0.9$\pm$0.2 &  - \\
$\nu_{\rm AME}$ [GHz] & 21.0$\pm$1.4 & 20.0$\pm$1.4 &  - \\
$\Delta T_{\rm CMB}$ [$\mu$K] & 98$\pm$67 & 174$\pm$88 &  - \\
$\tau_{250}\times10^3$ & 1.52$\pm$0.26 & 1.53$\pm$0.27 & 0.03$\pm$0.01\\
$\beta_{\rm dust}$ & 1.75$\pm$0.09 & 1.75$\pm$0.10 & 1.64$\pm$0.13\\
$T_{\rm dust}$ [K] & 20.4$\pm$0.9 & 20.4$\pm$0.9 & 20.4\\
\hline
$\chi^2$ & 12.3 & 8.9 & 16.8\\
$\chi^2_{\rm red}$ & 0.69 & 0.52 & 1.20\\
\hline
\hline
\end{tabular}
\end{table}

%% file: tables/w51_fitres.tex
\begin{table}
\centering
\caption{Same as in Table~\ref{tab:w49_fit}, but for the W51 region.}
\label{tab:w51_fit}
\setlength{\tabcolsep}{0.8em}
\renewcommand{\arraystretch}{1.1}
\begin{tabular}{cccc}
\hline
\hline
 \multicolumn{4}{c}{W51}   \\ 
\hline
 Parameter & I & I (fixed $\alpha_{\rm s}$) & P \\ 
\hline
\hline
\\[-1em]
$A_{\rm s}$ [Jy] &  -  & 359.0$\pm$140.3 & 19.8$\pm$4.6\\
$\alpha_{\rm s}$ &  -  & -0.51 & -0.51$\pm$0.07\\
EM [cm$^{-6}$pc] & 2828$\pm$178 & 428$\pm$1071 &  - \\
$A_{\rm AME}$ [Jy] & 105.6$\pm$26.3 & 359.3$\pm$121.7 &  - \\
$W_{\rm AME}$ & 0.9$\pm$0.3 & 1.7$\pm$0.4 &  - \\
$\nu_{\rm AME}$ [GHz] & 17.7$\pm$3.6 & 15.8$\pm$3.0 &  - \\
$\Delta T_{\rm CMB}$ [$\mu$K] & -51$\pm$105 & 94$\pm$181 &  - \\
$\tau_{250}\times10^3$ & 1.08$\pm$0.20 & 1.10$\pm$0.21 & 0.03$\pm$0.02\\
$\beta_{\rm dust}$ & 1.65$\pm$0.10 & 1.66$\pm$0.11 & 2.35$\pm$0.33\\
$T_{\rm dust}$ [K] & 23.1$\pm$1.2 & 22.9$\pm$1.3 & 23.1\\
\hline
$\chi^2$ & 22.2 & 19.2 & 29.2\\
$\chi^2_{\rm red}$ & 1.23 & 1.13 & 2.25\\
\hline
\hline
\end{tabular}
\end{table}

%% file: tables/ic443_fitres.tex
\begin{table}
\centering
\caption{Same as in Table~\ref{tab:w49_fit}, but for the IC443 region. In this case, the additional last column reports the results of the fit which adopts a broken power-law synchrotron model, and which results in the best modelling for IC433 intensity SED.}
\label{tab:ic443_fit}
\setlength{\tabcolsep}{0.5em}
\renewcommand{\arraystretch}{1.1}
\begin{tabular}{ccccc}
\hline
\hline
 \multicolumn{5}{c}{IC443}   \\ 
\hline
 Parameter & $I $& $I$ (fixed $\alpha_{\rm s}$) & $P$ & $I$ (curv. s.) \\ 
\hline
\hline
\\[-1em]
$A_{\rm s}$ [Jy] & 96.5$\pm$10.6 & 71.9$\pm$10.9 & 2.6$\pm$1.0 & 106.7$\pm$5.8\\
$\alpha_{\rm s}$ & -0.29$\pm$0.03 & -0.37 & -0.37$\pm$0.13 & -0.21$\pm$0.04\\
$\nu_{\rm 0,s}$ [GHz] &  -  &  -  &  -  & 114$\pm$73\\
EM [cm$^{-6}$pc] & 69$\pm$60 & 235$\pm$39 &  -  &  - \\
$\Delta T_{\rm CMB}$ [$\mu$K] & -58$\pm$37 & -65$\pm$35 &  -  & 29$\pm$54\\
$\tau_{250}\times10^3$ & 0.09$\pm$0.03 & 0.09$\pm$0.03 & 0.002$\pm$0.006 & 0.11$\pm$0.04\\
$\beta_{\rm dust}$ & 1.31$\pm$0.25 & 1.28$\pm$0.23 & 2.12$\pm$1.49 & 1.51$\pm$0.29\\
$T_{\rm dust}$ [K] & 20.6$\pm$2.0 & 20.7$\pm$1.9 & 20.0 & 19.4$\pm$1.9\\
\hline
$\chi^2$ & 13.1 & 13.8 & 12.7 & 11.0\\
$\chi^2_{\rm red}$ & 0.73 & 0.72 & 0.90 & 0.61\\
\hline
\hline
\end{tabular}
\end{table}

%% file: tables/uchii.tex
\begin{table}
\centering
\caption{Results for the contribution from UCHII regions in W49 and W51. We report the AME residual amplitude at 15\,GHz,  $S_{\rm 15 GHz}^{\rm AME}$, the measured UCHII total flux at 100\,$\mu$m from IRAS data, $S_{100\mu\text{m}}^{\rm IRAS}$, and its corresponding upper and lower limit extrapolations at 15\,GHz, $S_{15\text{GHz}}^{\rm uplim}$ and $S_{15\text{GHz}}^{\rm lowlim}$. The last column reports the total UCHII flux measured with CORNISH data at 5\,GHz.}
\label{tab:uchii}
\setlength{\tabcolsep}{1.em}
\renewcommand{\arraystretch}{1.1}
\begin{tabular}{cccccc}
\hline
\hline
 Source & $S_{\rm 15 GHz}^{\rm AME}$ & $S_{100\mu\text{m}}^{\rm IRAS}$ & $S_{15\text{GHz}}^{\rm uplim}$ & $S_{15\text{GHz}}^{\rm lowlim}$  &  $S_{\rm 5 GHz}^{\rm CORNISH}$  \\ 
\\[-1em]
  & [Jy] & [Jy] & [Jy] & [Jy] & [Jy] \\ 
\hline
\hline
W49 & 81.4 & 45098 & 45.1 & 0.1 & 8.5 \\
W51 & 103.9 & 33368 & 33.4 & 0.1 & 5.3 \\
\hline
\hline
\end{tabular}
\end{table}

%% file: tables/polame.tex
\begin{table}
\centering
\caption{AME polarisation constraints from 16 to 61\,GHz. For each frequency point we report the local residual AME amplitude $I_{\rm AME}$, the colour-corrected polarised flux $P$, the resulting polarisation fraction $P/I_{\rm AME}$ and the associated upper limit for the AME polarisation fraction $\Pi_{\rm AME}$, computed as the 95\% confidence level upper boundary on the fraction $P/I_{\rm AME}$. When no uncertainty on $P/I_{\rm AME}$ is quoted it means the debiased $P$ flux is null and the reported upper limit is already the top boundary of the 95\% confidence level.}
\label{tab:polame}
\setlength{\tabcolsep}{1.em}
\renewcommand{\arraystretch}{1.1}
\begin{tabular}{ccccc}
\hline
\hline
\multicolumn{5}{c}{W49}  \\ 
\hline
  Freq. &  $I_{\rm AME}$ &  $P$ &  $P/I_{\rm AME}$ & $\Pi_{\rm AME}$ \\ 
\\[-1em]
  [GHz] &  [Jy]        & [Jy] & [\%] & [\%] \\ 
\hline
\hline
16.8 & 84.2$\pm$17.3 & 1.2$\pm$0.3 & 1.4$\pm$0.4 & $\lesssim$2.3 \\
18.8 & 85.7$\pm$17.5 & 1.2$\pm$0.3 & 1.5$\pm$0.4 & $\lesssim$2.3 \\
22.8 & 84.9$\pm$17.5 & 1.0$\pm$0.2 & 1.2$\pm$0.4 & $\lesssim$1.9 \\
28.4 & 79.2$\pm$17.2 & 1.0$\pm$0.2 & 1.3$\pm$0.4 & $\lesssim$2.0 \\
33.0 & 72.8$\pm$16.9 & 0.8$\pm$0.3 & 1.0$\pm$0.5 & $\lesssim$2.0 \\
40.6 & 61.7$\pm$16.6 & 0.4$^{+0.2}_{-0.4}$ & 0.6$^{+0.4}_{-0.7}$ & $\lesssim$1.4 \\
44.1 & 56.8$\pm$16.5 & 0.3$\pm$0.2 & 0.6$^{+0.3}_{-0.4}$ & $\lesssim$1.2 \\
60.8 & 38.0$\pm$15.0 & $\lesssim$1.2 & $\lesssim$3.1 & $\lesssim$3.1 \\
\hline
\hline
\multicolumn{5}{c}{W51}  \\ 
\hline
  Freq. &  $I_{\rm AME}$ &  $P$ &  $P/I_{\rm AME}$ & $\Pi_{\rm AME}$ \\ 
\\[-1em]
  [GHz] &  [Jy]        & [Jy] & [\%] & [\%] \\ 
\hline
\hline
16.8 & 105.5$\pm$27.5 & 4.4$\pm$0.4 & 4.1$\pm$1.2 & $\lesssim$6.4 \\
18.8 & 105.4$\pm$28.0 & 5.1$\pm$0.5 & 4.9$\pm$1.4 & $\lesssim$7.6 \\
22.8 & 101.7$\pm$29.9 & 3.8$\pm$0.3 & 3.8$\pm$1.1 & $\lesssim$6.0 \\
28.4 & 92.7$\pm$32.1 & 3.9$\pm$0.3 & 4.2$\pm$1.5 & $\lesssim$7.1 \\
33.0 & 84.2$\pm$32.6 & 3.4$\pm$0.4 & 4.1$\pm$1.6 & $\lesssim$7.3 \\
40.6 & 70.6$\pm$32.0 & 3.2$\pm$0.5 & 4.5$\pm$2.1 & $\lesssim$8.7 \\
44.1 & 64.9$\pm$31.4 & 3.3$\pm$0.3 & 5.1$\pm$2.5 & $\lesssim$10.1 \\
60.8 & 43.4$\pm$27.6 & 2.3$\pm$0.7 & 5.4$\pm$3.8 & $\lesssim$12.9 \\
\hline
\hline
\end{tabular}
\end{table}

%% file: acknowledgements.tex
\section*{Acknowledgements}

We thank the staff of the Teide Observatory for invaluable assistance in the commissioning and operation of QUIJOTE.
The {\it QUIJOTE} experiment is being developed by the Instituto de Astrofisica de Canarias (IAC),
the Instituto de Fisica de Cantabria (IFCA), and the Universities of Cantabria, Manchester and Cambridge.
Partial financial support was provided by the Spanish Ministry of Science and Innovation 
under the projects AYA2007-68058-C03-01, AYA2007-68058-C03-02,
AYA2010-21766-C03-01, AYA2010-21766-C03-02, AYA2014-60438-P,
ESP2015-70646-C2-1-R, AYA2017-84185-P, ESP2017-83921-C2-1-R,
AYA2017-90675-REDC (co-funded with EU FEDER funds),
PGC2018-101814-B-I00, 
PID2019-110610RB-C21, PID2020-120514GB-I00, IACA13-3E-2336, IACA15-BE-3707, EQC2018-004918-P, the Severo Ochoa Programs SEV-2015-0548 and CEX2019-000920-S, the
Maria de Maeztu Program MDM-2017-0765, and by the Consolider-Ingenio project CSD2010-00064 (EPI: Exploring
the Physics of Inflation). We acknowledge support from the ACIISI, Consejeria de Economia, Conocimiento y 
Empleo del Gobierno de Canarias and the European Regional Development Fund (ERDF) under grant with reference ProID2020010108.
This project has received funding from the European Union's Horizon 2020 research and innovation program under
grant agreement number 687312 (RADIOFOREGROUNDS).
DT acknowledges the support from the Chinese Academy of Sciences (CAS) President's International 
Fellowship Initiative (PIFI) with Grant N. 2020PM0042; DT also acknowledges the support from the South African Claude Leon Foundation, that partially funded this work.
EdlH acknowledges partial financial support from the \textit{Concepci\'on Arenal Programme} of the Universidad de Cantabria. 
FG acknowledges funding from the European Research Council (ERC) under the European Union’s Horizon 2020 research and innovation programme (grant agreement No 101001897). 
FP acknowledges the European Commission under the Marie Sklodowska-Curie Actions within the \textit{European Union's Horizon 2020} research and innovation programme under Grant Agreement number 658499 (PolAME). FP acknowledges support from the Spanish State Research Agency (AEI) under grant numbers PID2019-105552RB-C43.
BR-G acknowledges ASI-INFN Agreement 2014-037-R.0.

%% file: tables/reg_contrib.tex
\begin{table*}
\centering
\caption{Evaluation of the region contribution in our apertures at different frequencies, towards W49 and W51. For each case we report the considered survey and frequency, the original map resolution, the  flux density $S_{\rm or}$ from the original maps, the flux density $S_{\rm 1d}$ measured on the 1-degree smoothed maps, and the fractional contribution of the region flux density $\Delta S/S_{\rm 1d}$ to the reference measurement on the smoothed maps. For the measurement of the original resolution fluxes we adopt an aperture radius $r_{\rm ap}=20\,\text{arcmin}$ for W49 and $r_{\rm ap}=30\,\text{arcmin}$ for W51.}
\label{tab:reg_contrib}
\setlength{\tabcolsep}{1.em}
\renewcommand{\arraystretch}{1.1}
\begin{tabular}{ccccccccccc}
\hline
\hline
 \multicolumn{3}{c|}{} & & \multicolumn{3}{c|}{W49} & & \multicolumn{3}{c|}{W51} \\ 
\hline
 Survey & Freq. & Reso. & &  $S_{\rm or}$ & $S_{\rm 1d}$ & $\Delta S/S_{\rm 1d}$ & &  $S_{\rm or}$ & $S_{\rm 1d}$ & $\Delta S/S_{\rm 1d}$ \\ 
\\[-1em]
  & [GHz] & [arcmin] & & [Jy] & [Jy] & [\%] & & [Jy] & [Jy] & [\%] \\ 
\hline
\hline

Urumqi & 4.8 & 9.5 &  & 90.1 & 150.5 & 40.1 &  & 473.6 & 657.9 & 28.0 \\
\textit{Planck} & 100 & 9.7 &  & 90.4 & 220.5 & 59.0 &  & 322.3 & 530.0 & 39.2 \\
\textit{Planck} & 143 & 7.3 &  & 150.9 & 493.8 & 69.4 &  & 431.8 & 957.7 & 54.9 \\
\textit{Planck} & 217 & 5.0 &  & 510.1 & 2021.9 & 74.8 &  & 1244.6 & 3545.3 & 64.9 \\
\textit{Planck} & 353 & 4.9 &  & 2326.0 & 9666.2 & 75.9 &  & 5300.1 & (1.6) $\times 10^{4}$ & 67.4 \\
\textit{Planck} & 545 & 4.8 &  & 8734.7 & (3.6) $\times 10^{4}$ & 75.6 &  & (2.1) $\times 10^{4}$ & (6.1) $\times 10^{4}$ & 65.5 \\
\textit{Planck} & 857 & 4.6 &  & (3.0) $\times 10^{4}$ & (1.2) $\times 10^{5}$ & 74.6 &  & (7.5) $\times 10^{4}$ & (2.1) $\times 10^{5}$ & 63.4 \\

\hline
\hline
\end{tabular}
\end{table*}